\documentclass[aps,prd,preprint,floatfix,superscriptaddress,nofootinbib,longbibliography,a4paper]{revtex4-1}
\usepackage{color}
\usepackage{graphicx}
\usepackage{textcomp}
\usepackage{subfigure}
\usepackage{feynmp}
\usepackage{amsmath, amssymb, array}
\usepackage{enumerate}
\usepackage{hhline}
\usepackage{rotating}
\usepackage{orcidlink} 
\usepackage{multirow}
\usepackage{hyperref}
\hypersetup{
    citecolor= red,
    colorlinks=true,
    linkcolor=blue,
    filecolor=magenta,      
    urlcolor=blue,}

\usepackage{float}
\usepackage{xfrac}
\usepackage{slashed}
\usepackage{ulem}
\newcommand{\nn}{\nonumber}
\newcommand{\beqa}{\begin{eqnarray}}
\newcommand{\eeqa}{\end{eqnarray}}
\newcommand{\be}{\begin{equation}}
\newcommand{\ee}{\end{equation}}
\newcommand{\ba}{\begin{array}} 
\newcommand{\ea}{\end{array}}

\newcommand{\msol}{\Delta m_{\rm sol}}
\newcommand{\mNO}{\mathbf{m}^{\rm Std-NO}_{\beta\beta}}
\newcommand{\matm}{\Delta m_{\rm atm}}
\newcommand{\ms}{\Delta m_{s}}
\newcommand{\ev}{\rm eV}

\newcommand{\mIH}{ \mathbf{m}^{\rm Std-IO}_{\beta\beta}}

\begin{document} 
\vspace*{0.5cm}
\title{Constraining the mass-spectra in the presence of a light sterile neutrino from absolute mass-related observables}
\bigskip
\author{Srubabati Goswami\orcidlink{0000-0002-5614-4092}}
\email{sruba@prl.res.in}
\affiliation{Physical Research Laboratory, Ahmedabad, Gujarat, 380009, India}
\affiliation{Northwestern University, Department of Physics and Astronomy, Evanston, IL 60208, USA}
\author{Debashis Pachhar\orcidlink{0000-0001-8931-5321}}
\email{debashispachhar@prl.res.in}
\affiliation{Physical Research Laboratory, Ahmedabad, Gujarat, 380009, India}
\affiliation{Indian Institute of Technology Gandhinagar, Gujarat, 382355, India}
\author{Supriya Pan\orcidlink{0000-0003-3556-8619}}
\email{supriyapan@prl.res.in}
\affiliation{Physical Research Laboratory, Ahmedabad, Gujarat, 380009, India}

\begin{abstract}
The framework of three-flavor neutrino oscillation is a well-established phenomenon, but results from the short-baseline experiments, such as the \textit{Liquid Scintillator Neutrino Detector} (\textit{LSND}) and \textit{MiniBooster Neutrino Experiment (MiniBooNE)}, hint at the potential
existence of an additional light neutrino state characterized by a mass-squared difference of approximately $1\,\rm eV^2$. The new neutrino state is devoid of all Standard Model (SM) interactions, commonly referred to as a “sterile” state. In addition, a sterile neutrino with a
mass-squared difference of $10^{-2}$ $\rm eV^2$ has been proposed to improve the tension between the results obtained from the \textit{Tokai to Kamioka (T2K)} and the \textit{NuMI Off-axis $\nu_e$ Appearance (NO$\nu$A)} experiments. Further, the non-observation of the predicted upturn in the solar neutrino spectra below 8 MeV can be explained by postulating an extra light sterile neutrino state with a mass-squared difference around $10^{-5}\, \rm eV^2$. The hypothesis of an additional light sterile neutrino state introduces four distinct mass spectra depending on the sign of the mass-squared difference.  In this paper, we discuss the implications of the above scenarios on the observables that depend on the absolute mass of the neutrinos, namely - the sum of the light neutrino masses $(\Sigma)$ from cosmology, the effective mass of the electron neutrino from beta decay $(m_{\beta})$, and the effective Majorana mass $( m_{\beta\beta})$ from neutrinoless double beta decay. We show that some scenarios can be disfavored by the current constraints of the above variables. The implications for projected sensitivity of \textit{Karlsruhe Tritium Neutrino Experiment (KATRIN)} and future experiments like \textit{Project-8, next Enriched Xenon Observatory (nEXO)} have been discussed. 
\end{abstract}

\maketitle
\section{Introduction}
\noindent The phenomena of neutrino oscillations, in which neutrino flavor states switch their identities while propagating, have been observed in several terrestrial experiments\cite{Super-Kamiokande:1998kpq, SNO:2002tuh, KamLAND:2002uet, MINOS:2006foh}. This requires at least two of the neutrinos to have small but non-zero masses and mixing between the different flavors. This, in turn, implies physics beyond the Standard Model (SM). Many BSM scenarios have been studied for generating neutrino masses. The smallness of the neutrino masses is often linked with lepton number violation through the dimension 5 Weinberg operator $\frac{LLHH}{\Lambda}$ \cite{Weinberg:1979sa}. This operator violates the lepton number, which signifies the Majorana nature of the neutrinos. 


\noindent Neutrino oscillation experiments are sensitive to two mass-squared differences and mixing angles of the neutrinos. However, they cannot shed light on the absolute mass scale or the nature of neutrinos. If neutrinos are considered to be Majorana in nature, a rare and slow nuclear decay, known as neutrinoless double beta decay ($0\nu\beta\beta$) \cite{Furry:1939qr}, can exist in nature. Several experiments aimed to observe this process, but there has not been any positive evidence so far. The \textit{KamLAND-Zen} experiment using the Xe$^{136}$ isotope as the decaying nucleus gives the lower bound on the half-live as $T_{1/2}^{0\nu\beta\beta} > 1.07 \times 10^{26}$ yr at 90\% confidence level \cite{Shirai:2018ycl} whereas the \textit{GERDA} experiment uses Ge$^{76}$ isotope and their latest limit on the half-life is  $T_{1/2}^{0\nu\beta\beta} > 1.8 \times 10^{26}$ yr at 90\% confidence level \cite{GERDA}. The lower bounds on half-lives can be translated into upper bounds on the effective Majorana mass parameter ($m_{\beta\beta} $), which depends on the neutrino masses, mixing angles, and the Majorana phases.


The information about the absolute mass scale of neutrinos can also come from tritium beta decay. 
The \textit{KATRIN} experiment sets the current limit on the mass parameter, $m_{\beta} \lesssim 0.8 \, \rm eV$ at 90\% confidence level \cite{KATRIN:2021uub}. 

Cosmological observations like CMB anisotropies, large-scale structure formation, etc., can also put bound on the absolute mass scale of neutrinos. The most stringent bound on the sum of the light neutrino masses ($\Sigma$) $<\, 0.12 \rm \, eV$ comes from the Planck collaboration by considering three degenerate neutrino mass eigenstates \cite{Planck:2018vyg}. 

Although the three-generation paradigm is well established, there are experimental anomalies that indicate the presence of an extra light sterile neutrino of mass of the order of eV. The short baseline experiments, \textit{LSND}\cite{LSND:2001aii} and \textit{MiniBooNE} \cite{MiniBooNE:2020pnu}, showed an excess signature of electron neutrinos coming from a muon neutrino beam. Gallium-based solar neutrino experiments \textit{GALLEX}\cite{GALLEX:1997lja}, \textit{SAGE} \cite{Abdurashitov:1996dp}, \& as well as the \textit{BEST}\cite{Barinov:2021asz} experiments found the deficit in electron neutrinos while calibrating the detector using the neutrinos from $^{51}Cr$ source. One possible resolution of the results from these experiments is provided by incorporating an additional light neutrino state with mass  $\sim 1$ \rm eV. 
There are also motivations for considering sterile neutrinos lower than the eV scale. The inclusion of a sterile neutrino with mass squared difference ($\ms^2$ )$\sim 10^{-5}\, \rm eV^2$ has been postulated \cite{deHolanda:2010am} to explain the absence of the upturn of solar neutrino probability below 8 MeV. Additionally, it is also shown that the tension between $NO\nu A$ and $T2K$ data can be reduced in the presence of a sterile neutrino with $\ms^2 \sim ( 10^{-4}:10^{-2})$ eV$^2$. Recently, the signatures of the sub-eV sterile neutrinos in future experiments have been studied in the references \cite{KumarAgarwalla:2019blx, Agarwalla:2018nlx, Chatterjee:2023qyr, Chatterjee:2022pqg} in the context of future long baseline atmospheric neutrino experiments.

In this paper, we study the implication of a very light sterile neutrino with $\ms ^2$ in the range ($10^{-4}:10^{-2}$) eV$^2$ on the mass-related variables such as $m_{\beta\beta}, m_{\beta},$ and $\Sigma$. Such investigations in the context of an eV scale sterile neutrino have been explored in \cite{Goswami:2005ng}. In our work, along with the sub-eV scale sterile neutrino we also present the results for an eV scale sterile neutrino with the current constraints on the mixing the parameters. We consider the 3+1 picture with a single sterile neutrino added to the three sequential neutrinos. In this case, there can be four mass possible spectra; two each with $\ms^2>0$ and $\ms^2<0$. We explore the implication of the cosmological constraint on the sum of light neutrino masses for these spectra. We also discuss the constraints on the possible mass spectra in the light of \textit{KATRIN} results on $m_\beta$ and \textit{KamLAND-Zen} results on $m_{\beta\beta}$. Additionally, we examine the implications of the future measurements by proposed experiments \textit{Project8, nEXO}.

The plan of the paper is as follows. Section 2 gives a brief overview of the neutrino mass and mixing scenarios in the standard three-generation and 3+1 framework. In section 3, we study the implications of the various mass spectra for $\Sigma$, $m_{\beta\beta}$, and $m_{\beta},$. Section 4 presents an analysis on the correlation between $m_{\beta\beta}, m_{\beta},$ and $\Sigma$. Finally, we summarize the results in section 5.

\section{Neutrino Masses and Mixing}\label{sec:mass-mixing}

\subsection{The Standard framework}

Neutrino oscillation is governed by the Pontecorvo-Maki-Nakagawa-Sakata (PMNS) matrix (U), which describes the relationship between the neutrino flavor and mass eigenstates \cite{Maki:1962mu}. The mass matrix in the flavour basis $\mathcal{M}_{\nu}$ and the mass matrix in the mass basis $M_{\nu}^{\rm diag} $ are related as,
\begin{eqnarray}
    \mathcal{M}_{\nu} &=& U \, M_{\nu}^{\rm diag} \, U^{T} ,\\
    \text{where } \quad M_{\nu}^{\rm diag} &=& \rm{diag}\, \left(m_1, m_2, m_3 \right) \, ,
\end{eqnarray}
The PMNS matrix is parameterized by three mixing angles $(\theta_{12}\, , \, \theta_{13}\, , \, \theta_{23})$ and one CP Phase $(\delta_{13})$ for Dirac neutrinos, whereas Majorana nature of neutrino adds two extra phases $(\alpha, \, \beta)$ along with it. Various oscillation experiments provide information about the mixing angles ($\theta_{12},\theta_{13},\theta_{23}$) and mass-squared differences ($\msol^2\, ,\,\matm^2 $). Here $\msol^2 \, > \, 0$ and defined as $m_2^2 -m_1^2 $.
Depending on the sign of $\matm^2 $, the masses in the three flavor framework are categorized into two mass orderings 

\begin{itemize}
    \item Normal Ordering (NO): In NO, $\matm^2 \equiv m_3^2 -m_1^2 > 0$. The mass ordering in this scenario is $m_1 <m_2<m_3$, and the mass relations can be expressed as 
    \beqa
    m_{\rm lightest } &=& m_1, \quad m_2 = \sqrt{m_1^2 + \msol^2}, \quad m_3 = \sqrt{m_1^2 + \matm^2} \label{eeq:std_NO}
    \eeqa
    \item Inverted Ordering (IO): In this case,  the mass ordering is $m_3 <m_1<m_2$ and $\matm^2 \equiv m_3^2-m_2^2 \,< \, 0$. In this ordering, the mass relation ns are written as, 
    \beqa
        m_{\rm lightest} &=& m_3, \quad m_2= \sqrt{m_3^2 + \matm^2}, \quad m_1 = \sqrt{m_3^2 +\matm^2 - \msol^2}\label{eq:std_IO}
    \eeqa
    \item Quasi Degenerate Spectrum (QD): Apart from NO and IO, there might be a scenario where $m_1 \approx m_2 \approx m_3 $. This scenario is generally referred to as quasi degenerate spectrum. In this scenario, the value of the lightest mass is greater than $\sqrt{\matm^2}$.
    \end{itemize}
\noindent The current best fit and $3\sigma$ range of these parameters, determined from various experiments, are given in table (\ref{tab:osc}).

\begin{table}[H]
        \caption{$3\sigma$ ranges and best fit values extracted of three neutrino oscillation parameters \cite{Esteban:2020cvm}. Here, $ \msol^2 \, \equiv \, m_2^2 -m_1^2$ and $\matm^2 \, \equiv \, m_3^2 - m_1^2 $ for NO and $ m_2^2 - m_3^2 $ for IO.  }
        \label{tab:osc}
        \centering
        \begin{tabular}{|c||c|c||c|c|}
        \hline Parameters & \multicolumn{2}{c||}{Normal Ordering} & \multicolumn{2}{c|}{Inverted Ordering}\\
         \cline{2-5}  & $3\sigma$ range & Best Fit &  $3\sigma$ range & Best Fit \\ \hline  \hline $\sin^2 \theta_{12}$ & $0.270 : 0.341$ & 0.303 & $0.270 : 0.341$ & 0.303\\ $\theta_{12}$ & $31.31^\circ:35.74^\circ$ & $33.41^\circ$ & $31.31^\circ:35.74^\circ$ & $33.41^\circ$ \\\hline $\sin^2 \theta_{13}$ & $0.0202:0.0239$ & 0.0220 & $0.0202:0.0239$ & 0.0220\\ $\theta_{13}$ & $8.19^\circ - 8.89^\circ$ & $8.54^\circ$ & $8.23^\circ : 8.90^\circ$ & $8.57^\circ$\\ \hline
         $\sin^2 \theta_{23}$ & $0.406:0.620$ & 0.572 & $0.412:0.623$ & 0.578\\ $\theta_{23}$ & $39.6^\circ :51.9^\circ$ & $49.1^\circ$ & $39.9^\circ : 52.1^\circ$ & $49.5^\circ$\\ \hline
         $\delta_{13}$  & $197^\circ$ & $108^\circ:404^\circ$ & $286^\circ$ & $192^\circ:360^\circ$\\ \hline
         $\Delta m^2_{\rm sol}/10^{-5}{\rm eV}^2$ & $6.82 : 8.03$ & $7.41$ & $6.82 : 8.03$ & $7.41$ \\ \hline
         $\Delta m^2_{\rm{atm}}/10^{-3}{\rm eV}^2$ & $2.428 : 2.597$ & $2.511$ & $(-2.581 : -2.408)$ & $ -2.498$ \\ \hline
        \end{tabular}
\end{table}

\subsection{The 3+1 framework}
\noindent In this case, we have one extra mass-squared difference ($\Delta m_s^2 \equiv m_{4}^2-m_1^2$), three new mixing angles $(\theta_{14}\, , \,\theta_{24}\, \,\theta_{34})$ and two new Dirac CP phases $(\delta_{14}\, , \,\delta_{24}) $ and one additional Majorana phase $(\gamma)$. The mass matrix in the flavor basis can be defined as,
\beqa
\mathcal{M}^{s}_{\nu} &=& U \, M_{\nu}^{\rm diag} \, U^{T} ,\quad  \text{where } \quad M_{\nu}^{\rm diag} = \rm{diag}\,\left( m_1,m_2,m_3,m_4\right)
\eeqa




\begin{figure}[H]
    \centering
    \includegraphics[width=0.87\linewidth]{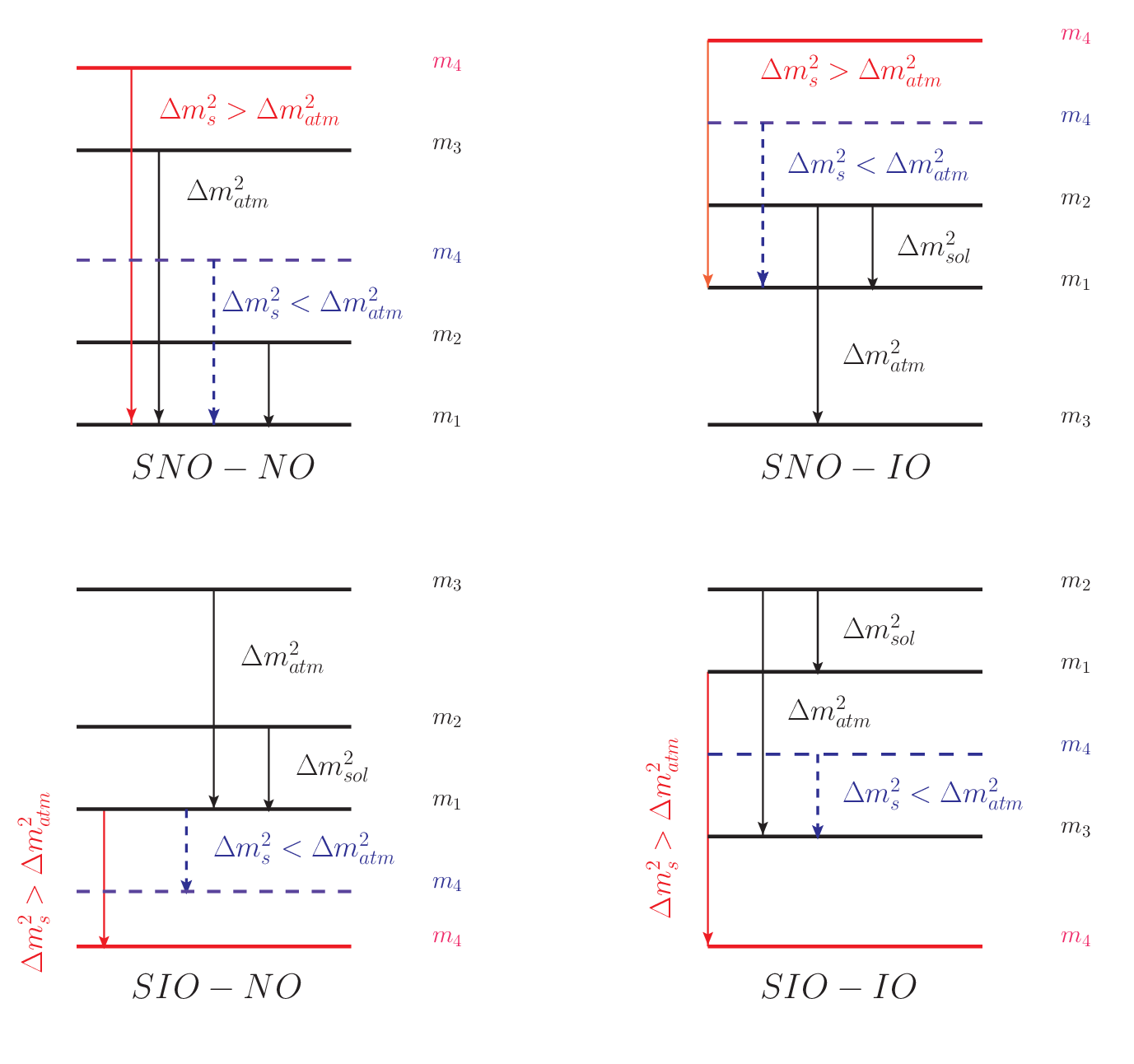}
    \caption{Possible mass spectra with the inclusion of a sterile neutrino. Here red solid line corresponds to the value of $m_4$ when $\ms^2 > \matm^2 $ whereas the blue dashed line indicates the same with $\ms^2 < \matm^2$}
    \label{fig:mass-spectrum}
\end{figure}

In the 3+1 framework, the mixing matrix $U$ can be parameterized as,
\beqa
U &=& R_{34}(\theta_{34})\,\Tilde{R}_{24}(\theta_{24},\delta_{24})\,\Tilde{R}_{14}(\theta_{14},\theta_{14})\, R_{23}(\theta_{23})\, \Tilde{R}_{13}(\theta_{13},\delta_{13}) \, R_{12}(\theta_{12})\, P \nn \\&=& \begin{pmatrix} 
U_{e1} & U_{e2} & U_{e3} & U_{e4} \\
U_{\mu1} & U_{\mu2} & U_{\mu3} & U_{\mu4} \\
U_{\tau1} & U_{\tau2} & U_{\tau3} & U_{\tau4}\\
U_{s1} & U_{s2} & U_{s3} & U_{s4}
\end{pmatrix}
\eeqa
where $R_{ij}$'s are the standard rotational matrices in the $i,j$ generational space. For instance 
\beqa
R_{34}(\theta_{34}) &=& \begin{pmatrix} 
1 & 0 & 0 & 0 \\
0 & 1 & 0 & 0 \\
0 & 0 & c_{34} & s_{34}\\
0 & 0 &-s_{34} & c_{34}
\end{pmatrix},
 \quad 
\Tilde{ R}_{14}(\theta_{14},\delta_{14}) = \begin{pmatrix} 
c_{14} & 0 & 0 & s_{14} \, e^{-i \, \delta_{14}} \\
0 & 1 & 0 & 0 \\
0 & 0 & 1 & 0\\
-s_{14}\, e^{i \, \delta_{14}} & 0 & 0 & c_{14}
\end{pmatrix}
\eeqa

\begin{table}[H]
   \caption{Allowed values of the sterile neutrino parameters $\Delta m^2_s , \, \sin^2\theta_{14}$ in the 3+1 scenario for three different mass squared differences ($\ms^2 = 10^{-4} \, \ev^2\, , \,  0.01\,   \ev^2 $ and $1.3 \, \ev^2$) are given. The value of the $\sin^2 \theta_{14}$ is chosen to be consistent with \textit{MINOS , MINOS$^{+}$,  Daya-Bay and Bugey-3} data \cite{MINOS:2017cae} }
    \label{tab:cosmo_table}
    \centering
    \begin{tabular}{|c|c|c|c|}
    \hline 
        Parameters & Case I & Case II & Case III\\
        \hline 
        $\Delta m_s^2$ &   $ 10^{-4} \,\rm eV^2$ & $10^{-2} \,\rm eV^2$  & $1.3\,\rm eV^2$ \\
        \hline
        $\sin ^2 \theta_{14} $  & $ 0.1\,:\,0.2 $ & $5\times 10^{-4}\,:\,5\times 10^{-3}$ & $0.001\, : \,0.01$\\
        \hline
    \end{tabular}
\end{table}

\noindent Here, $c_{ij} (s_{ij})$ stands for $\cos\theta_{ij} \,(\sin\theta_{ij})$ and $P$ is the diagonal matrix containing the Majorana phases, defined as $P=\text{diag}\, \left( 1,\, e^{i\frac{\alpha}{2}},\, e^{ i\left( \frac{\beta}{2} + \delta_{13}\right)},\, e^{ i\left( \frac{\gamma }{2} + \delta_{14} \right)} \right)$. 
In table (\ref{tab:cosmo_table}), we present three representative values of $\ms^2$ and $\sin^2 \theta_{14}$ extracted from the allowed region from \textit{MINOS , MINOS$^+$,  Daya-Bay\,  and Bugey-3} experiments\cite{MINOS:2017cae,Acero:2022wqg}.  
The value of $\sin^2\theta_{14}$ analysing the \textit{LSND and MiniBooNE} data is in the range $ (0.01:0.02)$ for $\ms^2 $ = 1.3 $\ev^2$,  whereas the \textit{MINOS , MINOS$^{+}$} data allows the region with $\sin^2\theta_{14} $ is  $<0.01$.


In the 3+1 framework, the sign and the magnitude of $\ms^2$ lead to different mass spectra. 


\begin{enumerate}
\item {\textbf{SNO-NO $ \bf  (\ms^2>0 \, , \, \matm^2 >0 \, )$:}}\\
In this scenario, mass ordering is different for $\ms^2 > \matm^2$ and $\ms^2 < \matm^2$ which is depicted in the top left corner of Fig. (\ref{fig:mass-spectrum}) with a red solid line and a blue dashed line respectively. For $\ms^2 > \matm^2$, the mass ordering is $m_1 < m_2 <m_3 < m_4 $, given in the top left corner of Fig. (\ref{fig:mass-spectrum}). Whereas for $\ms^2 < \matm^2$, the ordering is $m_1 <m_2 <m_4<m_3$. In both cases, the mass relations are expressed as 
    \beqa
m_{\text{lightest}} =m_1, \quad \quad  m_2 &= &\sqrt{m_1^2 + \msol^2} \nn \\
m_{3} =\sqrt{m_1^2 +\matm ^2}, \quad \quad m_4 &= &\sqrt{m_1^2 + \ms ^2} \label{eq:mrel-snono}
\eeqa

\item \textbf{SNO-IO $ \bf  (\ms^2>0 \, , \, \matm^2 < 0 \,) $ : }\\
In this case, the mass ordering is the same for both $\ms^2 > \matm^2$ and $ \ms^2 < \matm^2$ and is delineated as $m_3 < m_1 < m_2 < m _4 $. The mass relations are expressed as 
    \beqa
        m_{\text{lightest}} =m_3, \quad \quad  m_2 &= &\sqrt{m_3^2 + \matm ^2 } \nn \\
        m_{1} =\sqrt{m_3^2 +\matm ^2+ \msol ^2}, \quad \quad m_4 &= &\sqrt{m_3^2 + \matm ^2- \msol ^2+ \ms ^2} \label{eq:mrel-snoio}
    \eeqa

\item  \textbf{SIO-NO $ \bf  (\ms^2 < 0 \, , \, \matm^2 > 0 \, )$ : } \\
The mass ordering in this scenario is defined as $m_4 < m_1 < m_2 < m_3$, and it is the same for both the $\ms^2$ ranges. The mass relations can be written as, 
\beqa
    m_{\text{lightest}} =m_4, \quad \quad  m_2 &= &\sqrt{m_4^2 +\ms ^2 + \msol ^2} \nn \\
    m_{1} =\sqrt{m_4^2 +\ms ^2}, \quad \quad m_3 &= &\sqrt{m_4^2 +\ms ^2 +\matm ^2} \label{eq:m_siono}
\eeqa 

\item \textbf{SIO-IO $ \bf ( \ms^2 < 0 \, , \, \matm^2 < 0 \, )$ : }\\

\begin{itemize}
    \item For $\ms^2 > \matm^2 $, the mass ordering is $m_4 < m_3 < m_1 <m_2$ and the mass relations are defined as :
    \beqa
m_{\text{lightest}} =m_4, \quad \quad  m_2 &= &\sqrt{m_4^2 +\ms ^2 + \msol ^2} \nn \\
m_{1} =\sqrt{m_4^2 +\ms ^2}, \quad \quad m_3 &= &\sqrt{m_4^2 +\ms ^2+ \msol ^2 -\matm ^2} \label{eq:m_rel_sioio1}
\eeqa
\item For $\ms^2 < \matm^2 $, the mass ordering is $m_3 < m_4 < m_1 <m_2$ and the mass relations can be expressed as :
\beqa
m_{\text{lightest}} =m_3, \quad \quad  m_2 &= &\sqrt{m_3^2 +\matm^2} \nn \\
m_{1} =\sqrt{m_3^2 +\matm^2-\msol^2}, \quad \quad m_4 &= &\sqrt{m_3^2 +\matm^2-\msol^2-\ms^2} \nn \\ \label{eq:m_rel_sioio2}
\eeqa
In the appendix, we have given the variation of masses ($m_i$) with respect to the lightest mass for all the scenarios.
    
\end{itemize}

\end{enumerate}

\section{Neutrino mass variables}

In this section, we study the implications of adding an additional sterile neutrino for the mass variables $m_{\beta\beta}, m_\beta, \Sigma$.

\subsection{Bound from cosmology }

   Light sterile neutrinos can have a significant impact on the evolution of the universe, and thus, their presence can be investigated using cosmological observations.  If sterile neutrinos are massless, they contribute to the light relativistic degrees of freedom in the early universe, quantified as $N_{\rm eff}$, which can be directly constrained from Cosmic Microwave Background (CMB) and Large Scale Structure (LSS) data. The Standard Model of particle physics predicts $N^{\rm SM}_{\rm eff} = 3.044^{+0.0002}_{-0.0002}$ \cite{Bennett:2020zkv}, assuming only three degenerate light active neutrinos, but can increase in general when the sterile neutrino contribution is added\footnote{ However, $N_{\rm eff}$ can be decreased in certain scenarios like very low-reheating in sterile neutrinos \cite{Yaguna:2007wi,Abazajian:2017tcc} or self-interacting sterile neutrinos \cite{Dasgupta:2013zpn,Chu:2015ipa}}. 
   
   \begin{figure}[H]
    \centering
    \includegraphics[width=0.48\linewidth]{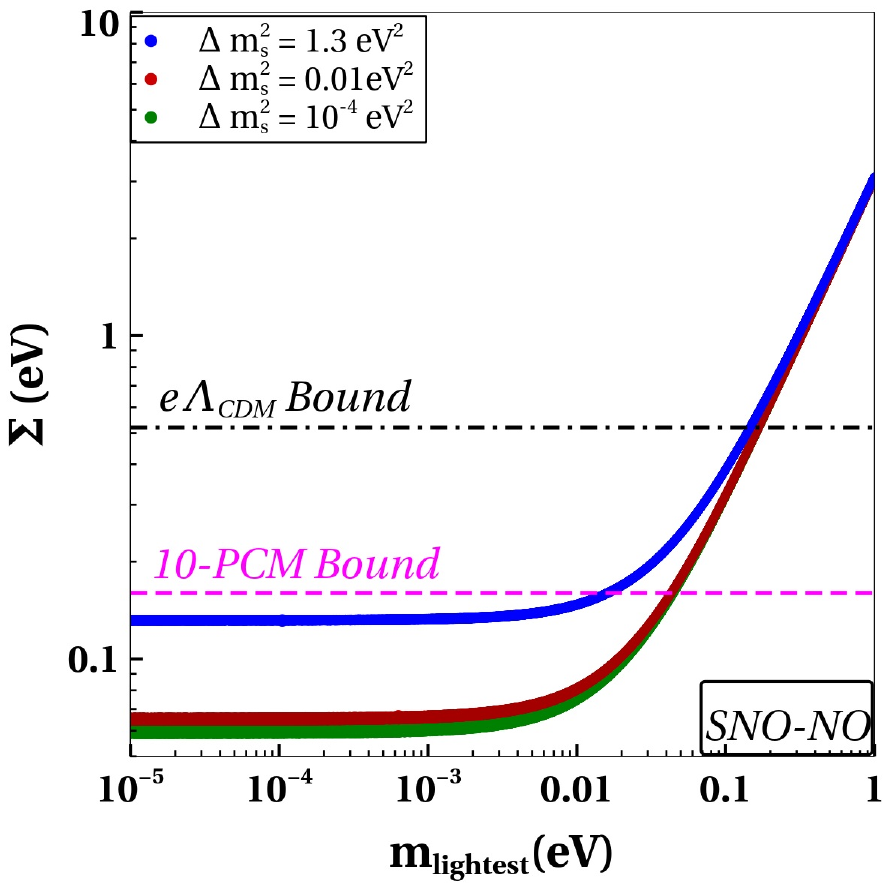}
    \includegraphics[width=0.48\linewidth]{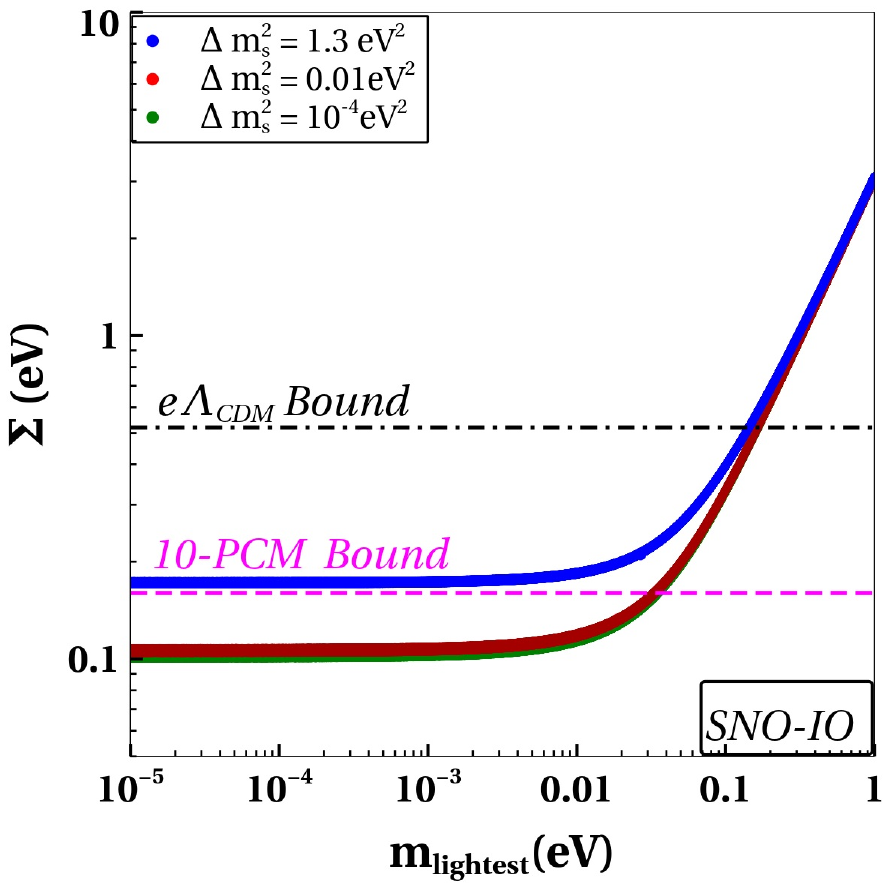}
    \includegraphics[width=0.48\linewidth]{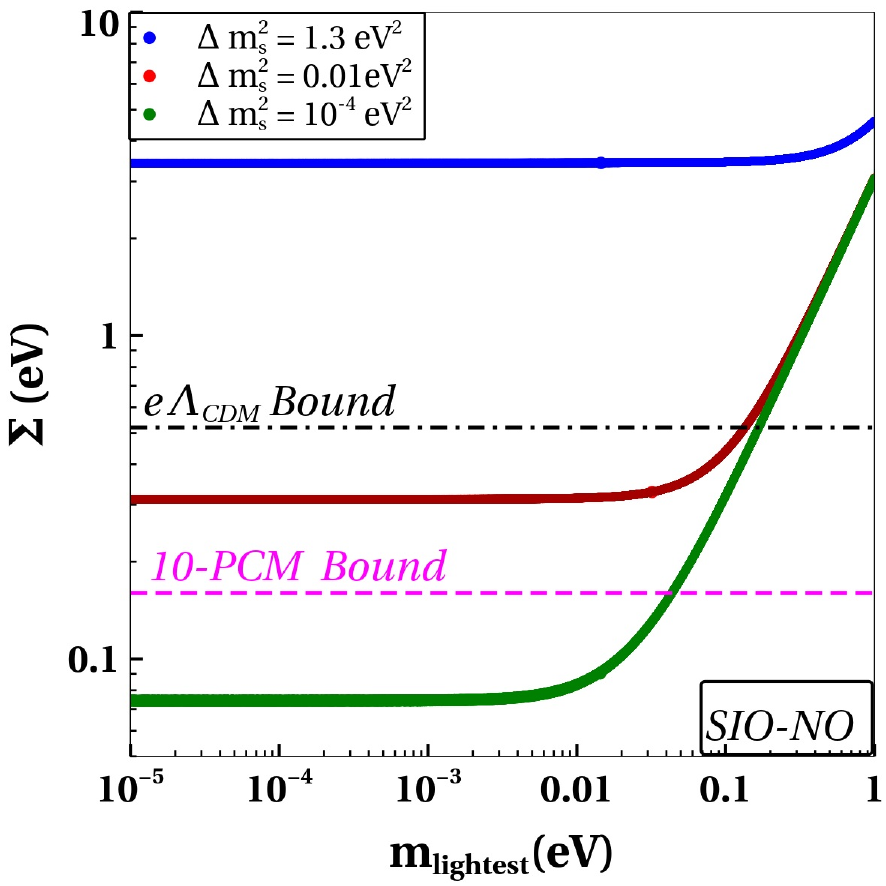}
     \includegraphics[width=0.48\linewidth]{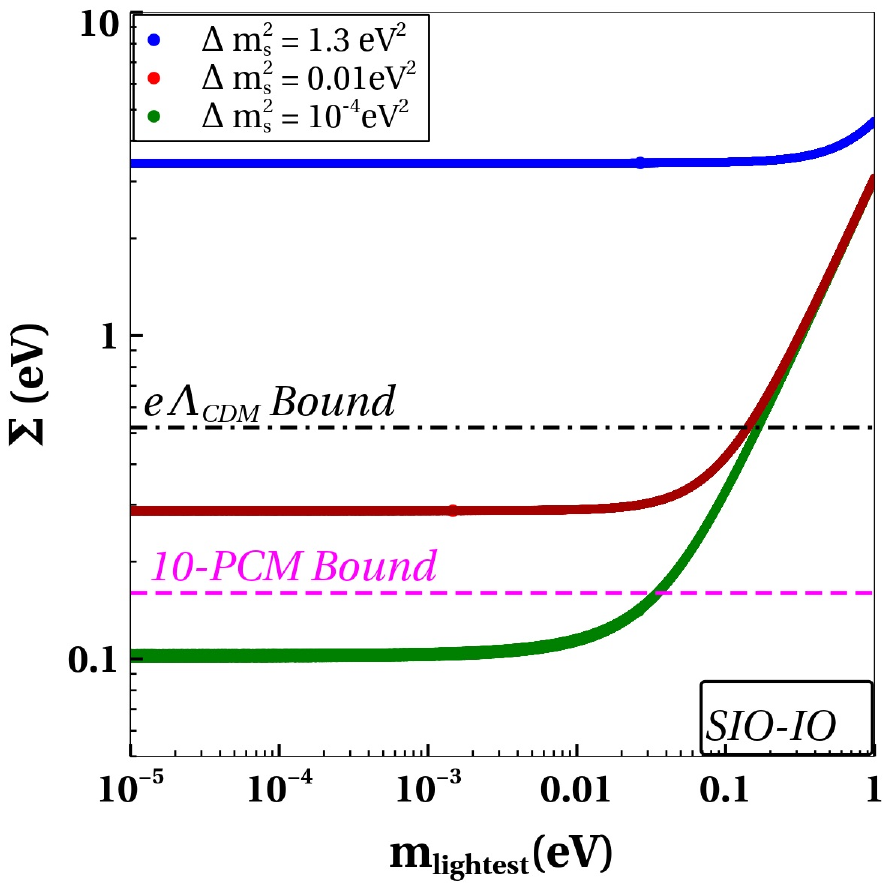}
    \caption{The variation total effective mass $\Sigma$ with the lightest neutrino mass $m_{\rm lightest}$ in different scenarios SNO-NO (top-left), SNO-IO (top-right), SIO-NO (bottom-left), and SIO-IO (bottom-right). The green, red, and blue colors correspond to $\ms^2=10^{-4}\, \rm eV^{2},0.01\, eV^{2},1.3$ eV$^2$, respectively. The magenta dashed line corresponds to the 10 parameter cosmological (\textit{10-PCM}) model and the black dashed-dot line corresponds to the extended $\Lambda_{CDM}$ ($e\Lambda_{CDM}$)bound.}
    \label{fig:mtotal}
\end{figure}

   In the case of massive sterile neutrinos, one needs to add one more free parameter, $m_{s}^{\rm eff.}$, the effective sterile neutrino mass in the cosmological models along with $N_{\rm eff} $. The effective sterile neutrino mass is different from its physical mass ($m^{\rm ph}_s$) but can be related as   $m_{s}^{\rm eff} \, = \, \Delta N_{\rm eff}^{3/4} \, m_s^{\rm ph}$ if the neutrinos are fully thermalized with active neutrinos and $m_{s}^{\rm eff } \, = \, \Delta N_{\rm eff} \, m_s^{\rm ph}$ for the partially thermalized sterile neutrinos where $\Delta N_{\rm eff}  = N_{\rm eff} - N_{\rm eff }^{\rm SM}$.

   When PLANK 2018 data is fitted with standard $\Lambda_{\rm CDM}$ cosmological model, it tends to disfavor the presence of extra light relativistic degrees of freedom \cite{Planck:2018vyg}. 
   However, with the inclusion of more parameters with the standard $\Lambda_{\rm CDM}$ cosmological model and fitting more data from different cosmological observations, the cosmological constraints can be relaxed. 
   For example, in a recent analysis, the Plank + BAO + Hubble parameter measurement \cite{Riess:2018uxu} + Supernova Ia \cite{Pan-STARRS1:2017jku} data fitted with a 10 parameter cosmological model (\textit{10-PCM}) i.e $\Lambda_{CDM}+N_{\rm eff} + m^s_{\rm eff} +w_{0} +n_{run}$, gives the constraints on $N_{\rm eff}$ and $\Sigma$ as follows \cite{Acero:2022wqg},
    \beqa
        N_{\rm  eff} = 3.11^{+0.37}_{-0.36}, \,\,\,
        \Sigma = 0.16 \, \rm eV \label{eq:cosmo2}
    \eeqa
    where $\omega_{0}$ is the equation of state parameter of the
dark energy and $n_{run}$ is the running of the scalar spectral index, a parameter related to the initial conditions of the universe.
   Another model with 12 parameters, called extended $\Lambda_{\rm CDM}$ ($e\Lambda_{\rm CDM}$) gives bound as ,
   \beqa
   N_{\rm eff} = 3.11^{+0.52}_{-0.48}, \,\,\,
   \Sigma = 0.52 \rm \, eV \label{eq:cosmo3}
   \eeqa
where, $\Sigma$ is defined as \cite{Hagstotz:2020ukm} 
    \beqa
    \Sigma = m_1 + m_2 + m_3 + m_s^{\rm eff}
    \eeqa

    A fully thermalized neutrino implies $\Delta N_{\rm eff} \approx 1$, which is ruled out from the cosmological data. Here, we have considered the sterile neutrino to be produced non-thermally which means that $m_{s}^{\rm eff} = \Delta N_{\rm eff} \, m_4$, $m_4$ is the physical mass of the sterile neutrino. 
    We have plotted $ \Sigma$ as a function of the lightest neutrino mass for different mass schemes in Fig. (\ref{fig:mtotal}) assuming the value of $N_{\rm eff} = 3.11$ from Eqn. (\ref{eq:cosmo2}), (\ref{eq:cosmo3}). The pink dashed line indicates the limit $\Sigma = \, 0.16 \, \rm eV$, and the black dashed-dot line corresponds to $\Sigma = 0.52 \, \rm eV$. 
    
 The important features observed from Fig. (\ref{fig:mtotal}) are as follows,
 \begin{itemize}
     \item The SNO-NO scenario is favored by $e\Lambda_{\rm CDM}$ model up to  $m_{\rm lightest} \sim 0.15$ eV for all the three mass-squared differences. Whereas the \textit{10-PCM} is more constraining and disfavor $\ms^2 = 1.3 \, \ev^2$ above $m_{\rm lightest}> 0.01\, \ev$  and $\ms^2= 10^{-4} \, \ev^2 \,,\, 0.01 \, \ev^2$ above $m_{\rm lightest} > 0.04 \, \ev$.

     \item For SNO-IO, $e\Lambda_{\rm CDM}$ model allows all values of $\ms^2$ up to  $m_{\rm lightest} \sim 0.15$ eV. However,  $\ms^2=1.3 $ eV$^2$ is disfavored by the \textit{10-PCM} for the entire range of $m_{\rm lightest}$. The lower values of $\ms^2$ are still allowed up to $m_{\rm lightest} \sim 0.04 \, \ev$.

     \item For SIO-NO and SIO-IO, the \textit{10-PCM} disfavors $\ms^2 = 0.01 \, \rm eV^2 $ and $1.3\, \rm eV^2$ for the entire range of $m_{\rm lightest}$ but $\ms^2=10^{-4} \, \rm eV^2 $ is still allowed up to $m_{\rm lightest} \sim 0.03\, \ev$. However, if we consider $e\Lambda_{\rm CDM}$ model, then $\ms^2= 0.01\, \ev^2$ gets allowed up to $m_{\rm lightest}\sim 0.1\, \ev$.
 \end{itemize}
The above discussion is summarised in table (\ref{tab:cosmo-table}). 

    \begin{table}[H]
 \caption{The table summarises the status of four mass spectra for three different $\ms^2$ in the light of different cosmological models. The limits correspond to the value of $m_{\rm lightest}$ up to which the scenario is allowed.}
         \label{tab:cosmo-table}
        \centering 
        \begin{tabular}{|p{25mm}||p{20mm}|p{20mm}||p{20mm}|p{20mm}||p{20mm}|p{20mm}|}
        \hline Mass ordering ($m_{\rm lightest})$ & \multicolumn{2}{c|}{ $\ms ^2 =10^{-4} \, \rm eV^2$} & \multicolumn{2}{c|}{ $\ms ^2 =0.01 \, \rm eV^2$}  & \multicolumn{2}{c|}{ $\ms ^2 =1.3 \, \rm eV^2$}  \\ 
        \cline{2-7} & Limit $\quad10-\rm PCM$ & Limit e$\Lambda_{\rm CDM}$ & Limit $\quad10-\rm PCM$ & Limit $e\Lambda_{\rm CDM}$ & Limit $\quad10-\rm PCM$ & Limit e$\Lambda_{\rm CDM}$ \\ \hline \hline
        SNO-NO ($m_1$) & $< 0.04$ & $< 0.15$ & $< 0.04$ & $< 0.15$ & $<0.01$  & $<0.15$ \\
        \hline
        SNO-IO ($m_3$) & $<0.03$ & $< 0.1$ & $<0.03 $ & $< 0.1$ & Disallowed  & $<0.1$ \\
        \hline
        SIO-NO ($m_4$)& $<0.04$  & $<0.1$   & Disallowed  & $<0.1$  & Disallowed & Disallowed  \\
        \hline
        SIO-IO ($m_3/m_4$) &$ < 0.04$ & $  <0.1$  & Disallowed & $ <0.1$   & Disallowed  & Disallowed  \\
        \hline
        \end{tabular}
       
\end{table}


\subsection{Bound from Tritium $\beta$ decay } 
A direct and model-independent constraint on the neutrino mass can be derived through the experimental analysis of the electron energy spectrum resulting from beta decay in atomic nuclei. In beta decay, the energy excess due to the nuclear mass difference is shared among the electron, (anti)neutrino and the daughter nucleus. If the energy resolution of the experiment exceeds the splittings of the neutrino mass states ($\Delta E >> m_i$) then the emitted electron's spectrum depends on a quantity called the ``kinematic mass" of the electron neutrino which is defined as
\beqa
m_{\beta} 
&=& \sqrt{ \left|U_{e1}\right|^2 m_1^2 +\left|U_{e2}\right|^2 m_2^2 +\left|U_{e3}\right|^2 m_3^2+\left|U_{e4}\right|^2 m_4^2} \nn \\
&=& \sqrt{c_{12}^2 c_{13}^2 c_{14}^2 m_{1}^2 + s_{12}^2 c_{13}^2 c_{14}^2 m_{2}^2 + s_{13}^2 c_{14}^2 m_{3}^2 + s_{14}^2 m_{4}^2} 
\label{eq:m_beta}
\eeqa

\begin{figure}[H]
    \centering
    \includegraphics[width=0.48\linewidth]{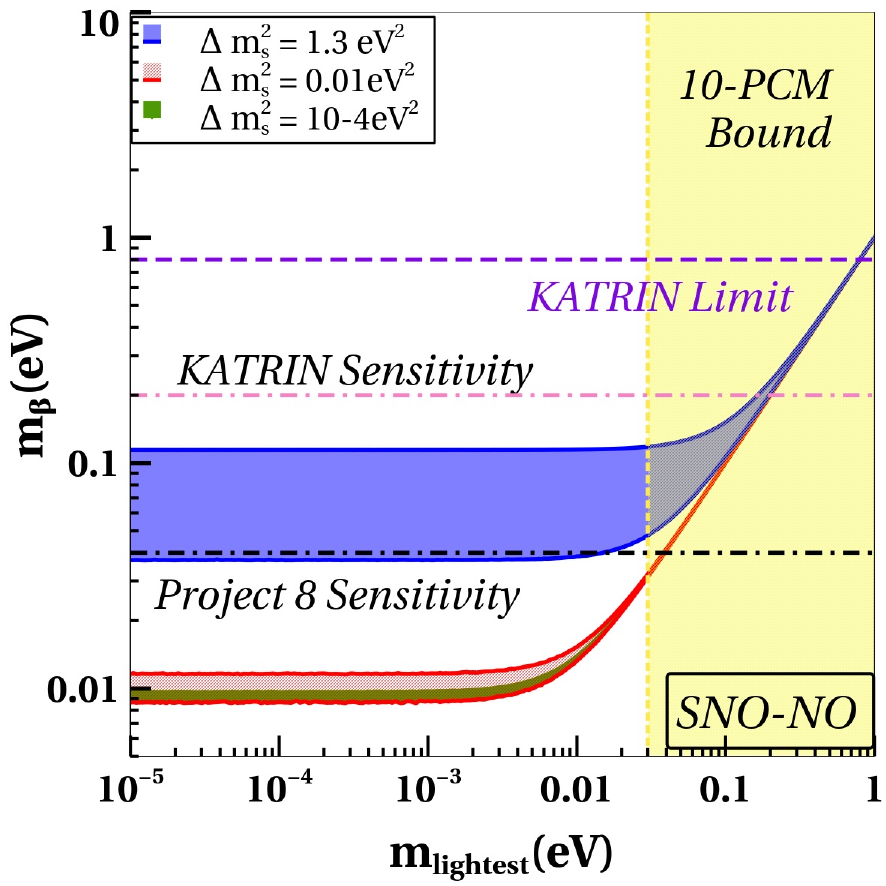}
    \includegraphics[width=0.48\linewidth]{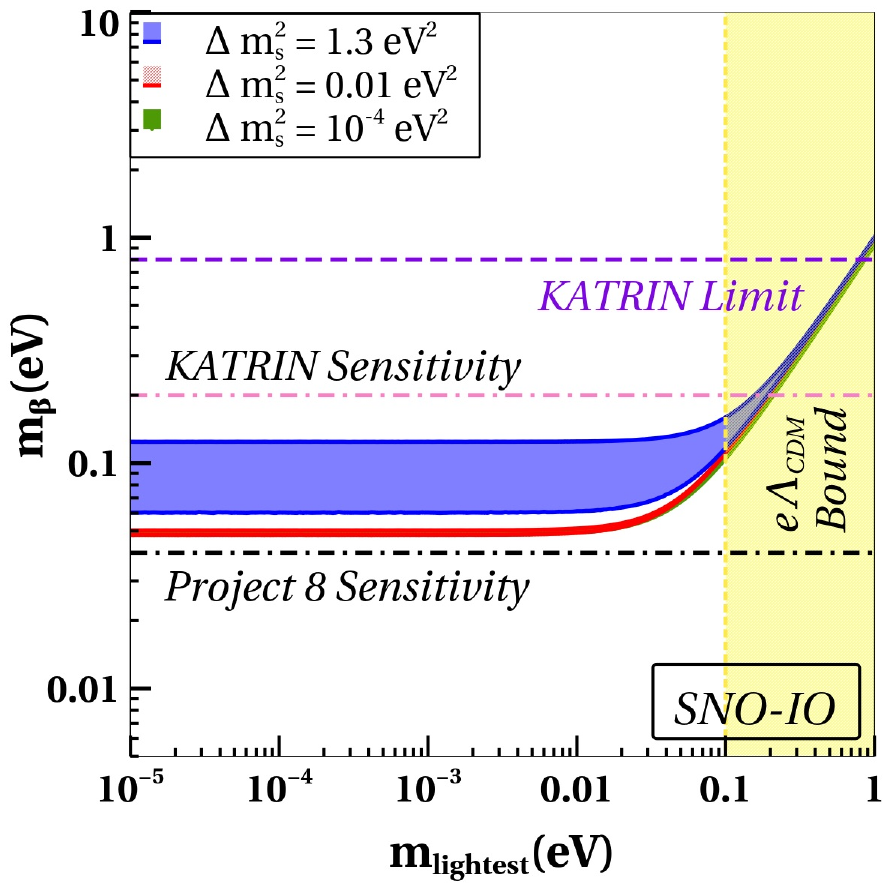}
    \includegraphics[width=0.48\linewidth]{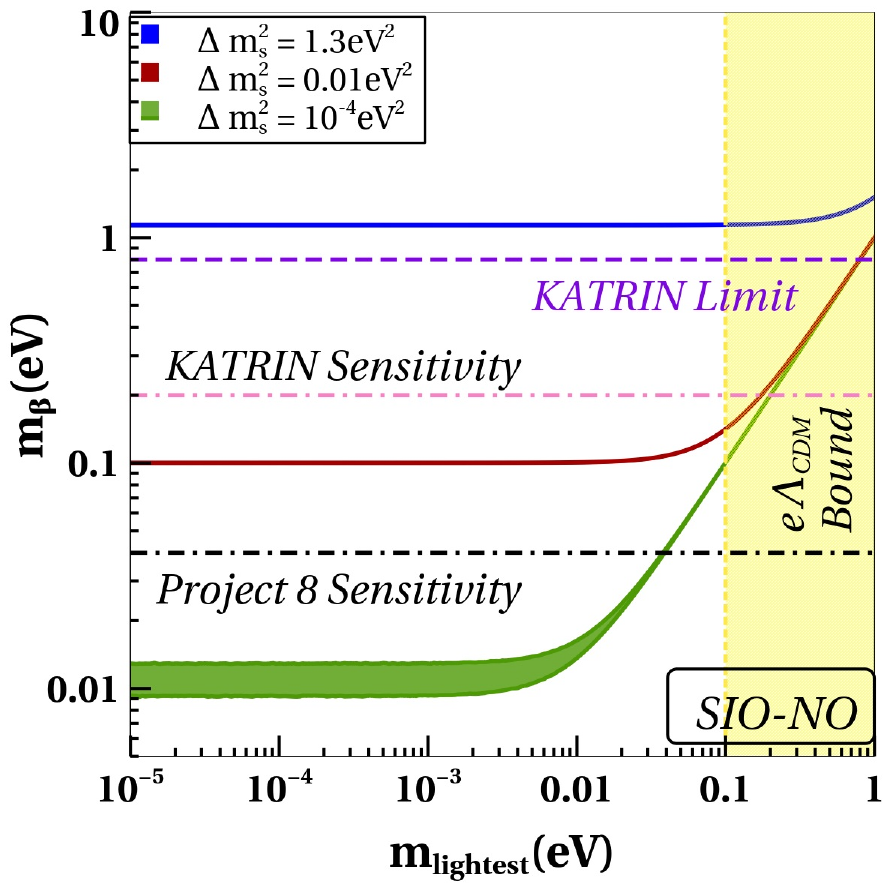}
     \includegraphics[width=0.48\linewidth]{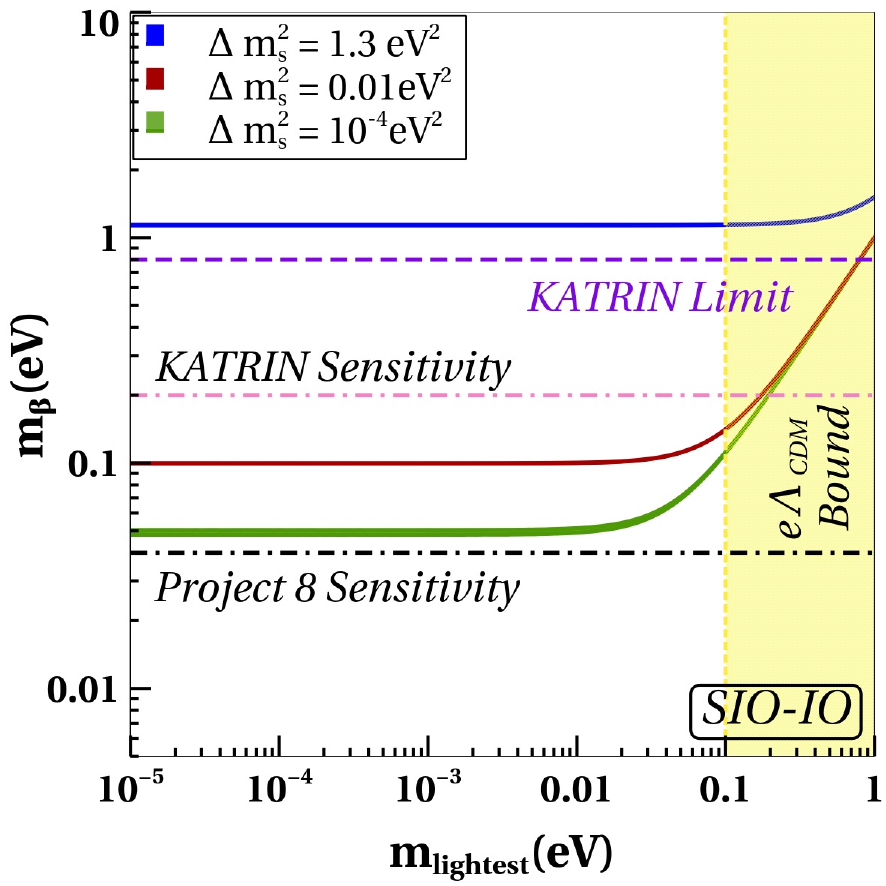}
     \caption{Kinematic mass $m_\beta$  from tritium $\beta$ decay in different scenarios of SNO-NO (top-left), SNO-IO (top-right), SIO-NO (bottom-left), and SIO-IO (bottom-right) for $\ms^2 = 10^{-4}\, \ev^2$ (green), $\ms^2 = 0.01\, \ev^2$ (red) and $\ms^2 = 1.3\, \ev^2$ (blue).}
    \label{fig:mb}
\end{figure}

\noindent The kinematic mass depends on the mixing parameters, mass squared differences, and the lightest neutrino mass. The current \textit{KATRIN} limit on $m_{\beta}$ is $\le \, 0.8\, \ev$ and the future sensitivity is quoted as $m_{\beta}\le \, 0.2\, \ev$. 
We have plotted $m_{\beta} $ as a function of the lightest neutrino mass in Fig. (\ref{fig:mb}) by varying all the parameters in their respective allowed intervals as given in table (\ref{tab:cosmo_table}). The cyan dashed lines in the figure show the projected sensitivity of the \textit{KATRIN} experiment of 0.2 eV. In this figure, we also show the sensitivity of future experiment \textit{Project 8} \cite{Project8:2022wqh}, by a dashed-dot black line, which plans to probe the lightest neutrino mass with a maximum sensitivity of up to 40 meV in a phased manner. In Fig. (\ref{fig:mb}), $|U_{e1}|^2, |U_{e2}|^2, |U_{e3}|^2$ are varied $(0.64:0.72), (0.26:0.33), (0.020:0.024) $ and the range of $|U_{e4}|^2$ as given in table (\ref{tab:cosmo_table}). In table (\ref{tab:comp-mb}), we provide the necessary values to explain the characteristics of Fig. (\ref{fig:mb}). 

\begin{table}[H]
        \caption{3$\sigma$ ranges of different combinations of oscillation parameters relevant to understanding kinematic mass ($m_{\beta}$) in the 3+1 scenario. }
        \label{tab:comp-mb}
        \centering
        \begin{tabular}{|c|c|c|c|c|c|}
            \hline   $|U_{e2}|^2 \, \msol^2 $ & $(1-|U_{e2}|^2) \, \msol^2 $ & $|U_{e3}|^2 \, \matm^2 $ & \multicolumn{3}{c|}{$|U_{e4}|^2 \, \ms^2$} \\
              \cline{4-6}    
              $\times 10^{-5}$ & $\times 10^{-5}$ & $\times 10^{-5}$ &  $\ms^2 = 10^{-4}\, \ev^2$& $ \ms^2=0.01 \, \ev^2$ & $\ms^2 = 1.3\, \ev^2$ \\ \hline 
               $(1.77 : 2.65)$ & $(4.57:5.95)$  & $(4.86:6.24) $ & $(1:2)\times 10^{-5}$ &  $ (0.5:5)\times 10^{-5}$ &$(0.13:1.3)\times 10^{-2}$ \\ \hline
        \end{tabular}
\end{table}
 
\pagebreak
The following observations can be made from Fig. (\ref{fig:mb}),

\begin{itemize}
    
    \item \textit{KATRIN}'s future sensitivity allows us to probe $m_\beta$ only above $m_{\rm lightest}\sim 0.2$ eV for SNO-NO, SNO-IO for all values of $\ms^2$. In case of SIO-NO, SIO-IO \textit{KATRIN} will be able to probe the entire spectrum of $m_{\rm lightest}$ for $\ms^2 =1.3$ eV$^2$, and above $m_{\rm lightest} \sim 0.02$ eV for $\ms^2 =10^{-4}, 0.01$ eV$^2$.
    
    \item The sensitivity of \textit{Project 8} allows us to probe $m_\beta$ only above $m_{\rm lightest}\sim 0.03$ eV for SNO-NO and SIO-NO of $\ms^2 = 10^{-4} \, \ev^2$.  However, \textit{Project 8} experiment can probe SNO-IO, SIO-NO, and SIO-IO for $\ms^2 = 0.01\, \ev^2 $ and $1.3\, \ev^2$ in the entire range of $m_{\rm lightest}$.

    \item \textbf{SNO-NO:} Using Eqn. (\ref{eq:mrel-snono}), $m_{\beta}$ can be approximated as 
    \beqa
    m_{\beta}^{SNO-NO} &=& \sqrt{ m_{\rm lightest}^2 + |U_{e2}|^2 \, \msol^2 + |U_{e3}|^2 \, \matm^2 + |U_{e4}|^2 \, \ms^2 } \label{eq:mb:snono}
    \eeqa
    \begin{itemize}
        \item For $m_{\rm lightest} < \sqrt{\msol^2} < \sqrt{\matm^2} $ , it is seen from table (\ref{tab:comp-mb}) that the second, third and the fourth term in Eqn. (\ref{eq:mb:snono}) varies in the similar range for $\ms^2=10^{-4} \, \ev^2$ and $ \ms^2 =0.01 \, \ev^2$. Hence $m_{\beta}^{SNO-NO}$ varies as $(0.009:0.01) $ and  $(0.008:0.011) \, \ev$. In the case of $\ms^2 =1.3 \, \ev^2$, $m_{\beta}^{SNO-NO} \approx |U_{e4}|\, \sqrt{\ms^2} $ and varies between $(0.036 : 0.114) \, \ev$.  
        \item For $ \sqrt{\msol^2} < m_{\rm lightest} <\sqrt{\matm^2} <\sqrt{\ms^2}$, $m_{\beta\beta}^{SNO-NO} \approx \, m_{\rm lightest} $ for $\ms^2 = 10^{-4}\, \ev^2 $ and $\ms^2 = 0.01 \, \ev^2$. Whereas $|U_{e4}|^2 \, \ms^2$ still dominates in this region for $\ms^2= 1.3 \, \ev^2$.
        \item For $\sqrt{\matm^2}<< |U_{e4}|\, \sqrt{\ms^2}  <<  m_{\rm lightest}$, $ m_{\beta}^{SNO-NO}$ is completely determined by the value of $m_{\rm lightest}$. 
    \end{itemize}  
    

    \item \textbf{SNO-IO:} 
    \beqa
    m_{\beta}^{SNO-IO} & \approx &  \sqrt{m_{\rm lightest}^2  + \matm^2  + |U_{e4}|^2\, \ms^2}\label{eq:mb_snoio}
    \eeqa
    \begin{itemize}
        \item For $m_{\rm lightest} << \msol^2 < \matm^2 $, $m_{\beta}^{SNO-IO}\approx \sqrt{\matm^2} \approx 0.05 \, \ev$ for $\ms^2 = 10^{-4}\, \ev^2 $ and $ 0.01\, \ev^2$ as $|U_{e4}|^2$ is very small. For $\ms^2 = 1.3 \, \ev^2$, the value of $m_{\beta}^{SNO-IO}\approx \,\sqrt{\matm^2+|U_{e4}|^2\,\ms^2}$. Thus, the value of $m_{\beta}$ for $\ms^2 = 1.3 \, \ev^2$ is greater than the $\sqrt{\matm^2}$ till $m_2 \approx 0.1 \, \ev$ . 
        \item $ 0.1<< m_{\rm lightest}$ , $m_{\beta}^{SNO-IO} \approx \, m_{\rm lightest}$ for the values of $\ms^2$. Hence, for higher $m_{\rm lightest}$, the behaviour of $m_{\beta}$  is fully characterised by $m_{\rm lightest}$.  
    \end{itemize}
    
    \item \textbf{SIO-NO:} 
    \beqa
    m_{\beta}^{SIO-NO} &=& \sqrt{m_{\rm lightest}^2 + \ms^2  + |U_{e2}|^2\, \msol^2  + |U_{e3}|^2\, \matm^2  }
    \eeqa
    \begin{itemize}
        \item For $ m_{\rm lightest} << \sqrt{\msol^2} << \sqrt{\matm^2}$, $m_{\beta}^{SIO-NO} \approx \sqrt{\ms^2} $ for $\ms^2 = 0.01 \, \ev^2$ and $1.3 \, \ev^2$. For $\ms^2 = 10^{-4} \, \ev^2$ , second and third term vary $\sim 10^{-5}$ , so we get a small variation due to that. 
        \item For $\sqrt{\ms^2} << m_{\rm lightest}$, $m_{\beta}^{SIO-NO} \approx m_{\rm lightest}$, and the value value of $m_{\beta}^{SIO-NO}$ depend on $m_{\rm lightest} $ only. 
    \end{itemize}
    
    \item \textbf{SIO-IO:} 
    \begin{itemize}
        \item For $\ms^2 > \matm ^2 $,  $m_{\beta}^{SIO-IO}$ can be written as 
    \beqa
    m_{\beta}^{SIO-IO} &=& \sqrt{m_{\rm lightest}^2 + \, \ms^2}  
    \eeqa
In this case, the conclusions are similar to SIO-NO for $\ms^2 = 0.01\, \ev^2 $ and $1.3\, \ev^2$.
\item For $\ms^2 < \matm^2 $, $m_{\beta\beta}^{SIO-IO}$ can be expressed as 
    \beqa
    m_{\beta}^{SIO-IO} &=& \sqrt{m_{\rm lightest}^2 + \matm^2 } 
    \eeqa
In this case, for lower $m_{\rm lightest} \, (< \sqrt{\matm^2)}$ region , $m_{\beta}^{SIO-IO} \approx \sqrt{\matm^2} \approx 0.05 \, \ev$. For higher values of $m_{\rm lightest} \,(>  \sqrt{\matm^2})$, the value of $m_{\beta}^{SNO-IO}$ is proportional to $m_{\rm lightest} $ which leads to a straight line behavior in the figures. 
    \end{itemize}
    
The expressions of $m_{\beta}^2$ in various $m_{lightest}$ limits are tabulated in table (\ref{tab:mbeta2}) in the appendix.
     
    
\end{itemize}


\subsection{Bound from neutrinoless double beta decay}
The cosmological observations and the tritium decay measurements are sensitive to the absolute neutrino mass scale, not to the nature of the neutrinos, i.e., whether the neutrinos are Dirac or Majorana. The neutrinoless double beta decay ($0\nu\beta\beta$) process can provide both pieces of information. The $0\nu\beta\beta $ decay process constrains the half-life of the decaying isotope, which can be expressed as,  
\beqa
T_{1/2} &=& \frac{m_e^2}{G_{0\nu}\, \left|\mathcal{M}_{0\nu}\right|^2 \, m_{{\beta\beta}}^2} ,
\eeqa
where $m_e$ is electron mass, $G_{0\nu}$ denotes the leptonic phase space and $\mathcal{M}_{0\nu}$ is the nuclear transition matrix element of the decay and $m_{\beta\beta}$ is the effective Majorana mass which can be expressed as 
\beqa
 \mathbf{m}_{\beta\beta} &=& \sum_{i} U_{ei}^2 \, m_i
 \label{eq:mbb-std} 
\eeqa
where $i$ runs over the light neutrino species.

The current upper limits are $m_{\beta\beta} \le (36-156)$ meV and $(79-180)$ meV as reported by the \textit{KamLAND-Zen} and \textit{GERDA} experiments respectively. But recently, it has been pointed out that the nuclear matrix element calculations should include a short-range contribution that originated from the hard-neutrino exchange mechanism described in \cite{Cirigliano:2018hja, Cirigliano:2019vdj}. Ref. \cite{Scholer:2023bnn} showed that the inclusion of the short-range contribution tightens the limit on $m_{\beta\beta}$ as $m_{\beta\beta}\le \, (25-68)$ meV for \textit{KamLAND-Zen}.

\subsubsection{\bf Standard three flavor framework}
In the standard three flavor framework Eqn. (\ref{eq:mbb-std}) can be expressed as 
\beqa
\mathbf{m}_{\beta\beta}^{\rm Std.} &=& m_1 \, c_{12}^2 \, c_{13}^2 +m_2 \,  s_{12}^2\, c_{13}^2 \, e^{i \alpha} + m_3 \, s_{13}^2\, e^{i \beta}\
\eeqa
Unlike neutrino oscillation experiments, the effective Majorana mass is sensitive to the Majorana phases of the neutrinos. In addition, the effective Majorana mass is also sensitive to the mass orderings.

In Figures (\ref{fig:mbb-comp-snono},\ref{fig:mbb-comp-snoio},\ref{fig:mbb-comp-siono}), grey and light brown regions display the effective mass governing $0\nu\beta\beta$ as a function of the lowest mass in the standard three-flavor framework for NO and IO respectively. In these figures, the oscillation parameters are varied over their $3\sigma$ ranges as tabulated in the table (\ref{tab:osc}), and Majorana phases ($\alpha,\beta$) are varied between $(0 \,:  \, \pi)$. 

\vspace{2 em}
{\underline{Normal Ordering ($ m_1< m_2< m_3$)}}\\
\begin{itemize}
    \item For $m_{\rm lightest}(m_1) << \sqrt{\msol^2}<<\sqrt{\matm^2}$,   $m_2 \approx \sqrt{\msol^2} \approx 0.01 \, \rm eV$ and $m_3 \approx \sqrt{\matm^2} \approx 0.05 \, \rm eV$. The effective Majorana mass can be approximated as 
    \beqa
    \mNO &=& \sqrt{\matm^2} c_{13}^2 \left( \sqrt{r}\, s_{12}^2 \, e^{i\, \alpha} + t_{13}^{2} \, e^{i\beta}\right),
    \eeqa 
    where $r=\frac{\msol^2}{\matm^2}$. Complete cancellation is possible if $ \sqrt{r}\,s_{12}^2 = t_{13}^{2}$. In table (\ref{tab:mbb-std}), we enlist different combinations of parameters appearing in the expression of $\mNO$. As can be seen from the table (\ref{tab:mbb-std}), the maximum value of $t_{13}^2$ is much less than $\sqrt{r}\, s_{12}^2$, so complete cancellation is not possible in this region. For $\alpha=\beta=0$, we get the highest value of $\mNO$, while the lowest value is obtained for $\alpha =0 \,,\, \beta=\pi$ or $\alpha=\pi \, , \, \beta=0$. In this region, the effective mass satisfies $0.001 \, \rm eV \lesssim \left|\mNO\right| \lesssim 0.004\, \rm eV$. 

    \item For $m_{\rm lightest} \approx \sqrt{\msol^2}$. Here, $\mNO$ can be expressed as 
    \beqa  \mNO &=& \sqrt{\matm^2} \, c_{13}^2 \left(\sqrt{r} c_{12}^2 + \sqrt{r} \,s_{12}^2 \, e^{i \, \alpha} + t_{13}^2 \, e^{i\, \beta} \right) 
    \eeqa  
    The effective mass attains minimum value for $\alpha=\beta=\pi$ and complete cancellation occurs when
    $\sqrt{r}\, \cos 2\theta_{12} = t_{13}^2$
    From table (\ref{tab:mbb-std}), it can be inferred that complete cancellation is not possible in this region, which is also observed in Figures (\ref{fig:mbb-comp-snono},\ref{fig:mbb-comp-snoio},\ref{fig:mbb-comp-siono}).
    
    \item From Fig. (\ref{fig:mbb-comp-snono}), it can seen that the value of $m_{\beta\beta}$ is very small in a region $0.002 \, \rm eV \,  \lesssim m_{\rm lightest} \, \lesssim \, 0.007 \, \rm eV$. This region is commonly referred to as the cancellation region. As an example, considering the mixing parameters equal to their best fit values and $m_{\rm lightest} = 0.005 \, $eV, we get $m_{\beta\beta} \approx \, 10^{-4} $ for the Majorana phases $\alpha =\beta=\pi$.  
    

\end{itemize}
\vspace{2 em}

{ \underline{Inverted Ordering ($ m_3 <m_1<m_2$)}}  \\

\begin{itemize}
    \item  In the limit $m_3 \, \approx 0$, $m_1\approx m_2 \approx \sqrt{\matm^2}$ and the effective mass can be expressed as, 
    \beqa 
    \mIH &=& \sqrt{\matm ^2} \, c_{13}^2 \left( c_{12}^2 + s_{12}^2 \, e^{i \, \alpha}\right)
    \eeqa 
    In this region, $\mIH$ is bounded from below and above by minimum and maximum values as, 
\beqa
\left|\mIH \right|_{\rm min} &=& \sqrt{\matm ^2} \, c_{13}^2 \cos2\theta_{12} = 0.02 \, \rm eV \nn\\
\left|\mIH \right|_{\rm max} &=& \sqrt{\matm ^2} \, c_{13}^2 = 0.05\,\rm eV
\eeqa
\end{itemize}  
These bounds are reflected also in Figures (\ref{fig:mbb-comp-snono},\ref{fig:mbb-comp-snoio},\ref{fig:mbb-comp-siono}).

\vspace{2 em }
   {\underline{Quasi Degenerate Spectrum ($ m_1 \approx m_2 \approx m_3 \, \gtrsim 0.05  \, \rm  eV$)}} \\
   
The region where $m_{\rm lightest} \gtrsim \sqrt{\matm^2}\gtrsim 0.05$ eV  (for both mass orderings) , $m_1\,,\,m_2\,,\,m_3$ are approximately equal. This region is called the quasi-degenerate region. Here the effective mass can be expressed as
\beqa
    \mathbf{m}_{\beta\beta}^{\rm QD} &=& m_0 \, c_{13}^2 \left( c_{12}^2 + s_{12}^2 \, e^{i \alpha} + t_{13}^2 \, e^{i \, \beta}\right)
\eeqa
In this region, cancellation is not possible, as $t_{13}^2\approx 0.02, s_{12}^2\approx 0.3$ will not be able to cancel out $c_{12}^2 \approx 0.7$ as can be seen from the Fig. (\ref{fig:mbb-comp-snono},\ref{fig:mbb-comp-snoio},\ref{fig:mbb-comp-siono}). This region is in serious tension with the cosmological observations because, for three degenerate neutrinos, the bound on $m_{\rm lightest}\, < \, 0.05$ eV considering $\sum m_{\nu} < 0.16$ eV (from Eqn. (\ref{eq:cosmo2})).

\begin{table}[H]
        \caption{3$\sigma$ ranges of different combinations of oscillation parameters relevant to understanding the effective Majorana mass in the standard three-flavor scenario. }
        \label{tab:mbb-std}
        \centering
        \begin{tabular}{|m{7em}|m{7em}|m{7em}|m{7em}|m{7em}| m{7em}|}
            \hline Param. &$\sqrt{r}$  & $\sqrt{r} \,  s_{12}^2$  &  $\sqrt{r} \, \cos 2\theta_{12}$ & $t_{13}^2$ &$ \sqrt{r}\, t_{13}^2$ \\
            & & &  &  & \\
        \hline
           Max & 0.18 &0.0614  & 0.0828  & 0.0246 & 0.00443 \\ \hline
           Min & 0.16 & 0.0432 & 0.0509 & 0.0204  & 0.00326 \\ 
          \hline
        \end{tabular}
\end{table}
\subsubsection{\textbf{3+1 framework}}
\begin{figure}[H]
    \centering
    \includegraphics[width=0.32\linewidth]{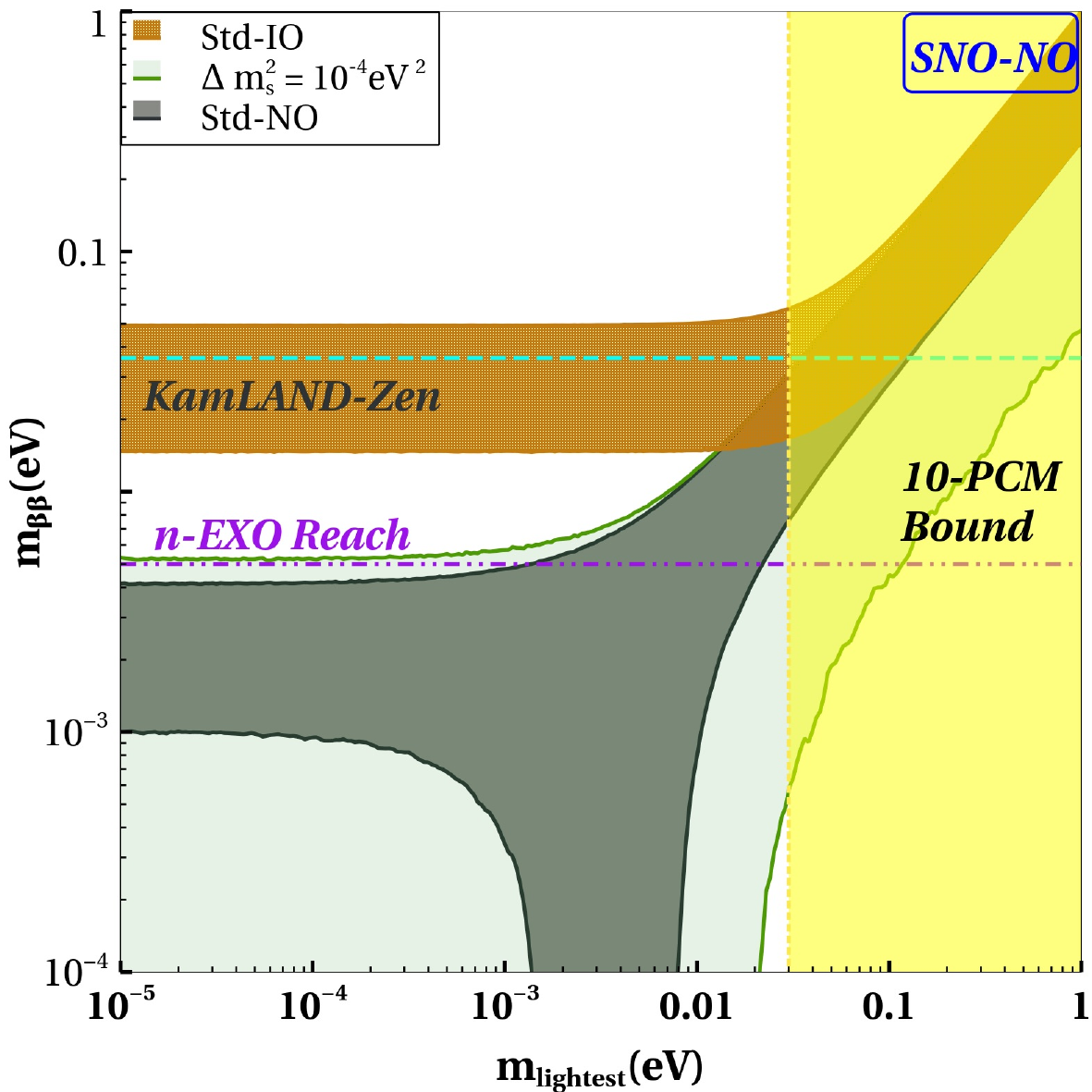}
    \includegraphics[width=0.32\linewidth]{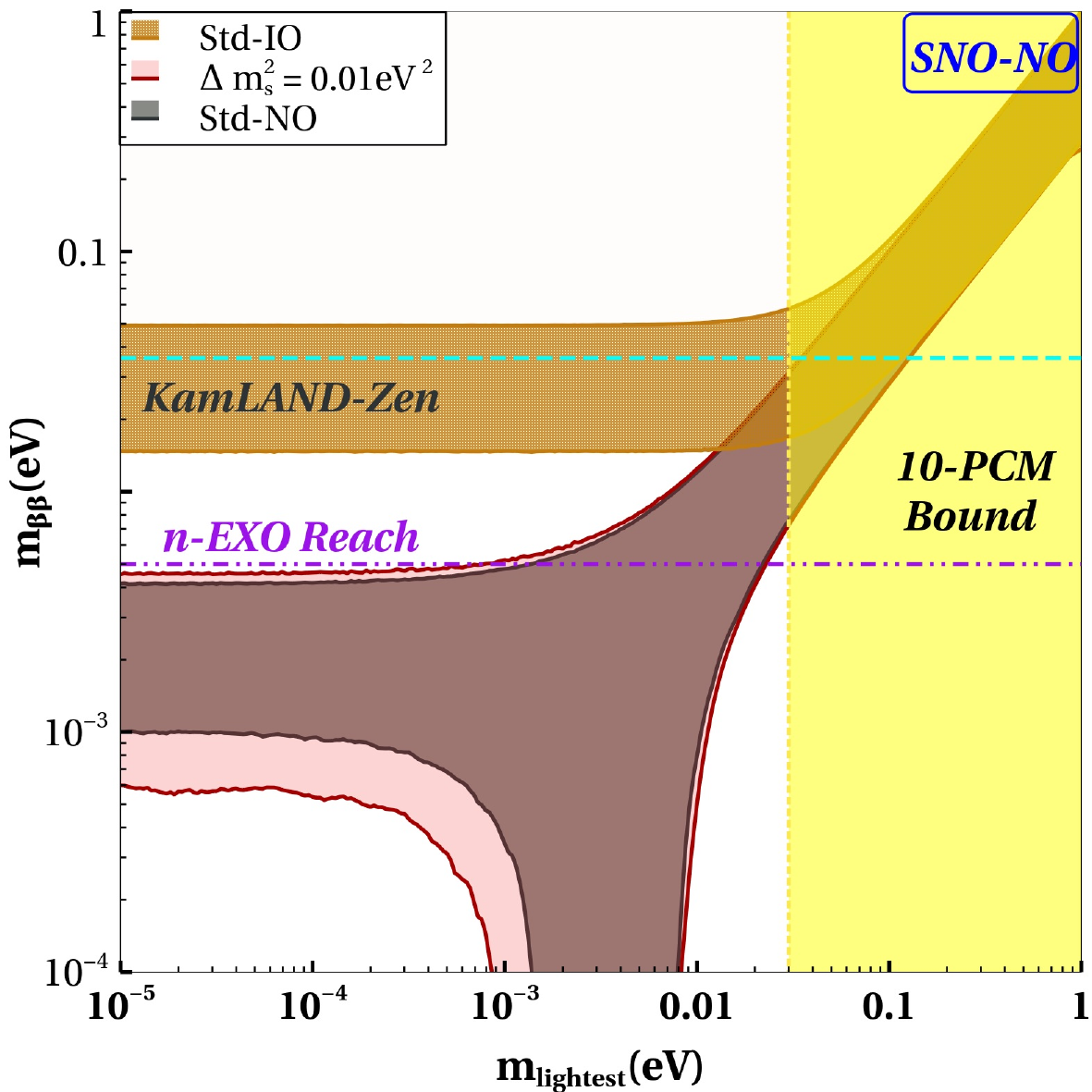}
    \includegraphics[width=0.32\linewidth]{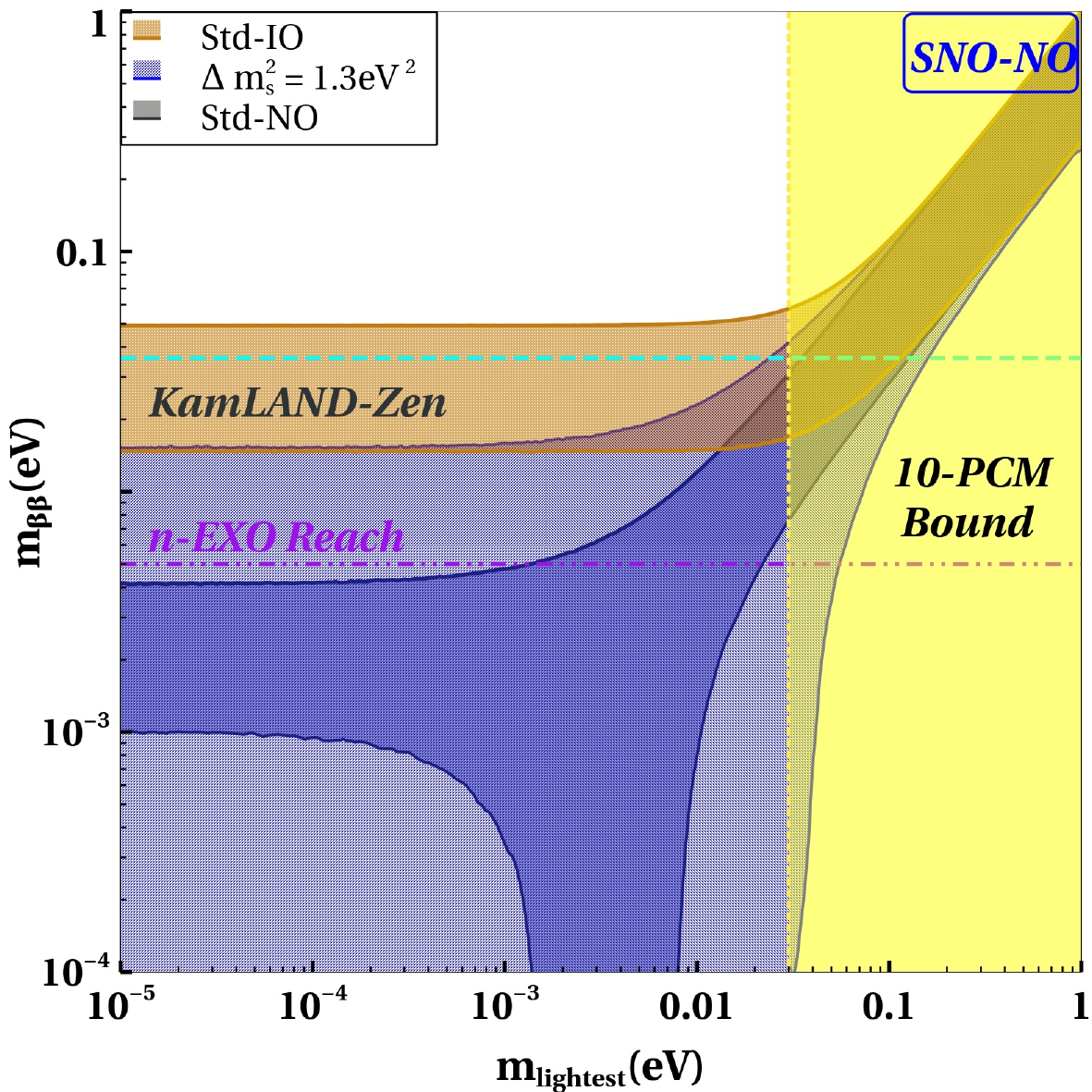}
    \caption{$m_{\beta\beta}$ is plotted for SNO-NO (green) scenario against the lightest neutrino mass  with  the mass squared difference ($\Delta m_s^2 $) = $ 10^{-4}\rm eV^2$ (green), 0.01 $\rm eV^2$ (red), and 1.3 $\rm eV^2$(blue) along with standard three flavor NO (grey) and IO (brown) .}
    \label{fig:mbb-comp-snono}
\end{figure}

\begin{figure}[H]
    \centering
     \includegraphics[width=0.32\linewidth]{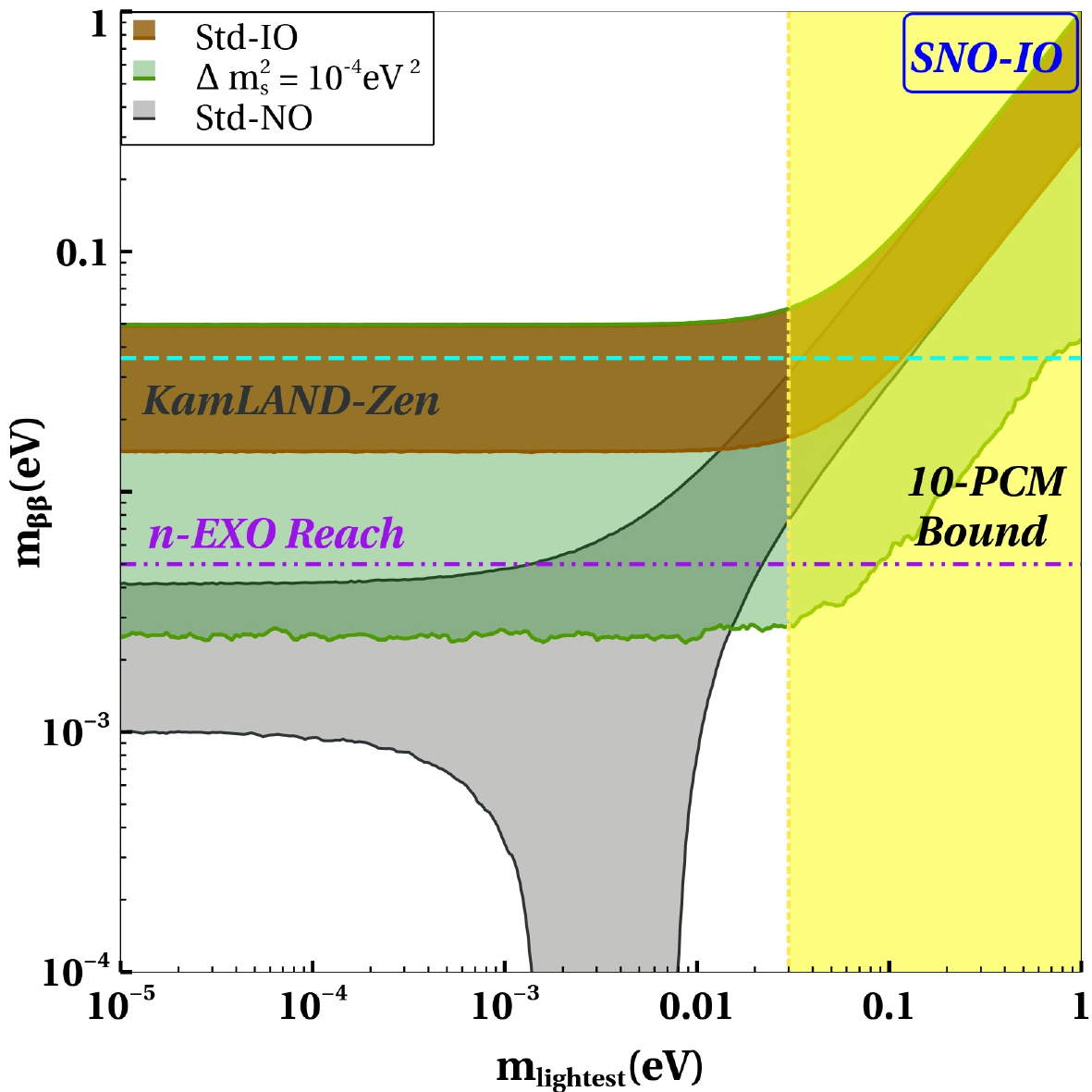}
    \includegraphics[width=0.32\linewidth]{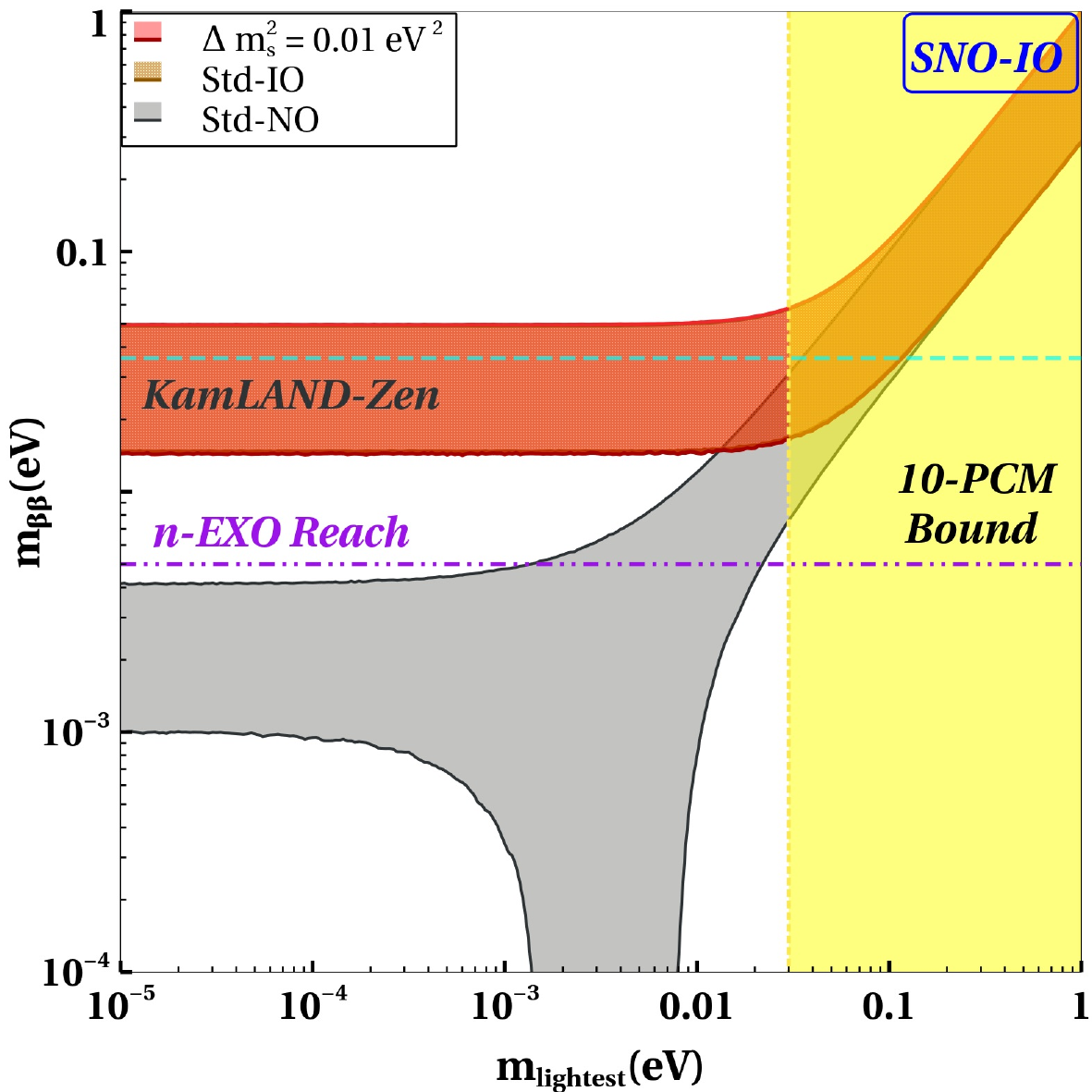}
    \includegraphics[width=0.32\linewidth]{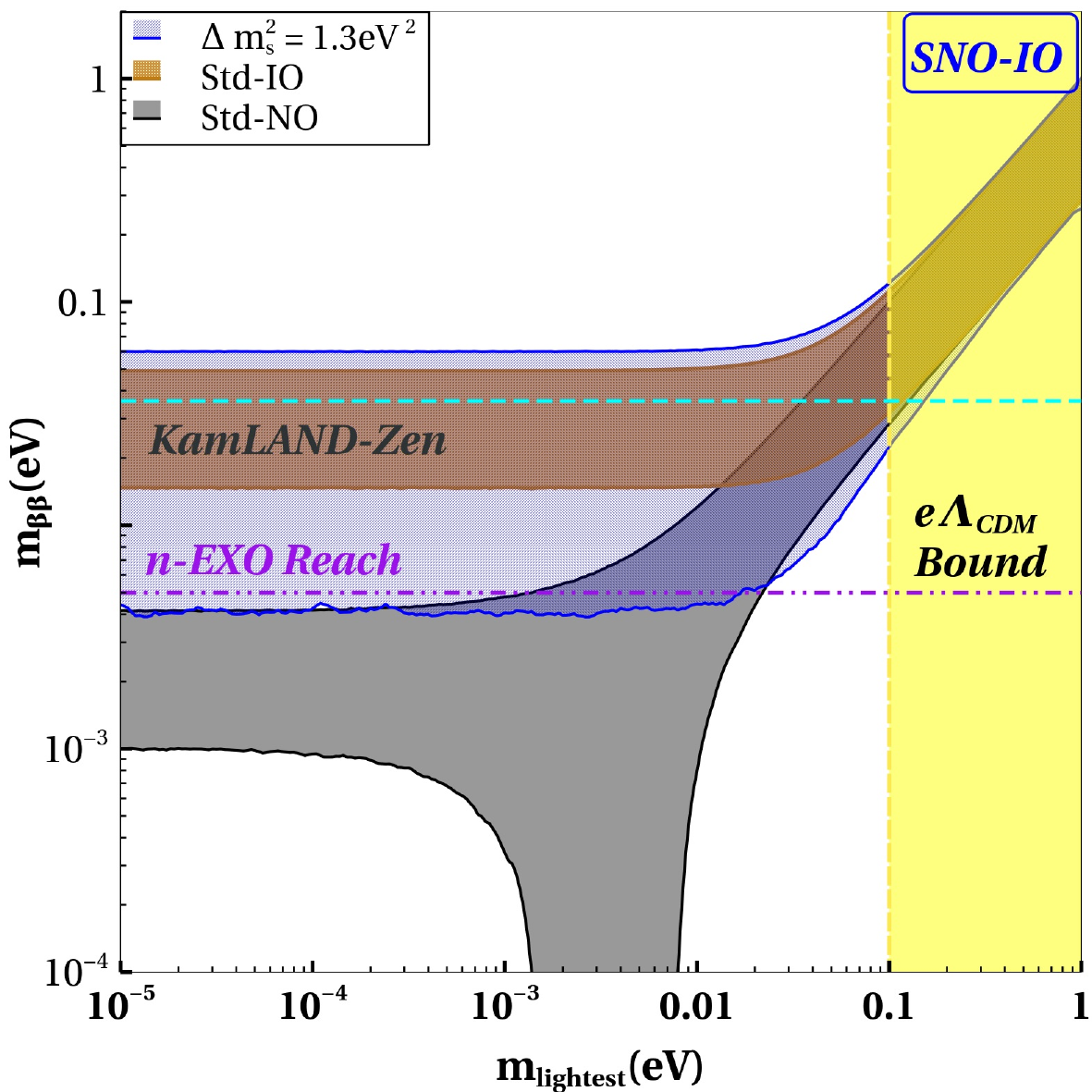}
    \caption{$m_{\beta\beta}$ is plotted for SNO-IO scenario against the lightest neutrino mass with the mass squared difference ($\Delta m_s^2 $) = $  10^{-4}\rm eV^2$ (green), 0.01 $\rm eV^2$ (red), 1.3 $\rm eV^2$ (blue) along with standard three flavor Normal Ordering (grey) and Inverted Ordering (brown) .}
    \label{fig:mbb-comp-snoio}
\end{figure}

\begin{figure}[H]
    \centering
    \includegraphics[width=0.32\linewidth]{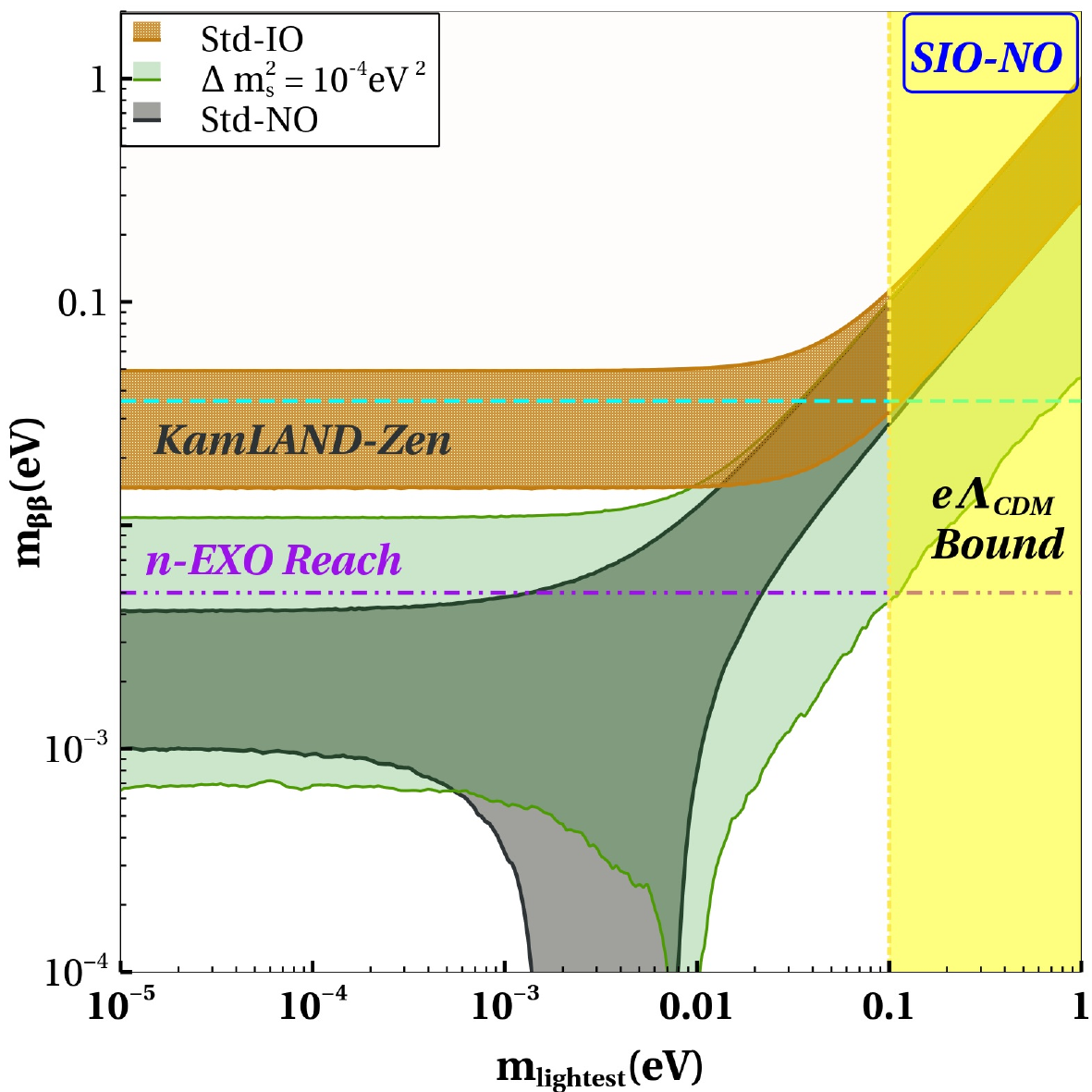}
    \includegraphics[width=0.32\linewidth]{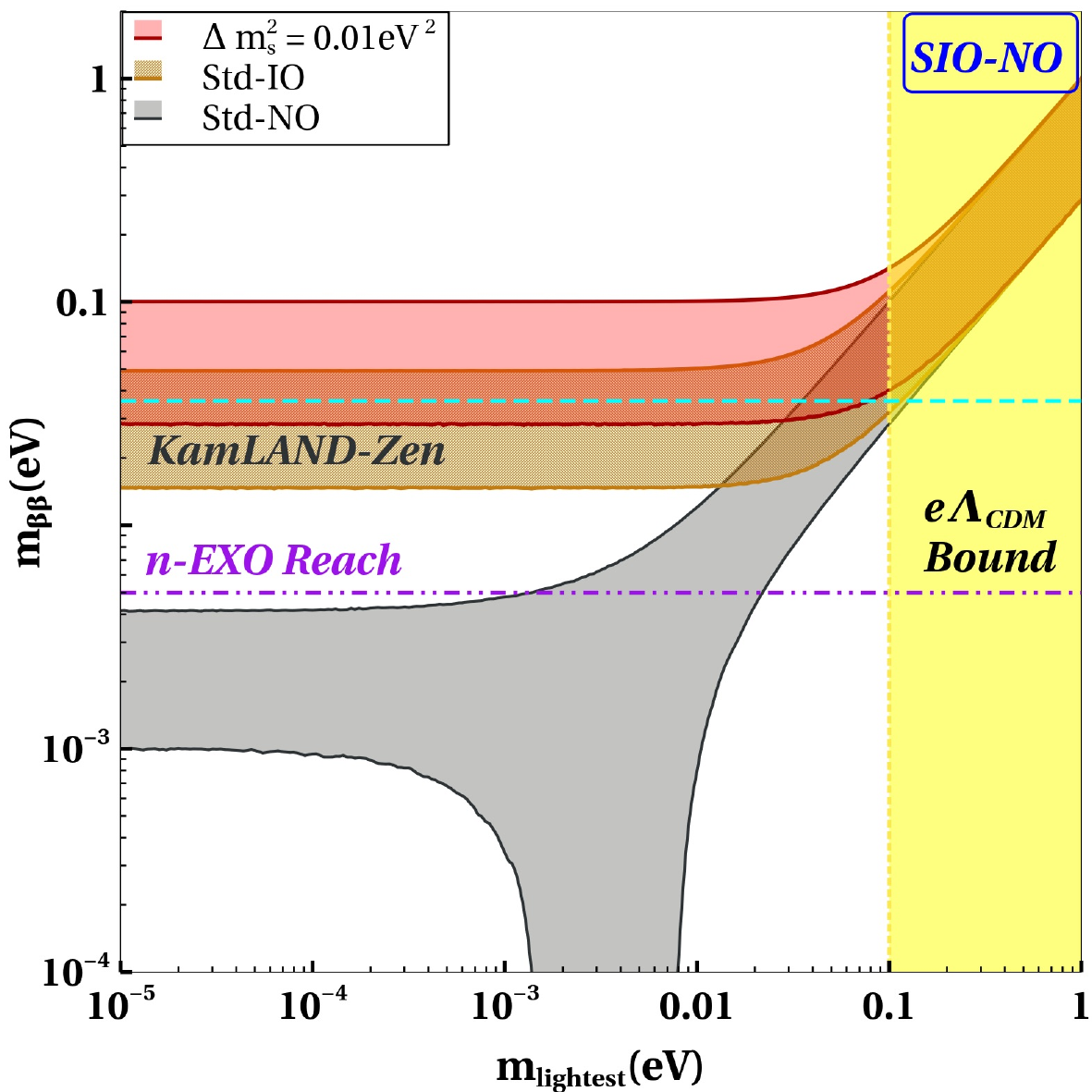}
    \includegraphics[width=0.32\linewidth]{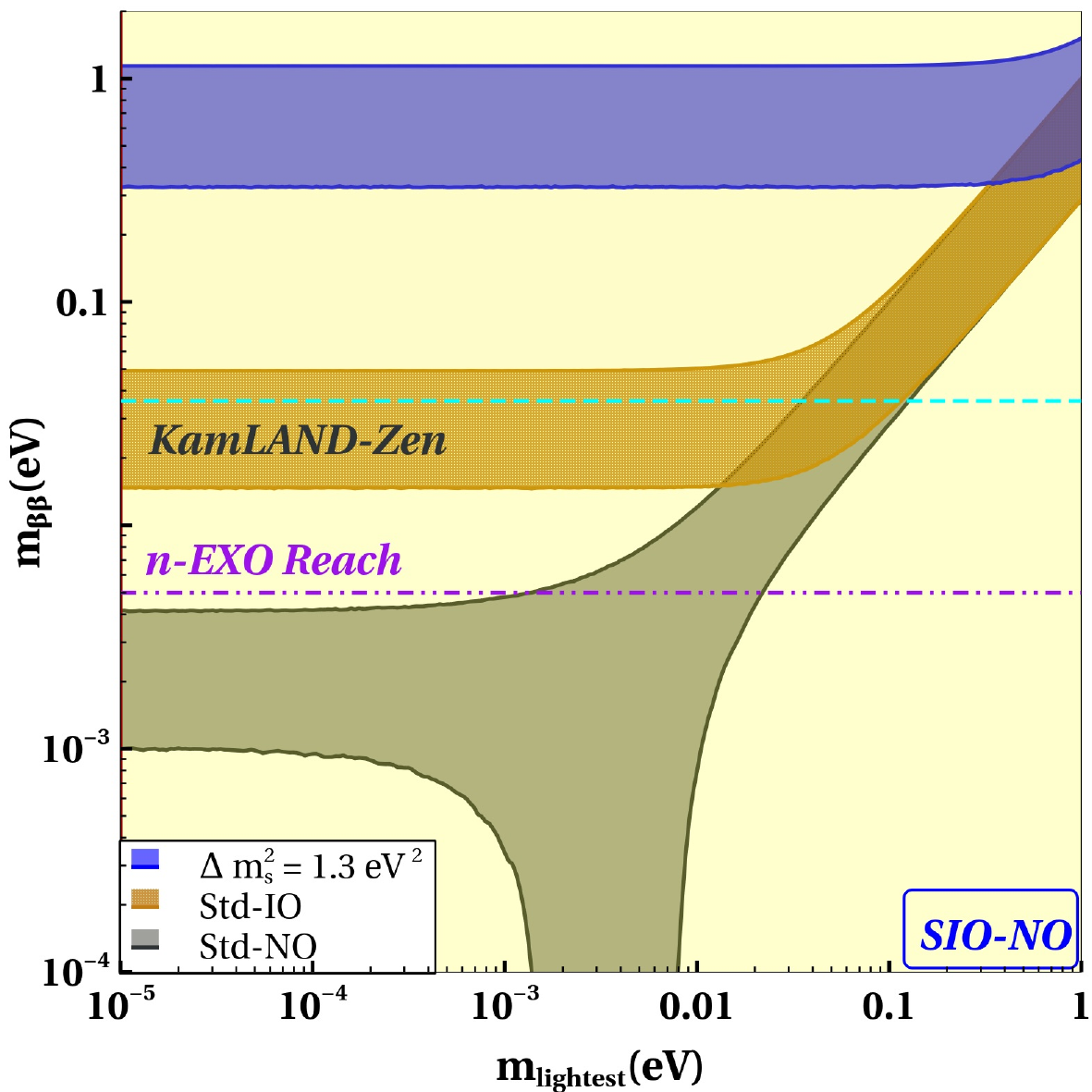}
    \caption{$m_{\beta\beta}$ is plotted for SIO-NO (green) scenario against the lightest neutrino mass  with  the mass squared difference ($\Delta m_s^2 $) = ($ 10^{-4}\rm \,eV^2$, 0.01 $\rm eV^2$, 1.3 $\rm eV^2$) along with standard three flavor Normal Ordering (red) and Inverted Ordering (yellow) .}
    \label{fig:mbb-comp-siono}
\end{figure}

\begin{figure}[H]
    \centering
    \includegraphics[width=0.32\linewidth]{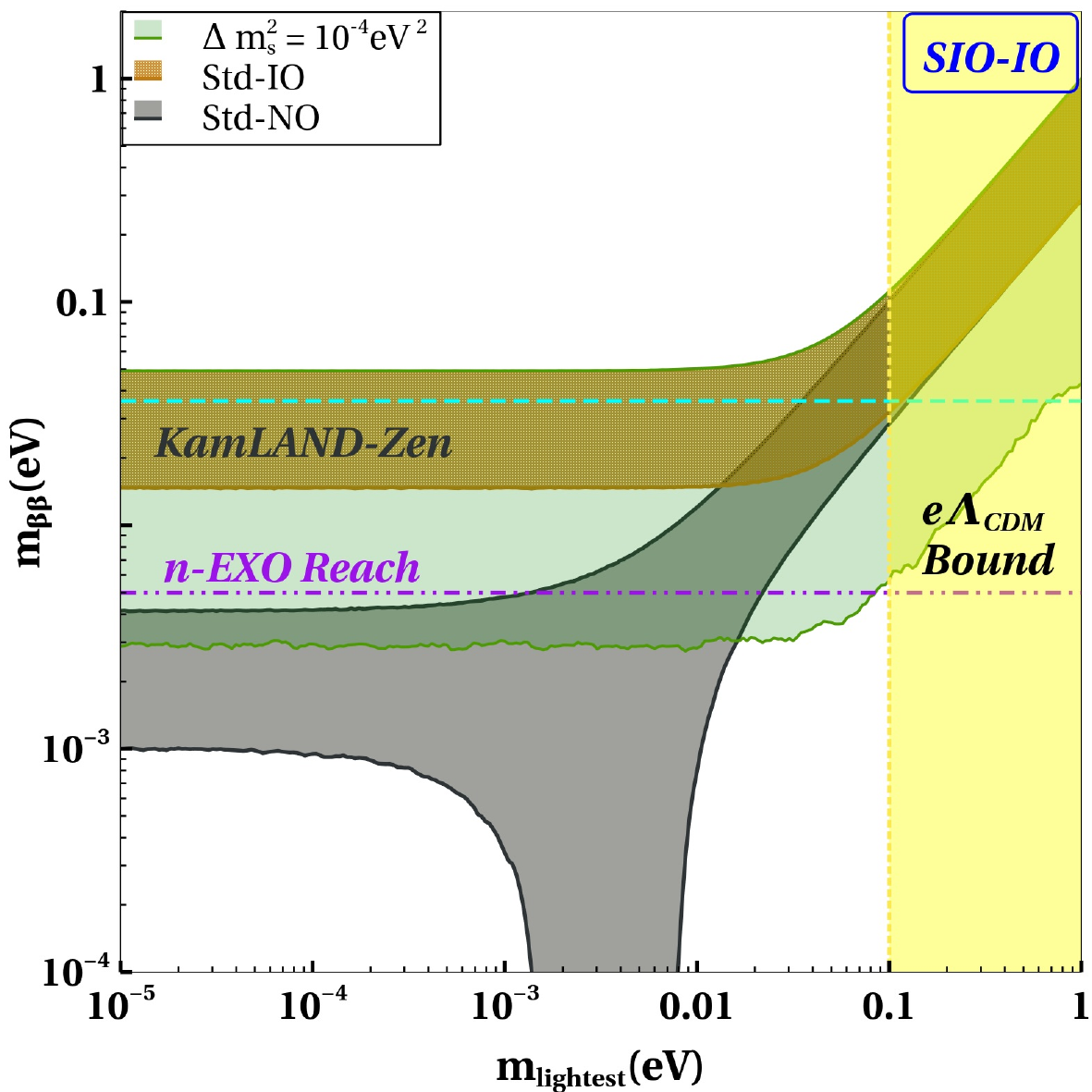}
    \includegraphics[width=0.32\linewidth]{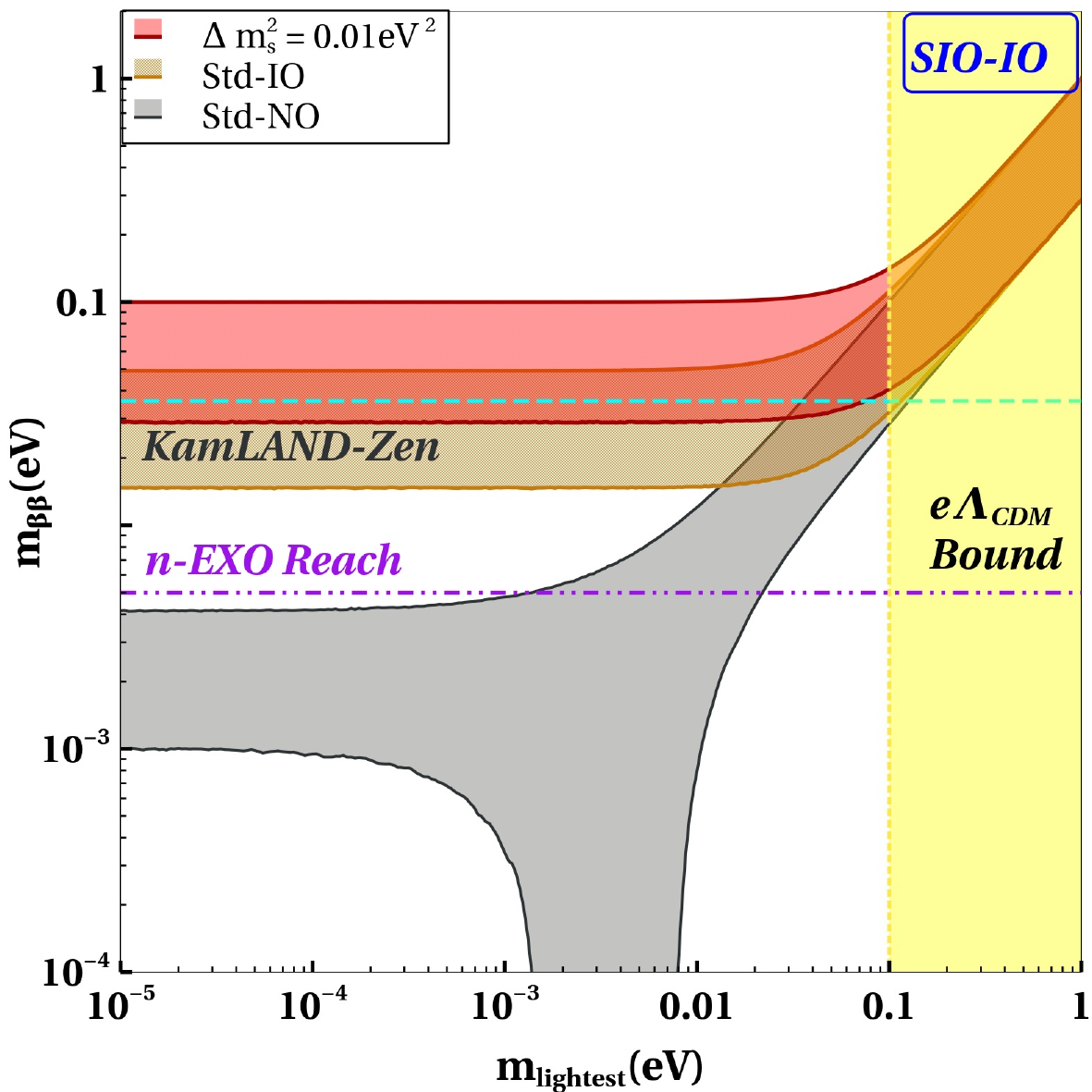}
    \includegraphics[width=0.32\linewidth]{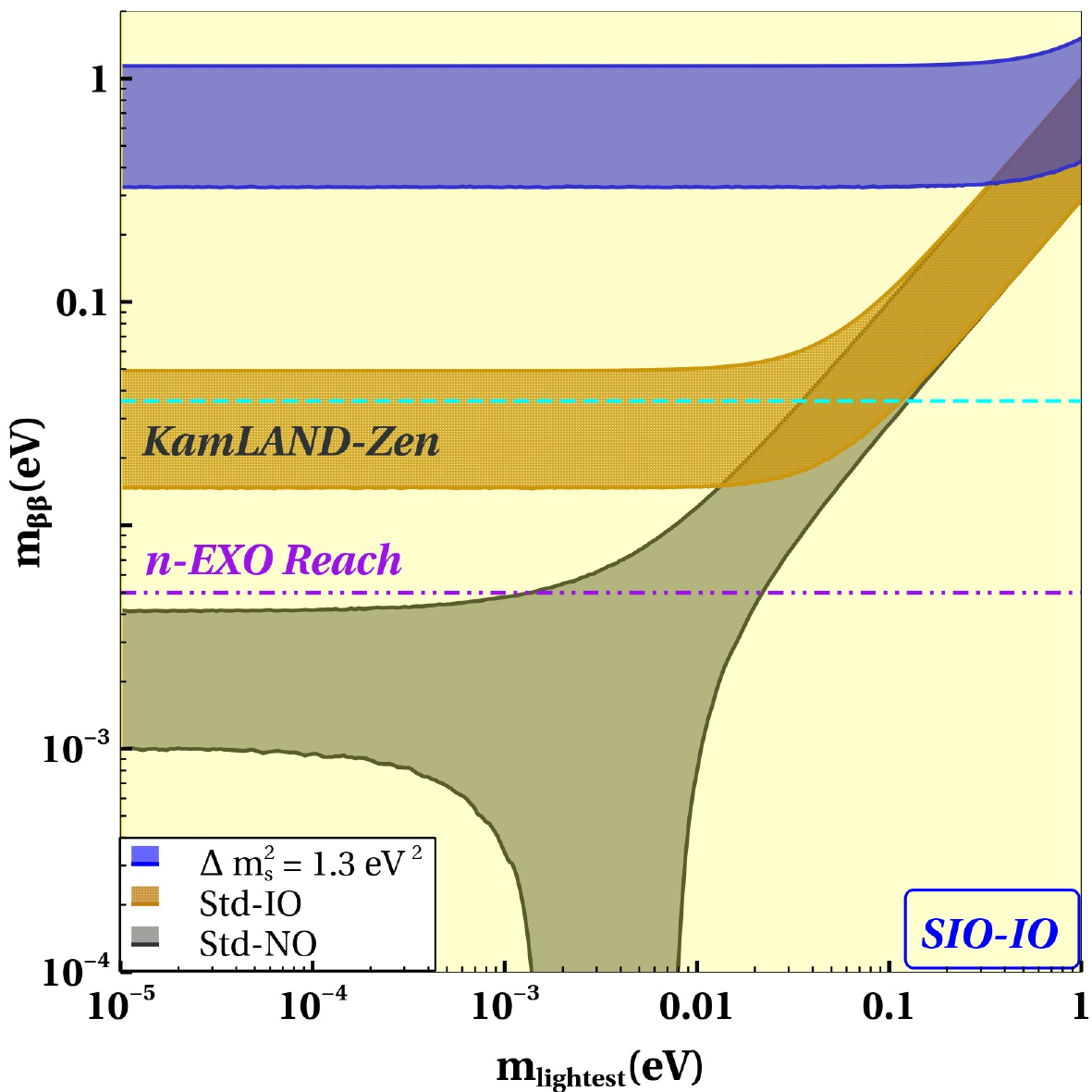}
    \caption{$m_{\beta\beta}$ is plotted for SIO-IO (green) scenario against the lightest neutrino mass  with  the mass squared difference ($\Delta m_s^2 $) = ($10^{-4}\rm \,eV^2$, 0.01 $\rm eV^2$, 1.3 $\rm eV^2$) along with standard three flavor Normal Ordering (red) and Inverted Ordering (yellow) .}
    \label{fig:mbb-comp-sioio}
\end{figure}

In this subsection, the behavior of $m_{\beta\beta}$ is studied in the context of various mass ordering schemes in the presence of a light sterile neutrino. The plots in Figs. (\ref{fig:mbb-comp-snono}, \ref{fig:mbb-comp-snoio}, \ref{fig:mbb-comp-siono}, \ref{fig:mbb-comp-sioio}) are generated by allowing all the oscillation parameters to vary in their $3\sigma$ range as mentioned in table \ref{tab:osc}, and the sterile parameters are varied according to the table (\ref{tab:cosmo_table}). 

\vspace{2mm}
$\star$ \uuline{\textbf{SNO-NO}}\\
The effective Majorana mass in this scenario can be written as 
\beqa
m_{\beta\beta}^{\rm SNO-NO} & = & c_{14}^2 \, \left| \mNO  + t^2_{14} \, m_4 \, e^{i\, \gamma}\right| \label{eq:mbb-SNONO}
\eeqa 
where $\mNO$ is the standard three flavor effective mass for normal ordering. In Fig. (\ref{fig:mbb-comp-snono}), we have plotted $m_{\beta\beta}^{\rm SNO-NO}$ as a function of the lightest neutrino mass ($m_{\rm lightest}=m_1$) for the three mass squared differences. To explain the behavior of $m_{\beta\beta}^{\rm SNO-NO}$ in Fig. (\ref{fig:mbb-comp-snono}), we consider different limits of $m_{\rm lightest}$.
\begin{table}[H]
    \caption{3$\sigma$ ranges of different combinations of oscillation parameters relevant to understanding the effective Majorana mass for SNO-NO in the 3+1 framework.}
    \label{tab:mbb-snono}
    \centering
    \begin{tabular}{|c|c|c|c|c|}
    \hline Regions & {$\mNO$} (eV) & \multicolumn{3}{c|}{ $ \left|m_4 \,t_{14}^{2} \right|$ (eV)} \\ 
    \cline {3-5} &    & $\ms^2 \ = 10^{-4}\, \rm eV^2$ & $\ms^2\, = 0.01\, \rm eV^2$ & $ \ms^2=1.3\, \rm eV^2$ \\
    \hline
    $m_1 \approx 0 $ & 0.001 : 0.004 & 0.001 : 0.002 & $5\times 10^{-5}:10^{-4} $ & 0.001 : 0.01 \\
    \hline 
    $m_1 \approx \sqrt{\msol^2}$ & 0.0018 : 0.018 & 0.0014 : 0.003 & $5\times 10^{-5}:10^{-4} $ & 0.001 : 0.01 \\
    \hline
    $m_1 \approx 0.1$ & 0.02 : 0.1 & 0.01 : 0.02 & $5\times 10^{-4}: 10^{-3}$ & 0.001 : 0.01\\
    \hline
    \end{tabular}
\end{table}
The values of different terms in Eqn. (\ref{eq:mbb-SNONO}) are mentioned for various limits of $m_1$ in the table (\ref{tab:mbb-snono}) where the maximum value of $m_{\beta\beta}^{\rm SNO-NO}$ corresponds to $\gamma=0$ and minimum is for $\gamma=\pi$. The important points are as follows:
\begin{itemize}
    \item For $m_1 << \sqrt{\msol^2} << \sqrt{\matm^2}<< \sqrt{\ms^2}$, it is seen from table (\ref{tab:mbb-snono}) that for $\ms^2 = 10^{-4}, 1.3$ eV$^2$ complete cancellation is possible between $\mNO$ and $m_4 \, t_{14}^2$ for $\gamma = \pi$. 
    \item For $m_1 \approx \sqrt{\msol^2}$, complete cancellations continue to occur for $\ms^2 = 10^{-4}, 1.3$ eV$^2$. 
    \item At higher values of $m_1 \approx 0.1$ eV, complete cancellation happens only for $\ms^2 = 10^{-4}$ eV$^2$ as seen from third row.
    \item In the 3+1 scenario quasi-degenerate (QD) condition will arise when $m_1 \approx m_2 \approx m_3 \approx m_4$. As seen in Fig.(\ref{fig:m_SNONO}) (\ref{app:appendix1}), the QD region occurs around $0.08, 0.2$ eV for $\ms^2=10^{-4},0.01$ eV$^2$. \textit{KamLAND-Zen} and \textit{nEXO} both can probe a fraction of the QD region for $\ms^2=10^{-4}$ eV$^2$ and the entire region for $\ms^2=0.01$ eV$^2$. However, cosmological bounds  ($m_1>0.03$ eV) reject the QD region for both values of $\ms^2$. 
\end{itemize}


$\star$ \uuline{\textbf{SNO-IO}}\\
Effective Majorana mass from double beta decay can be expressed as 
\beqa
    m_{\beta\beta}^{\rm SNO-IO} & = & c_{14}^2 \left|\mIH + t_{14}^2 m_4 \, e^{i \gamma} \right| \label{eq:mbb-snoio}
\eeqa 
We have plotted $m_{\beta\beta}^{\rm SNO-IO}$ as a function of the lightest neutrino mass ($m_{3}$) for the three mass squared differences in the Fig. (\ref{fig:mbb-comp-snoio}). The values of the terms in Eqn. (\ref{eq:mbb-snoio}) are enlisted in the table (\ref{tab:mbb-snoio}).
\begin{table}[H]
       \caption{The 3$\sigma$ ranges of different combinations of oscillation parameters relevant to understanding the effective Majorana mass for SNO-IO in the 3+1 framework.}
        \label{tab:mbb-snoio}
        \centering
        \begin{tabular}{|c|c|c|c|c|}
        \hline Regions & {$\mIH$} (eV) & \multicolumn{3}{c|}{ $m_4 \,t_{14}^{2}$ (eV)} \\ 
        \cline {3-5} &    & $\ms^2 \ = 10^{-4}\, \rm eV^2$ & $\ms^2\, = 0.01\, \rm eV^2$ & $ \ms^2=1.3\, \rm eV^2$ \\
          \hline
           $m_3 \approx 0 $ &$ 0.02  \, : \, 0.05$ & $0.005 \, : \,0.01$ &$ 5 \times 10^{-5}\, : \,5 \times 10^{-4} $& $0.001\,: \, 0.01$ \\
           \hline 
            $m_3\approx 0.1 $ & $0.03 \, : \,0.1$ & $0.01 \, : \,0.025$ &  $ 7.5\times 10^{-5} \, : \, 7.5\times 10^{-4}$ & $0.001 \, : \,0.01 $\\ \hline
        \end{tabular}
\end{table}
The notable points in the SNO-IO case are as follows,
\begin{itemize}
    \item It is evident from table (\ref{tab:mbb-snoio}), that the minimum value of $\mIH$ is always greater than the maximum value of $m_4 \, t_{14}^2$ for all the three mass squared differences. Hence, complete cancellation is not possible for the entire range of $m_{\rm lightest}.$
    
    \item The value of $m_4 \, t_{14}^2$ for $\ms^2 = 0.01 \, \rm eV^{2}$ is very small compared to $\mIH$. Therefore, $m_{\beta\beta}^{\rm SNO-IO}$ is  approximately equal to $\mIH$ which is visible from the middle panel of Fig. (\ref{fig:mbb-comp-snoio}).

    \item For $\ms^2 = 10^{-4}\, \ev^2$ and $1.3 \, \ev^2$, the minimum value of $m_{\beta\beta}^{\rm SNO-IO}  \approx 0.01\, \ev$ is attained for $\gamma=\pi$ which can be probed partially in the future experiment, \textit{nEXO}.

    \item The QD regions, as observed from Fig. (\ref{fig:m_SNOIO}), is occurred at $m_3 > 0.1, 0.2$ eV for $\ms^2=10^{-4},\,0.01$ eV$^2$ respectively. Although the QD region is disfavored by cosmology for both the $\ms^2$ values, \textit{KamLAND-Zen} and \textit{nEXO} can probe this region partially for $\ms^2=10^{-4}$ eV$^2$ and completely for $\ms^2=0.01$ eV$^2$.
    
\end{itemize}
\vspace{2mm}
$\star$ \uuline{\textbf{SIO-NO}}\\ 
The effective Majorana mass is expressed as,
\beqa\label{eq:mbb_siono}
    m^{\rm SIO-NO}_{\beta\beta} & \approx & c_{14}^2 \left( \sqrt{m_4^2+\ms^2} \left( c_{12}^2 + s_{12}^2 \, e^{i \alpha}\right) + m_4 \, t_{14}^2 \, e^{i \, \gamma} \right) \hspace{3em}  \left[ \ms^2 > \matm^2\right] \nn \\
    m^{\rm SIO-NO}_{\beta\beta} & \approx &  c_{14}^2\, \left( c_{13}^2\left( \sqrt{m_4^2+\ms^2} \left( c_{12}^2 + s_{12}^2 \, e^{i \alpha}\right) + \sqrt{m_4^2+\matm^2}\, s_{13}^2 \, e^{i \beta} \right)  + m_4 \, t_{14}^2 \, e^{i \, \gamma} \right)  \nn\\ 
    & & \hspace{25em} \left[ \ms^2 < \matm^2\right]  
\eeqa
Here, we have used the mass relations mentioned in Eqn. (\ref{eq:m_siono}). In Fig. (\ref{fig:mbb-comp-siono}), we have shown $m_{\beta\beta}$ as function of $m_{\rm lightest}$ ($m_4$) in three panels corresponding to different values of $\ms^2$. The table (\ref{tab:mbb-siono}) depicts the terms of Eqn. (\ref{eq:mbb_siono}).
\begin{table}[H]
       \caption{The 3$\sigma$ ranges of different combinations of oscillation parameters relevant to understanding the effective Majorana mass for SIO-NO in the 3+1 framework.}
        \label{tab:mbb-siono}
        \centering
        \begin{tabular}{|c|c|c|c|c|c|c|c|}
        \hline Regions & \multicolumn{3}{c|}{$\sqrt{\ms^2} \,\cos 2\theta_{12}$ (eV)} & {$\sqrt{\matm^2}\, t_{13}^2$ (eV)} & \multicolumn{3}{c|}{$m_4 \, t_{14}^2$ (eV)}\\ 
        \cline{2-4} \cline{6-8} & $ \Delta = 10^{-4} $& $ \Delta =0.01 $ &$ \Delta = 1.3$ &  &$ \Delta= 10^{-4} $&$ \Delta=0.01 $&$  \Delta= 1.3$\\
         \hline
         $m_4 \approx 0$ & $0.003$ & 0.03 & 0.33 & 0.001 &  0 & 0 & 0 \\ \hline
         $m_4 \approx 0.01$ & $0.003$ & 0.03 & 0.33 & 0.001 & 0.001 : 0.002 & $5.\,10^{-5}:5.\,10^{-4}$ & $10^{-5}:10^{-4}$ \\ \hline
         \end{tabular}
\end{table}
\begin{itemize}
    \item For the region where the lightest mass is negligible, Eqn. (\ref{eq:m_siono}) will be,
    \beqa
        m_{4} \approx 0, \quad   m_2 &\approx & \, m_1\, \approx m_{3} \, \approx \sqrt{\ms ^2 }  \hspace{6em} \left( \ms^2 > \matm^2\right) \nn \\
        m_4 \approx 0 ,\quad  m_1 &\approx & m_2 \approx \, \sqrt{\ms^2}, \, \,  m_3 \, \approx \sqrt{\matm^2} \hspace*{1em}  \left( \ms^2 < \matm^2\right) \label{eq:m_siono2}
    \eeqa 

    Effective Majorana mass from double beta decay
    \beqa
        m^{\rm SIO-NO}_{\beta\beta} & = & \sqrt{\ms^2} \left( c_{12}^2 + s_{12}^2 \, e^{i \alpha}\right) \hspace{13em} \left( \ms^2 > \matm^2\right) \nn \\
        m^{\rm SIO-NO}_{\beta\beta} & = & c_{13}^2 \, c_{14}^2 \left(\sqrt{\ms^2} \left( c_{12}^2 + s_{12}^2 \, e^{i \alpha}\right) + \sqrt{\matm^2}\, t_{13}^2 \, e^{i \beta} \right) \hspace{1em} \left( \ms^2 < \matm^2\right) \nn \\
    \eeqa 
    In the first case, complete cancellation can happen for $\alpha = \pi $ and $c_{12}^2 = s^2_{12}$. But, since $\theta_{12}$ is less than $45^\circ$, this cannot happen, as shown in Fig. (\ref{fig:mbb-comp-siono}) for $\ms=10^{-4}$ eV$^2$. In the second case, complete cancellation occurs for $\alpha= \beta = \pi$ and
    \beqa
        \sqrt{\ms^2}\, \cos 2\theta_{12} = \sqrt{\matm^2} \, t_{13}^2
    \eeqa
    This condition is not satisfied for $\ms^2=1.3, \, 0.01$ eV$^2$ as can be seen from table (\ref{tab:osc}) and table (\ref{tab:mbb-siono}). The value of $m_{\beta\beta}^{\rm SIO-NO}$ varies between $(0.3 \, :\,1)$ eV and $(0.001\, : 0.01 \, )$ eV for $\ms^2=1.3,10^{-4}$ eV$^2$ respectively, as seen in from Fig. (\ref{fig:mbb-comp-siono}).

    \item Around $m_4 \approx 0.01 \, $eV, in case of $\ms^2 = 0.01 \, \rm eV^2 \, and \, 1.3 \, eV^2$, the sterile contribution is negligible compared to other terms as the value of $\theta_{14}$ is small and thus no cancellation occurs. But due to large $\theta_{14} $ for $\ms^2 = 10^{-4}$, the value of $m_4 \,t_{14}^2$ varies between $(0.001\,:\,0.002)$ which allows us to have a narrow cancellation region for $\alpha=\beta=\gamma= \pi$.
 
    \item It is to be noted that the \textit{KamLAND-ZEN} experiment disallows the entire parameter space of $m_{\beta\beta}^{\rm SIO-NO}$ for $\ms^2 = 1.3 \rm\,  eV^2$. For $\ms^2=0.01$ eV$^2$ a part of the parameter space gets disfavored for all values of $m_{\rm lightest}$, whereas for $\ms^2=10^{-4}$ eV$^2$ regions with higher values of $m_{\rm lightest}(>0.3 \, \ev)$ are disfavored. For $10^{-4}$ eV$^2$, the allowed region of $m_{\beta\beta}^{\rm SIO-NO}$ can be partially probed by \textit{nEXO} experiment.
\end{itemize}
  
\vspace{2mm}
$\star$ \uuline{\textbf{ SIO-IO}}\\

In three panels of Fig. ({\ref{fig:mbb-comp-sioio}}), Majorana mass $m_{\beta\beta}$ in SIO-IO scenario has been plotted against $m_{\rm lightest}=m_4$.
\begin{itemize}
    \item For $\ms^2 > \matm^2$, $m_{\beta\beta}$ is exactly similar to the SIO-NO scenario ($\ms^2 > \matm^2$). Thus, the results and the conclusions remain identical. 
    
    \item For $\ms^2 < \matm^2$ , the value of $m_{\beta\beta}$ in a region where $m_{\rm lightest}$ is small, can be approximated as
    \beqa
    \left(m_{\beta\beta}^{\rm SIO-IO} \right)_{\ms^2<\matm^2} = \sqrt{\matm^2}\, c_{14}^2 \left( c_{12}^2 + s_{12}^2 \,e^{i \alpha} +t_{14}^2\, e^{i\gamma} \right) \label{eq:mbb_sioio_caseii}
        \eeqa
       Here, complete cancellation requires $\alpha = \gamma = \pi$ and \beqa \cos 2\theta_{12} = t_{14}^2 \label{eq:cancel_sioio}
    \eeqa
    In this region, complete cancellation is not possible as Eqn. (\ref{eq:cancel_sioio}) is not satisfied for the allowed range of mixing angle $\theta_{14}$ given in table (\ref{tab:cosmo_table}). 
        
    \item When $m_{\rm lightest} > \sqrt{\matm^2}$ , $m_1\approx m_2 \approx m_3 \approx m_4\approx m_0$ and the value of $m_{\beta\beta}$ can be written as 
    \beqa
        \left(m_{\beta\beta}^{\rm SIO-IO} \right)_{\ms^2<\matm^2} = m_0\, c_{14}^2 \left( c_{12}^2 + s_{12}^2 \,e^{i \alpha} +t_{14}^2\, e^{i\gamma} \right) 
    \eeqa
    In this region, cancellation is also not possible, and $m_{\beta\beta}$ is proportional to the value of the lightest mass.

    \item It can be seen from Fig. (\ref{fig:mbb-comp-sioio}), higher values of $m_{\beta\beta}$ are disfavored by \textit{KamLAND-Zen} for all values of $m_{\rm lightest}$ and \textit{nEXO} can rule out an even greater part of the parameter space in the absence of any signal.
\end{itemize}
The expressions of $m_{\beta\beta}$ in various $m_{lightest}$ limits are tabulated in table (\ref{tab:mbetambeta}) in the appendix.

\subsection{Correlations}

As discussed, cosmology, direct mass measurement experiment (single $\beta$ decay), and $0\nu\beta\beta$ decay put independent constraints on the absolute mass scale of the neutrino. In this section, we discuss the correlations of the mass observable amongst each other. We have plotted in Figs. (\ref{fig:corr-comp_SNONO}-\ref{fig:corr-comp-SIOIO}), the correlation of $m_{\beta}$ against $\Sigma$ (left), $m_{\beta\beta}$ against $\Sigma$ (middle), and $m_{\beta\beta}$ against $m_{\beta}$ (right) for all the mass spectra. The yellow-shaded and the brown-hatched regions correspond to cosmologically excluded regions mentioned in Eqn. (\ref{eq:cosmo2}) and Eqn. (\ref{eq:cosmo3}), respectively. The other horizontal and vertical lines are the current experimental limits (\textit{KamLAND-Zen} [Cyan]) and future sensitivity (\textit{KATRIN} [Pink], \textit{Project 8} [Black], \textit{nEXO} [Magenta]) with their respective color mention in brackets. Blue, red, and green regions in the plots of Fig. (\ref{fig:corr-comp_SNONO}-\ref{fig:corr-comp-SIOIO}) correspond to $\ms^2=1.3, \, 0.01, \, 10^{-4}$ eV$^2$ respectively.

\begin{enumerate}
    \item {\bf SNO-NO:}
    \begin{figure}[H]
        \centering
        \includegraphics[width=0.32\linewidth]{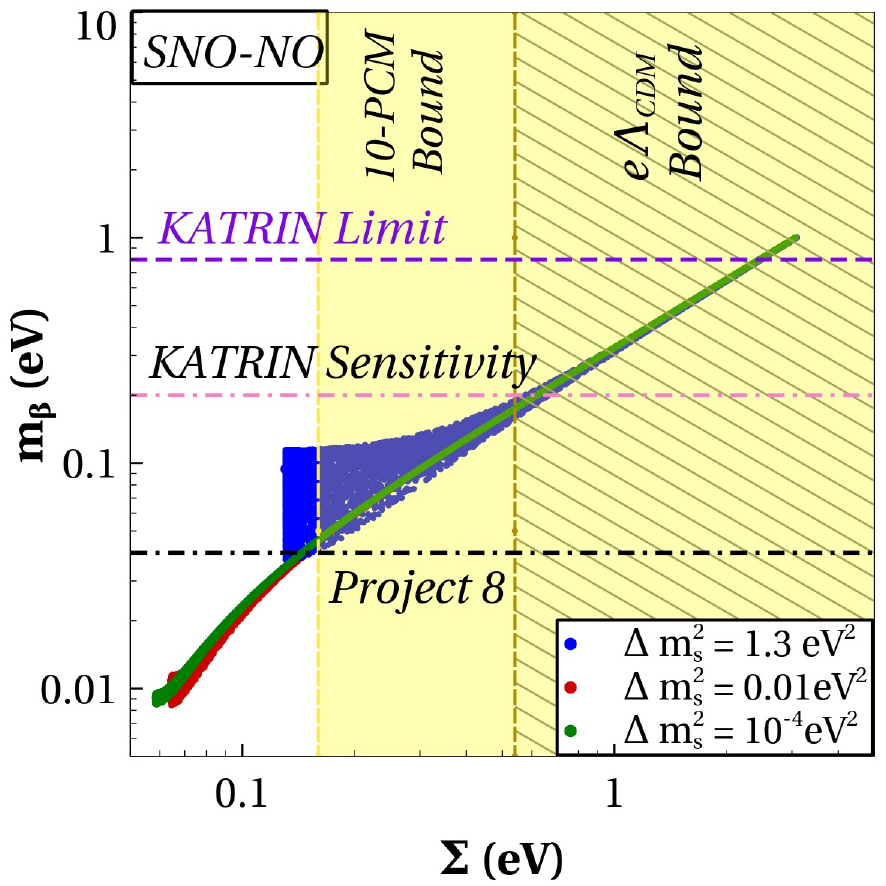}
        \includegraphics[width=0.32\linewidth]{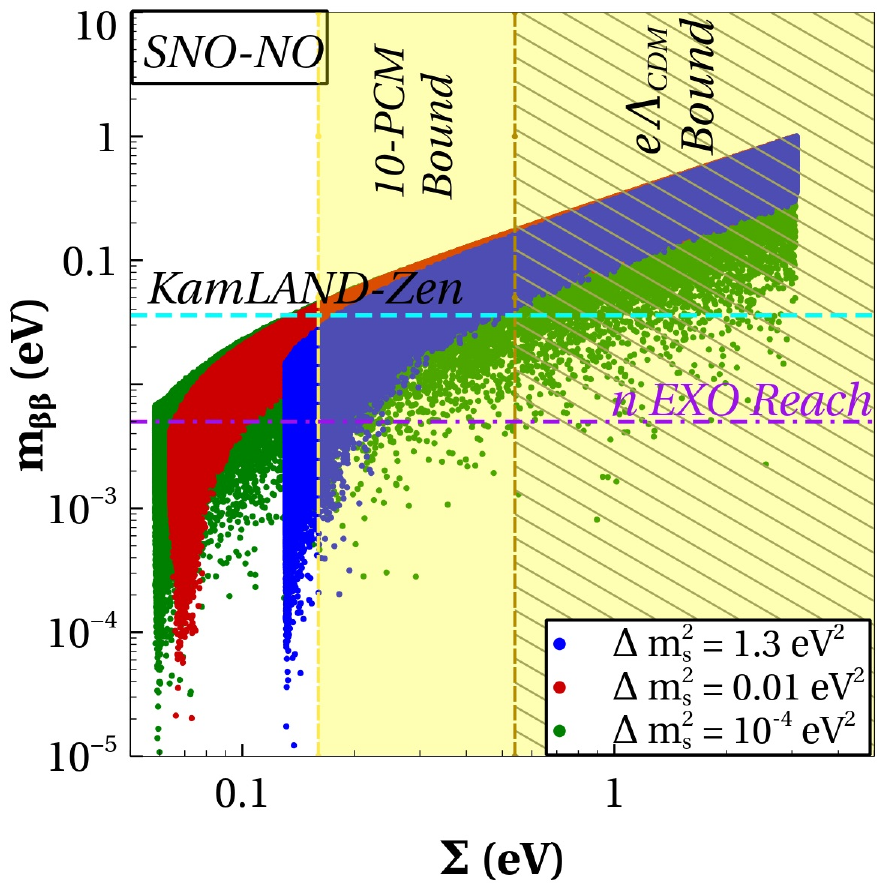}
        \includegraphics[width=0.32\linewidth]{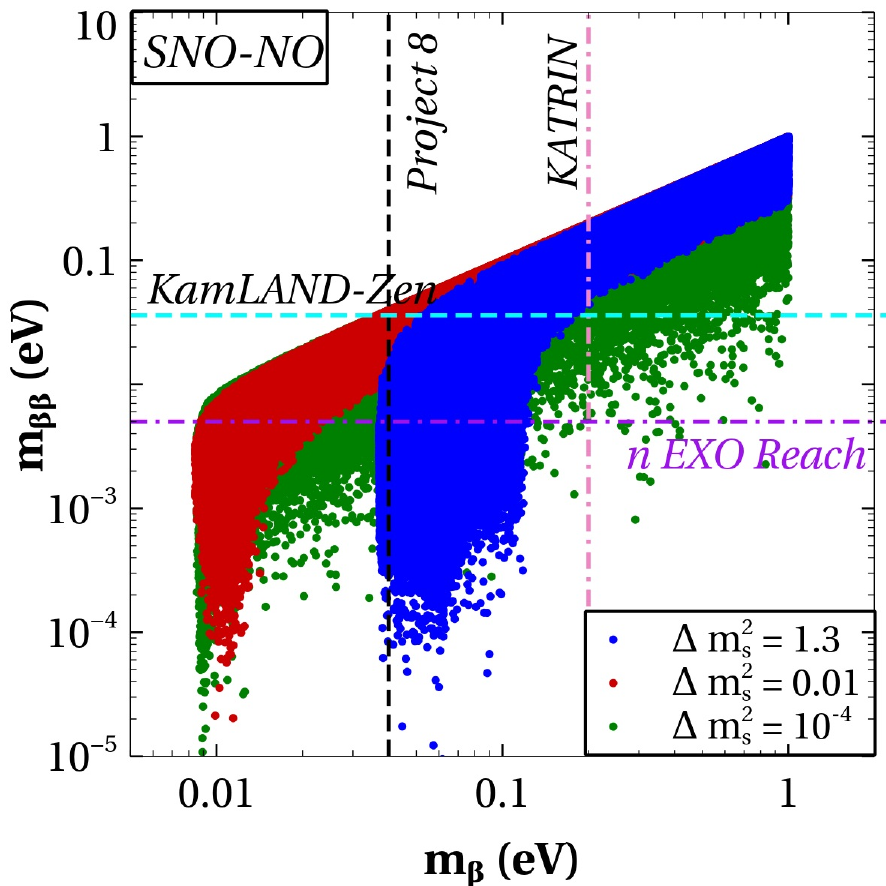}
        \caption{Correlations of $m_{\beta}$ against  and $\Sigma $(left) , $m_{\beta\beta}$ against $ \Sigma$ (middle) and $m_{\beta\beta}$ against $m_{\beta}$ (right) for SNO-NO is plotted here. The green, blue, and red regions describe the values for $\ms^2= 10^{-4} \rm \, eV^2$, $0.01 \rm \, eV^2$ and $1.3 \rm \, eV^2$ respectively. The yellow-shaded and brown-hatched regions correspond to the exclusion regions by Eqn. (\ref{eq:cosmo2}) and Eqn. (\ref{eq:cosmo3}) respectively.}
        \label{fig:corr-comp_SNONO}
    \end{figure}
   The correlation plots for SNO-NO are shown in  Fig. (\ref{fig:corr-comp_SNONO}).
   
    \begin{itemize}
        \item From the left panel, it is seen that the cosmological mass bound disfavors a large parameter space for all three mass-squared differences. The allowed region from cosmology will not be sensitive to \textit{KATRIN}'s projected limit, but the proposed \textit{Project 8} experiment can probe the parameter space for $\ms^2 = 1.3\, \ev^2$.

        \item From the middle panel, it is observed that some part of the parameter space disfavored by the cosmological bound is also disfavored by \textit{KamLAND-Zen}. In the region allowed by cosmology, $m_{\beta}$ can be very low. Therefore, \textit{KamLAND-Zen} can probe a very small part of it, and the projected sensitivity \textit{nEXO} experiment can only probe some parts of these regions for all the mass-squared differences. 

        \item From the right panel, it can be noted that the proposed experiments \textit{nEXO} and Project-8 together can rule out almost the entire parameter space for $\ms^2=1.3$ eV$^2$ in the absence of any signal. However, in the case of $\ms^2=0.01,10^{-4}$ eV$^2$, only parts of the parameter space can be probed by the upcoming above-cited experiments.
    \end{itemize}

    \item {\bf SNO-IO:}
    \begin{figure}[H]
        \centering
        \includegraphics[width=0.32\linewidth]{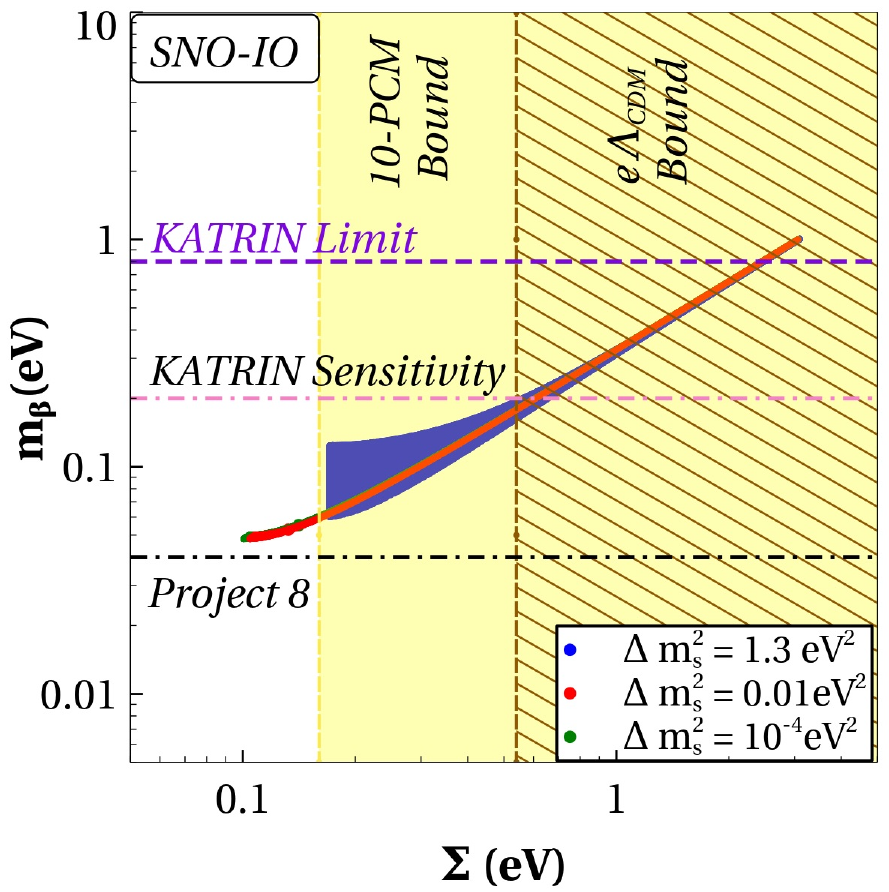}
        \includegraphics[width=0.32\linewidth]{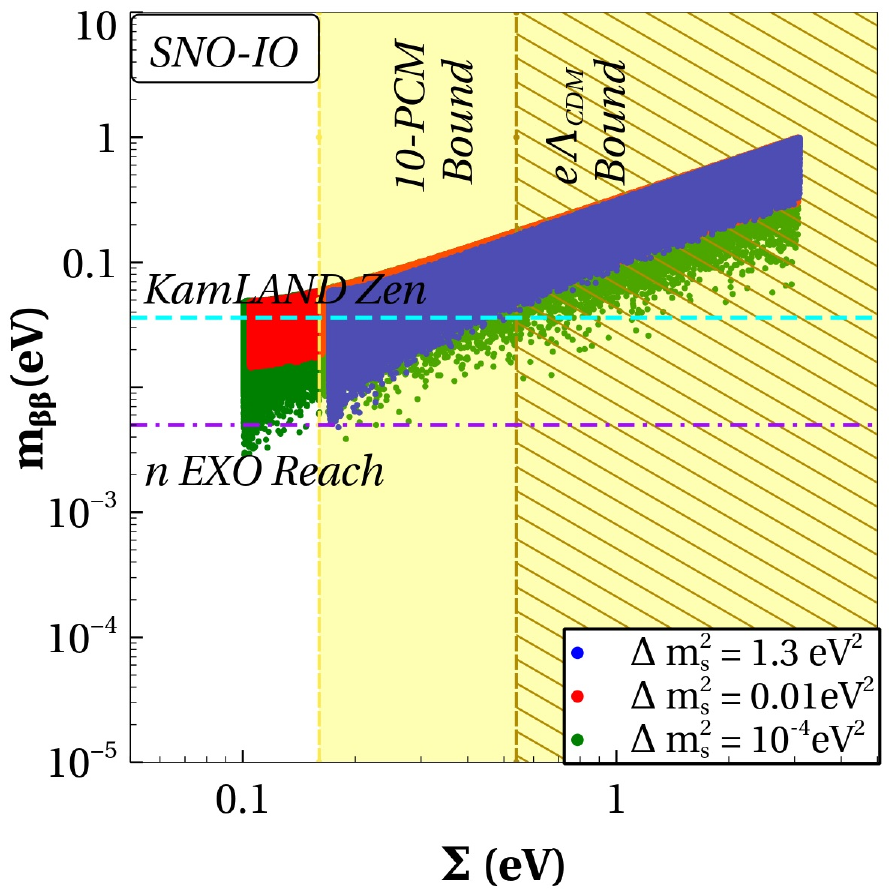}
        \includegraphics[width=0.32\linewidth]{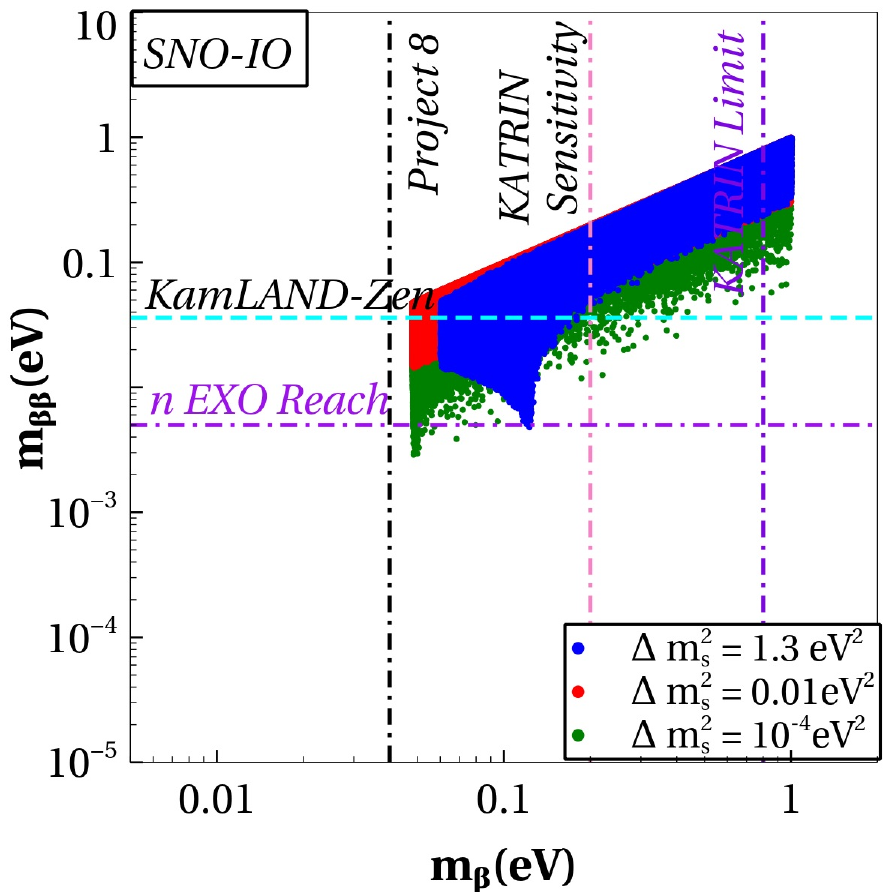}
        \caption{Correlations of $m_{\beta}$ against  and $\Sigma $(left) , $m_{\beta\beta}$ against $ \Sigma$ (middle) and $m_{\beta\beta}$ against $m_{\beta}$ (right) for SNO-IO is plotted here. The green, blue, and red regions describe the values for $\ms^2= 10^{-4} \rm \, eV^2$, $0.01 \rm \, eV^2$ and $1.3 \rm \, eV^2$ respectively. The yellow-shaded and brown-hatched regions correspond to the exclusion regions by Eqn. (\ref{eq:cosmo2}) and Eqn. (\ref{eq:cosmo3}) respectively.}
        \label{fig:corr-comp-SNOIO}
    \end{figure}

    Fig. (\ref{fig:corr-comp-SNOIO}) shows the correlation plots for SNO-IO.
    \begin{itemize}
        \item From the left panel, it is visible that $\ms^2 = 1.3\, \rm eV^{2}$ is ruled out by stringent cosmological limit. But for $\ms^2\, = 10^{-4} \,\ev^2$ and $ 0.01 \, \ev^2$ small parts of parameter space are allowed by cosmology and  \textit{KATRIN}'s projected sensitivity. These allowed regions can be completely probed in the proposed \textit{Project 8} experiment. 
        
        \item It can be noted from the middle panel that \textit{KamLAND-Zen} and cosmology rule out a large part of the parameter space for all the mass-squared differences. For $\ms^2 = 0.01, \, 10^{-4}\,\ev^2$, the region allowed by cosmology and \textit{KamLAND-Zen} can be probed in future experiment \textit{nEXO}.

        \item From the right panel, it is observed that \textit{Project 8} and \textit{nEXO} experiments together can probe the entire parameter space for $\ms^2 = \, 10^{-4},\, 0.01 $ and $ 1.3 \, \ev^2$.
        
    \end{itemize}
    \item  {\bf SIO-NO:}
    \begin{figure}[H]
        \centering
        \includegraphics[width=0.32\linewidth]{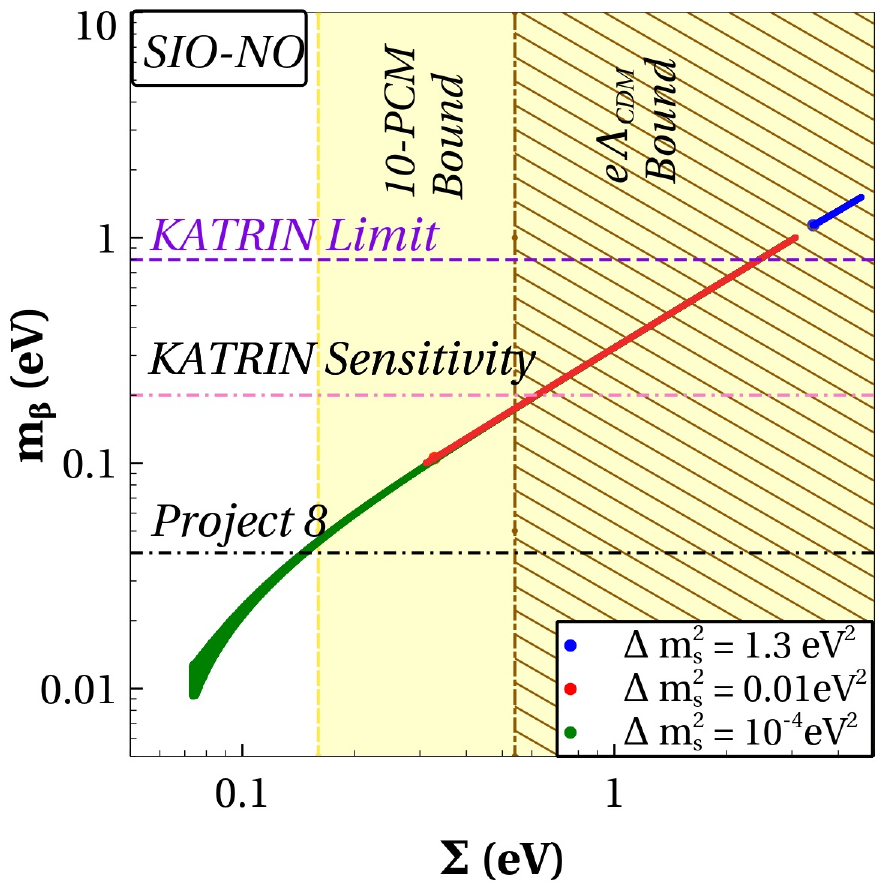}
        \includegraphics[width=0.32\linewidth]{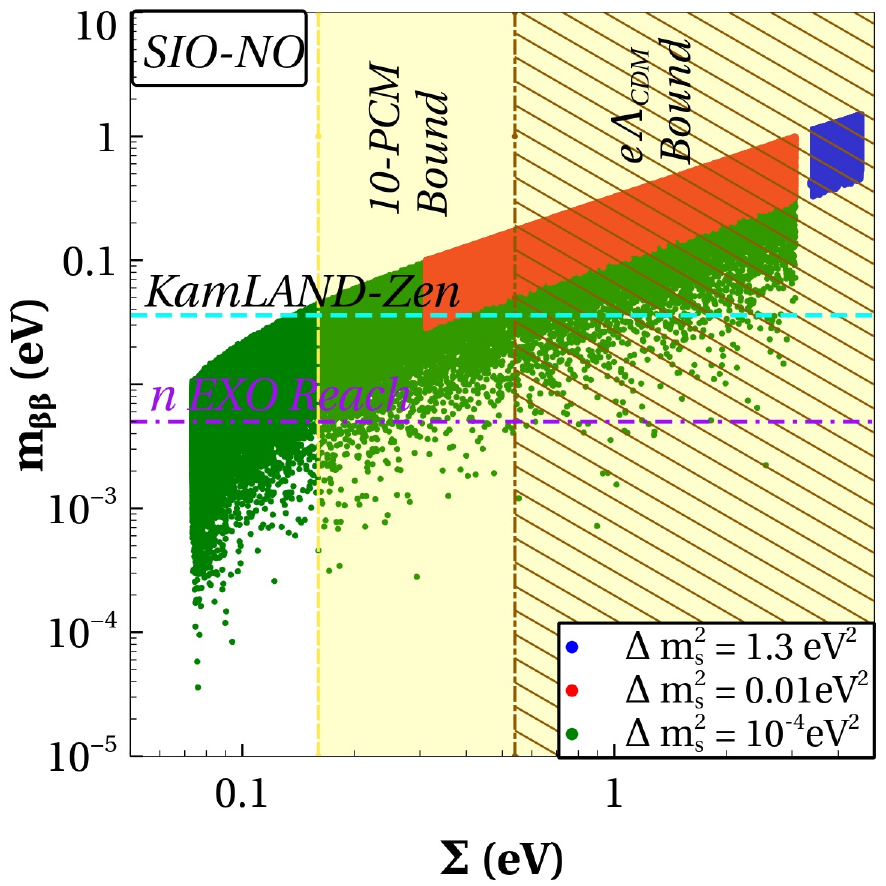}
        \includegraphics[width=0.32\linewidth]{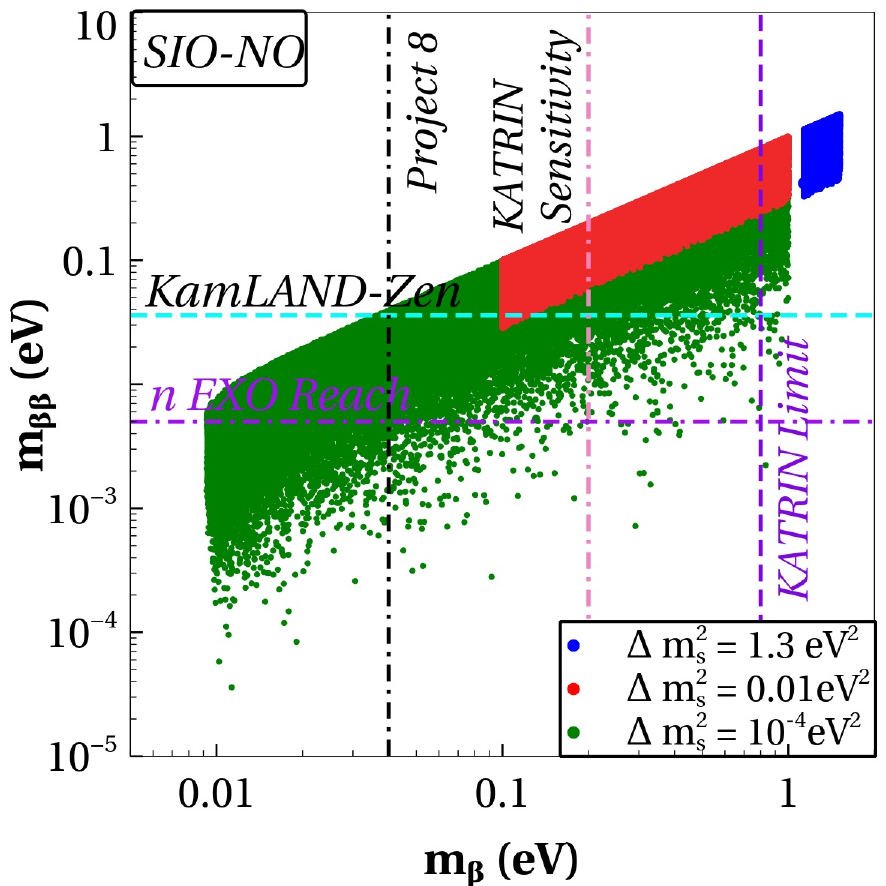}
        \caption{Correlations of $m_{\beta}$ against  and $\Sigma $(left) , $m_{\beta\beta}$ against $ \Sigma$ (middle) and $m_{\beta\beta}$ against $m_{\beta}$ (right) for SNO-IO is plotted here. The green, blue, and red regions describe the values for $\ms^2= 10^{-4} \rm \, eV^2$, $0.01 \rm \, eV^2$ and $1.3 \rm \, eV^2$ respectively. The yellow-shaded and brown-hatched regions correspond to the exclusion regions by Eqn. (\ref{eq:cosmo2}) and Eqn. (\ref{eq:cosmo3}) respectively.}
        \label{fig:corr-comp-SIONO}
    \end{figure}

    In Fig. (\ref{fig:corr-comp-SIONO}), correlations between the mass variables for the SIO-NO scenario are plotted. 
    \begin{itemize}
        \item The left and middle panels show that $e\Lambda_{CDM}$ model only allows a part of the parameter space for $\ms^2=0.01, 10^{-4}$ eV$^2$, however with the cosmological bound only $\ms^2=10^{-4} \, \ev^2$ is preferred.

        \item From the left panel, it is visible that the current \textit{KATRIN} bound can't probe the regions allowed by cosmological and $e\Lambda_{CDM}$ model. Only proposed \textit{Project 8} can probe allowed regions for $1.3\,\ev^2$. 

        \item The middle panel depicts that the \textit{KamLAND-Zen} sensitivity will be able to probe the $e\Lambda_{CDM}$ favored regions of $ 10^{-4} \, \ev^2$ and $0.01 \, \ev^2$ partially. The \textit{nEXO} can completely probe allowed regions of $0.01 \, \ev^2$.

        \item It is to be noted from the right panel that the future experiments \textit{nEXO} and \textit{Project 8} can together probe the entirety of the parameter space for  $\ms^2=1.3, \,0.01$ eV$^2$, and a fraction of the regions for $\ms^2=10^{-4}$ eV$^2$.
    \end{itemize}

    \pagebreak
    
    \item  {\bf SIO-IO:}
    \begin{figure}[H]
        \centering
        \includegraphics[width=0.32\linewidth]{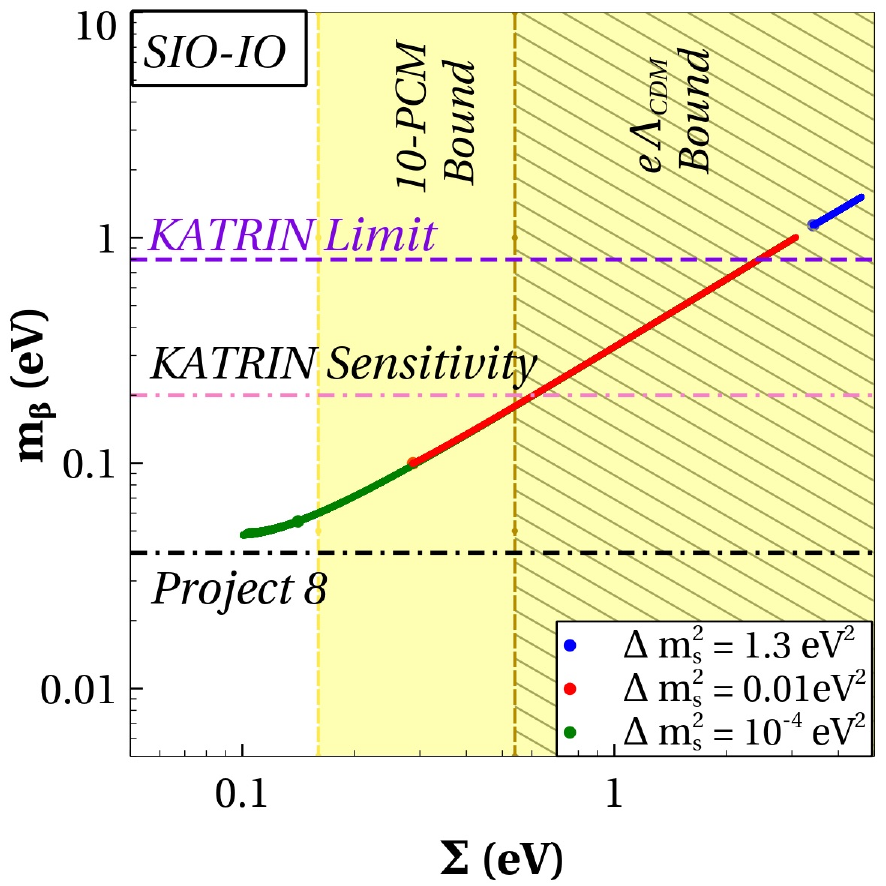}
        \includegraphics[width=0.32\linewidth]{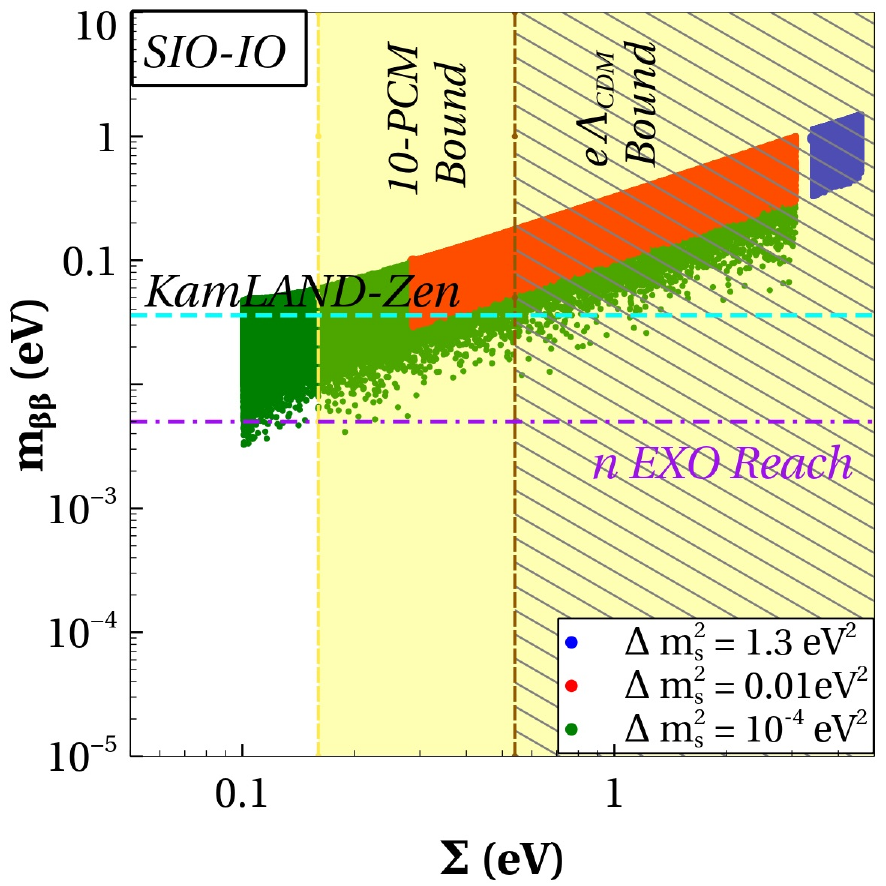}
        \includegraphics[width=0.32\linewidth]{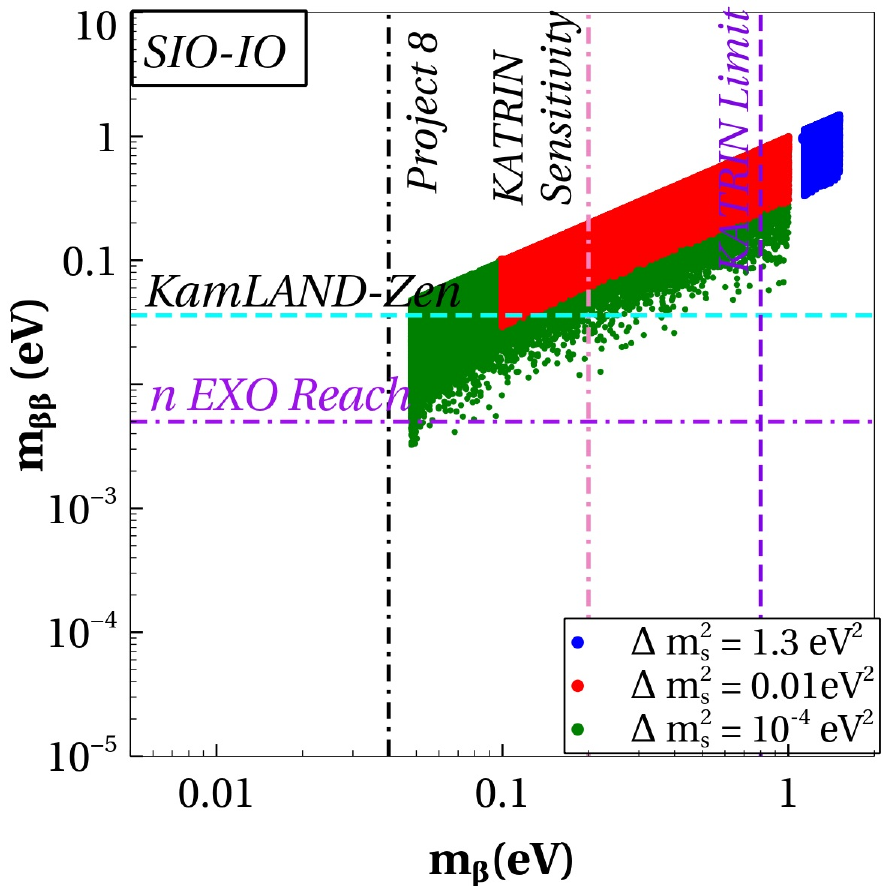}
        \caption{Correlations of $m_{\beta}$ against  and $\Sigma $(left) , $m_{\beta\beta}$ against $ \Sigma$ (middle) and $m_{\beta\beta}$ against $m_{\beta}$ (right) is plotted here. The green, blue, black, and red regions describe SNO-NO, SNO-IO, SIO-NO \& SIO-IO, respectively. The shaded regions correspond to the exclusion regions of the respective x-axis labels. }
        \label{fig:corr-comp-SIOIO}
    \end{figure}
    
    The correlations amongst the mass variables for the SIO-IO scenario are plotted in Fig. (\ref{fig:corr-comp-SIOIO}).

    \begin{itemize}
        \item From the left panel, we understand that the SIO-IO scenario is similar to the SIO-NO scenario. The only difference is that \textit{Project 8} will be able to probe the entire cosmologically allowed regions of $\ms^2=10^{-4}$ eV$^2$.
        
        \item The middle panel portrays similar observations to that of SIO-NO apart from the fact that now future experiment \textit{nEXO} can cover almost the total parameter space for all $\ms^2$ values considered by us.

        \item From the right panel, it can be seen that the proposed experiments \textit{Project 8} and \textit{nEXO} can together cover the entire parameter space for all the values of $\ms^2$. 
    \end{itemize}

\end{enumerate}

\section{Summary and Discussion}
The results from short baseline neutrino oscillation experiments e.g. \textit{LSND} and \textit{MiniBooNE} and radio-chemical experiments e.g \textit{GALLEX}, \textit{SAGE}, \& \textit{BEST} indicate the possibility of having an extra neutrino state with $\mathcal{O}( \,\rm eV)$ mass squared difference. Moreover, the tension between the results of T2K and NO$\nu$A experiments can be improved by invoking an additional state mass squared difference $\sim 10^{-2}\, \rm eV^2$ and lack of upturn events in the solar neutrino spectra below 8 MeV can be explained by an ultralight sterile neutrino. Thus, sterile neutrinos with a very wide range of mass differences ($\ms^2 = m_4^2 - m_1^2$) have been proposed in the literature. The addition of a sterile state implies four mass spectra, namely: SNO-NO ($\ms^2>0\,,\, \matm^2 >0$), SNO-IO ($\ms^2>0\,,\, \matm^2 <0$), SIO-NO ($\ms^2<0\,,\, \matm^2 >0$), and SIO-IO ($\ms^2<0\,,\, \matm^2 <0$) where NO (IO) stands for +ve (-ve) value of $\Delta m^2_{31}$ and SNO(SIO) stand for +ve (-ve) value of $\Delta m^2_{41}$. The mass spectra are depicted in Fig. (\ref{fig:mass-spectrum}). We explore the implications of the mass spectra for sum of light neutrino masses from cosmology, beta decay, and $0\nu\beta\beta$ decay.  
\begin{itemize}
 \item    The scenario of $\ms^2 = 1.3 \, \ev^2$ with $\ms^2<0$ is known to be in conflict with the cosmological bound on the sum of neutrino masses. The specific bounds depend on the chosen data sets and the cosmological models used for fitting. Here we consider two different cosmological models: a 10 parameter cosmological model (\textit{10-PCM}) and a 12 parameter cosmological model $(e\Lambda_{CDM})$ which provide the limit on the total mass of the light neutrino species as $\sum < 0.16 \, \ev$ and $\sum<0.52 \, \ev$ respectively. We find that SIO-NO and SIO-IO is completely ruled out by cosmology. Moreover, such scenarios are disfavored from the current limit on $m_{\beta}$ by \textit{KATRIN} experiment and also from the upper limit on $m_{\beta\beta} $ by \textit{KamLAND-Zen} experiment. We want to emphasize that SIO-NO and SIO-IO scenarios for $\ms^2 = 1.3 \, \ev^2$ are not only disfavored by cosmology but also by \textit{KATRIN} and \textit{KamLAND-Zen}. However, we see that SNO-NO and SNO-IO for $\ms^2=1.3\, \ev^2 $ is still allowed below $m_{\rm lightest} \approx 0.1\, \ev $, in the limit of $e\Lambda_{CDM}$ model, \textit{KATRIN} and \textit{KamLAND-Zen} but proposed experiment \textit{Project 8 } will be able to probe the scenarios with the projected limit of $m_{\beta}$. 
    \item It is often believed that sterile neutrinos with mass-squared difference smaller than 1.3 $\ev^2$ can be allowed by cosmology. Here we find that, for $\ms^2 = 0.01 \, \ev^2$, all mass spectra are allowed in $e\Lambda_{CDM}$ model up to a value of $m_{\rm lightest} \approx 0.1 \, \ev$ but SIO-NO and SIO-IO is disfavored when \textit{10-PCM} model is considered whereas SNO-NO and SNO-IO scenarios remain valid up to $m_{\rm lightest} \sim 0.03 \, \ev$. It is also noted that projected sensitivity from \textit{KATRIN} experiments will not be able to probe the mass spectra, but SNO-IO, SIO-NO, and SIO-IO scenarios can be probed completely with \textit{Project 8}'s proposed sensitivity. In the case of neutrinoless double decay measurements, \textit{KamLAND-Zen} experiment ruled out most of the parameter space of SIO-NO and SIO-IO scenario for $\ms^2 = 0.01\, \ev^2$ and next generation experiment $nEXO$ will be able to probe the parameter space completely. Moreover, \textit{nEXO} will also be able to probe the SNO-IO scenario completely for $\ms^2=0.01\, \ev^2$. 
    \item It is seen from Fig. (\ref{fig:mtotal}) that $\ms^2 = 10^{-4} \, \ev^2$ i.e sterile neutrino with very small mass-squared difference is allowed up to $m_{\rm lightest} \approx 0.03 \, \ev$  and up to $m_{\rm lightest} \approx 0.1 \, \ev$ from $e\Lambda_{CDM}$ model. In case of direct mass measurement, \textit{KATRIN}'s projected limit can probe the mass spectra up to $m_{\rm lightest} \approx \, 0.2\, \ev$ whereas \textit{Project 8 } will be able to probe SNO-IO, SIO-IO scenarios completely and SNO-NO, SIO-NO scenarios up to $m_{\rm lightest}\approx 0.04 \, \ev$. We also find that neither \textit{KamLAND-Zen} nor \textit{nEXO} can completely probe the mass spectra, but they rule out some parameter space for SNO-IO, SIO-NO and SIO-IO scenarios. 
\end{itemize}


\noindent In conclusion, in the presence of a light sterile state, mass-related observables can provide constraints on the possible spectra and can disfavor some of these depending on the mass of the sterile state. 

\section*{Acknowledgement}
SG acknowledges the J.C. Bose Fellowship (JCB/2020/000011) of the Science and Engineering Research Board of the Department of Science and Technology, Government of India. She also acknowledges Northwestern University (NU), where the majority of this work was done, for hospitality and Fullbright-Neheru fellowship for funding the visit to NU. The computations were performed on the Param Vikram-1000 High Performance Computing Cluster of the Physical Research Laboratory (PRL). We also acknowledge Arup Chakraborty for his help in learning the use of HPC. 

\appendix
\section{Mass-spectrum\label{app:appendix1}}

\begin{figure}[H]
    \centering
     \includegraphics[width=0.32\linewidth]{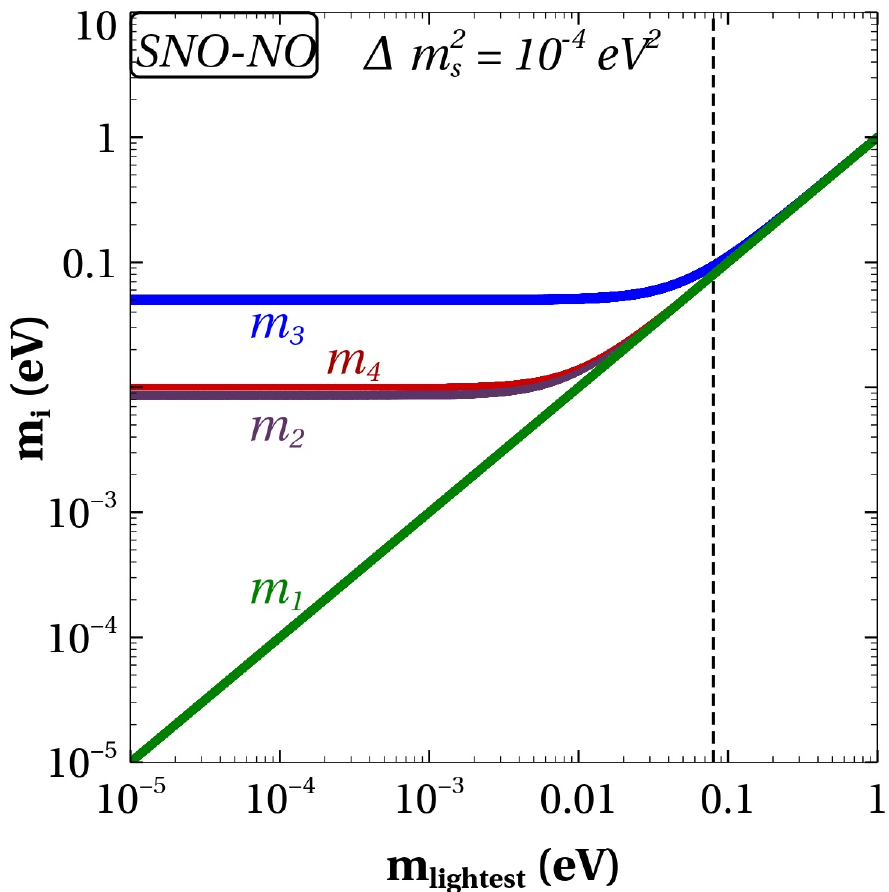}
    \includegraphics[width=0.32\linewidth]{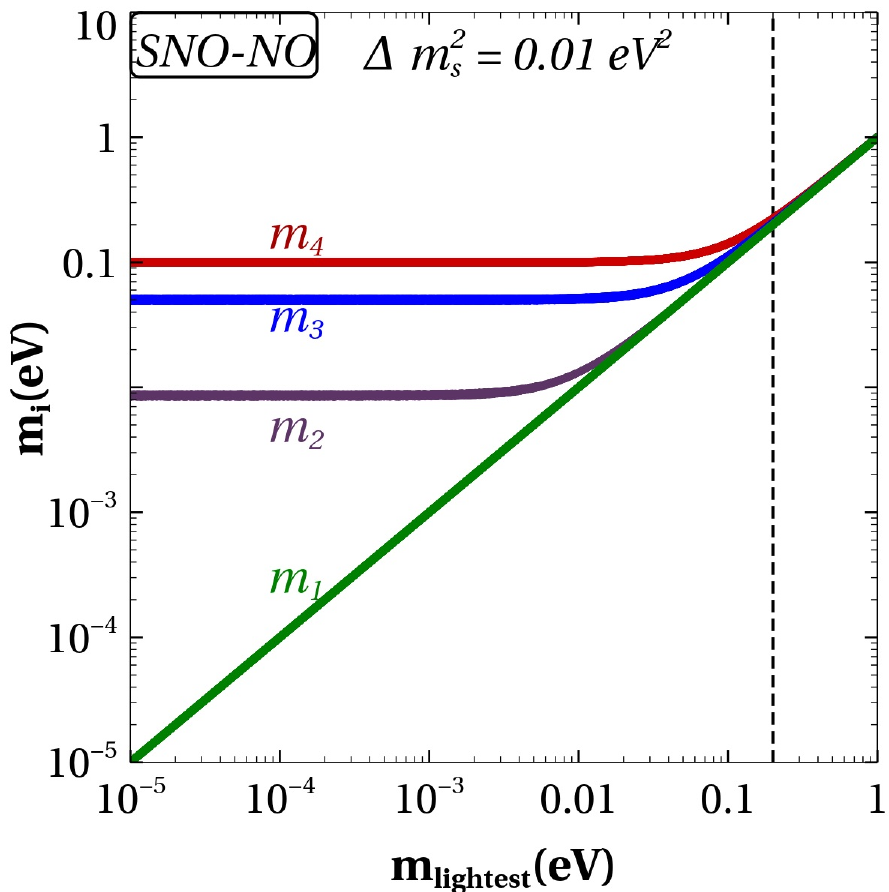}
   \includegraphics[width=0.32\linewidth]{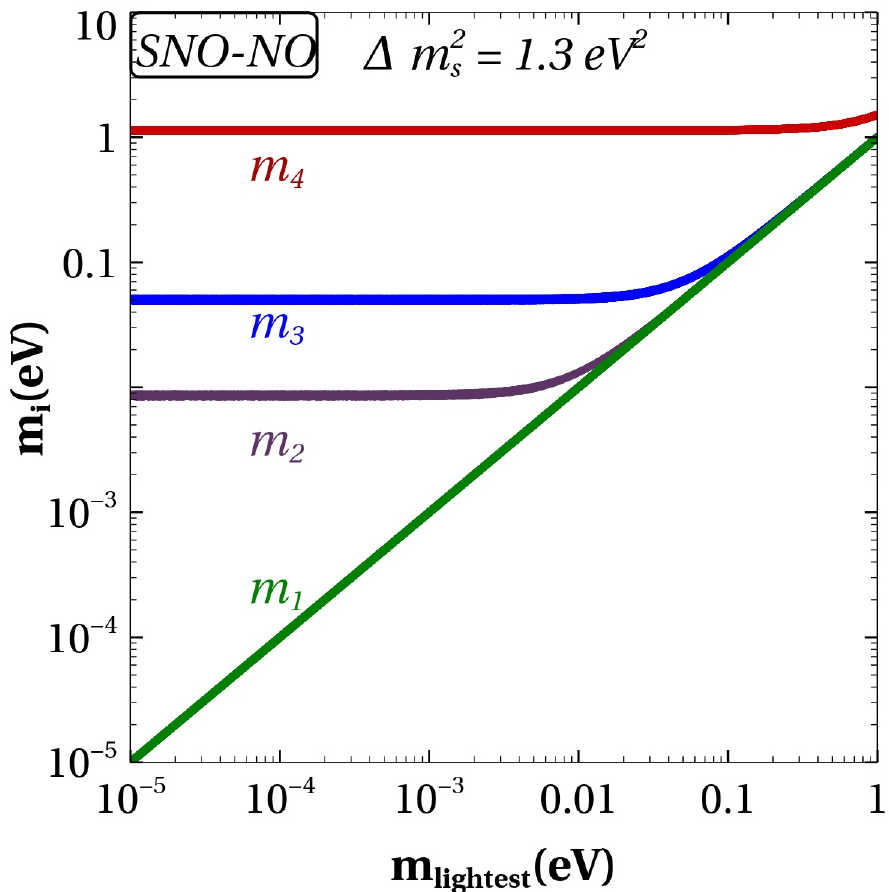}
    \caption{variation of masses with respect to the lightest neutrino mass for SNO-NO}
    \label{fig:m_SNONO}
\end{figure}

\begin{figure}[H]
    \centering
     \includegraphics[width=0.32\linewidth]{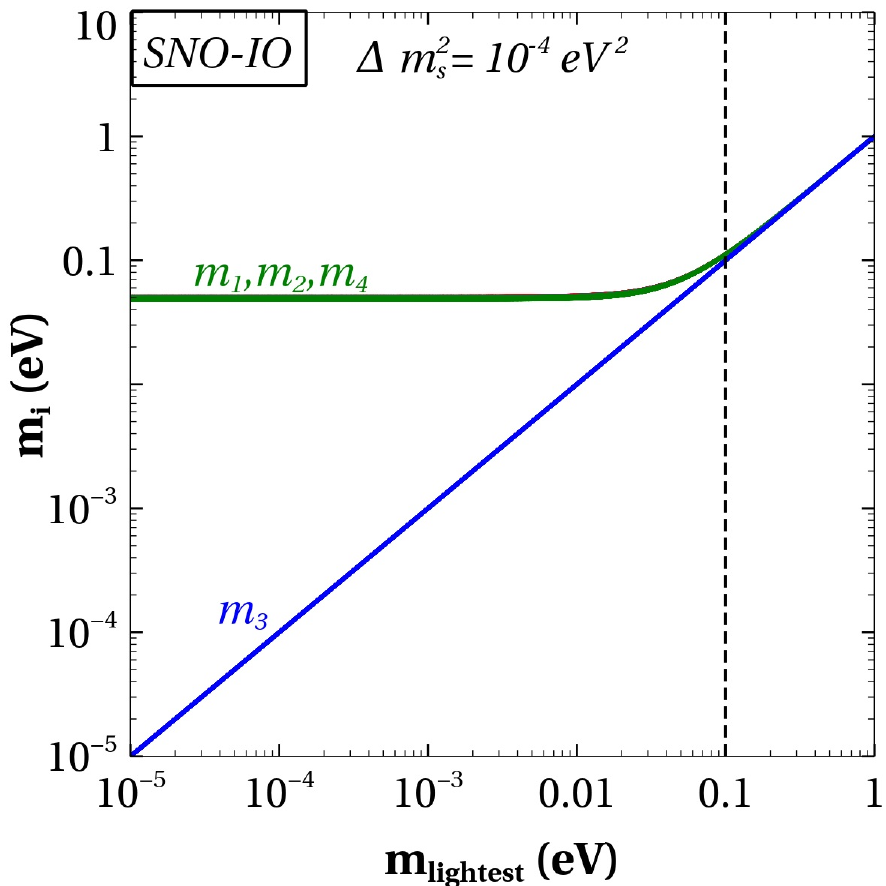}
    \includegraphics[width=0.32\linewidth]{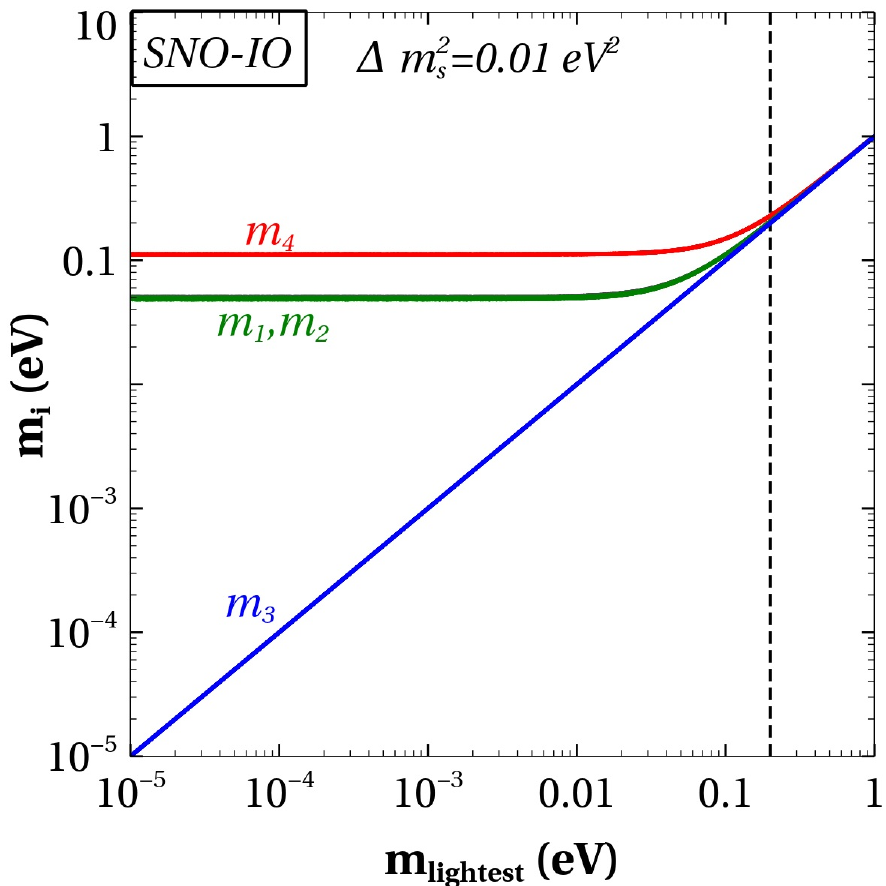}
   \includegraphics[width=0.32\linewidth]{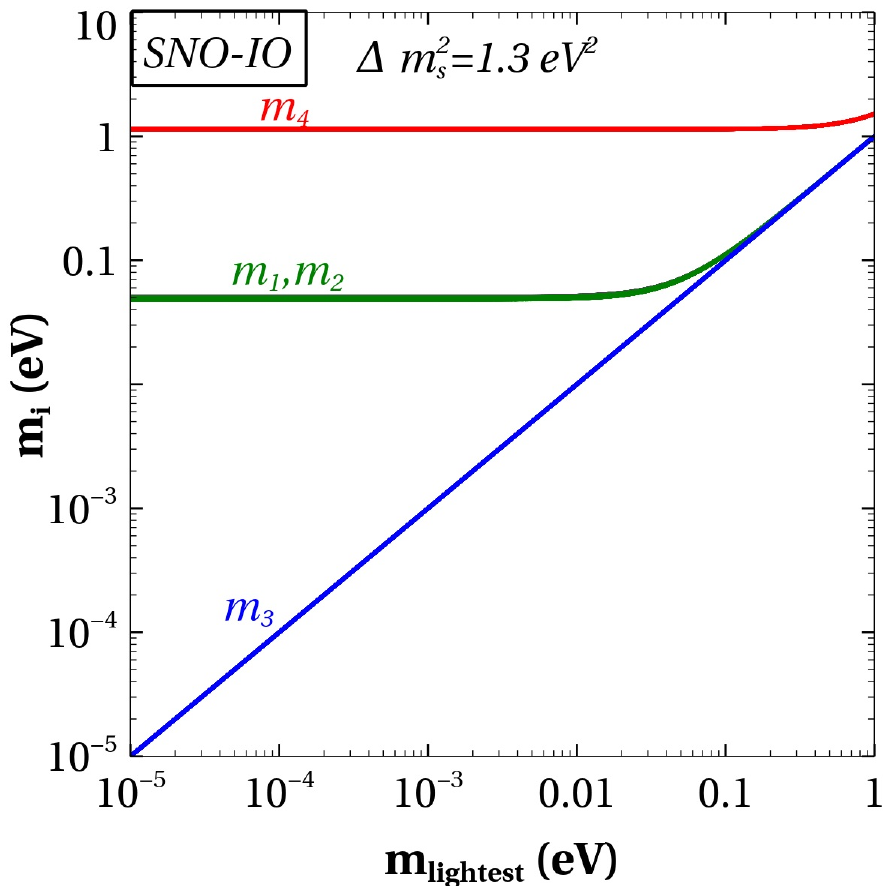}
    \caption{variation of masses with respect to the lightest neutrino mass for SNO-IO}
    \label{fig:m_SNOIO}
\end{figure}

\begin{figure}[H]
    \centering
     \includegraphics[width=0.32\linewidth]{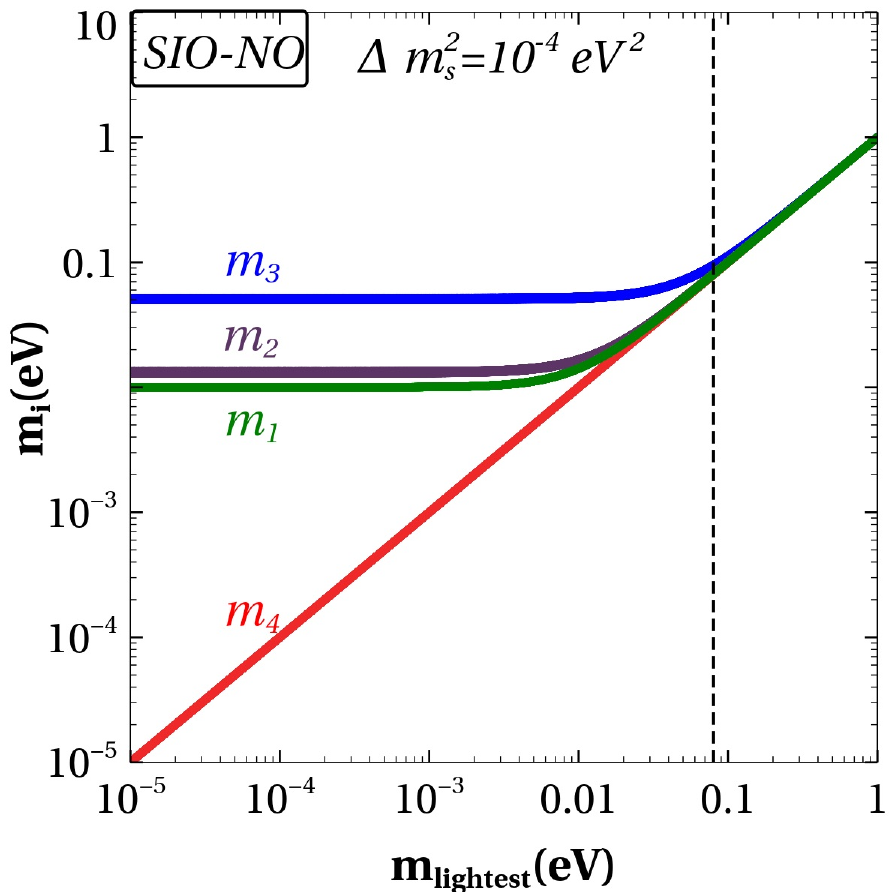}
    \includegraphics[width=0.32\linewidth]{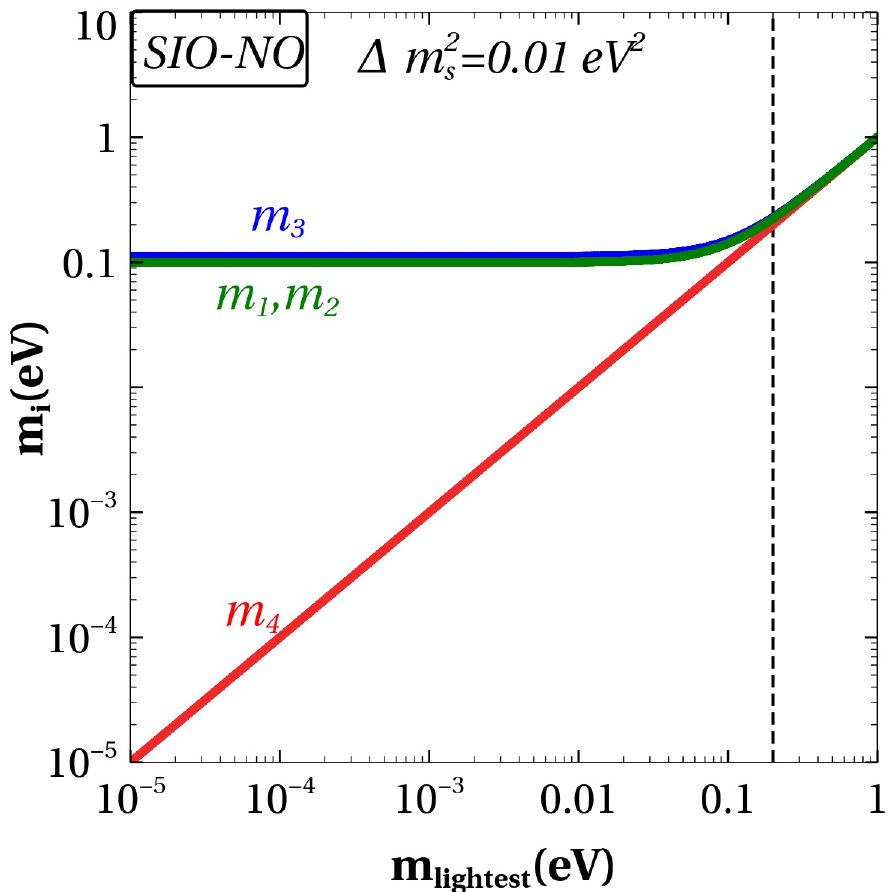}
   \includegraphics[width=0.32\linewidth]{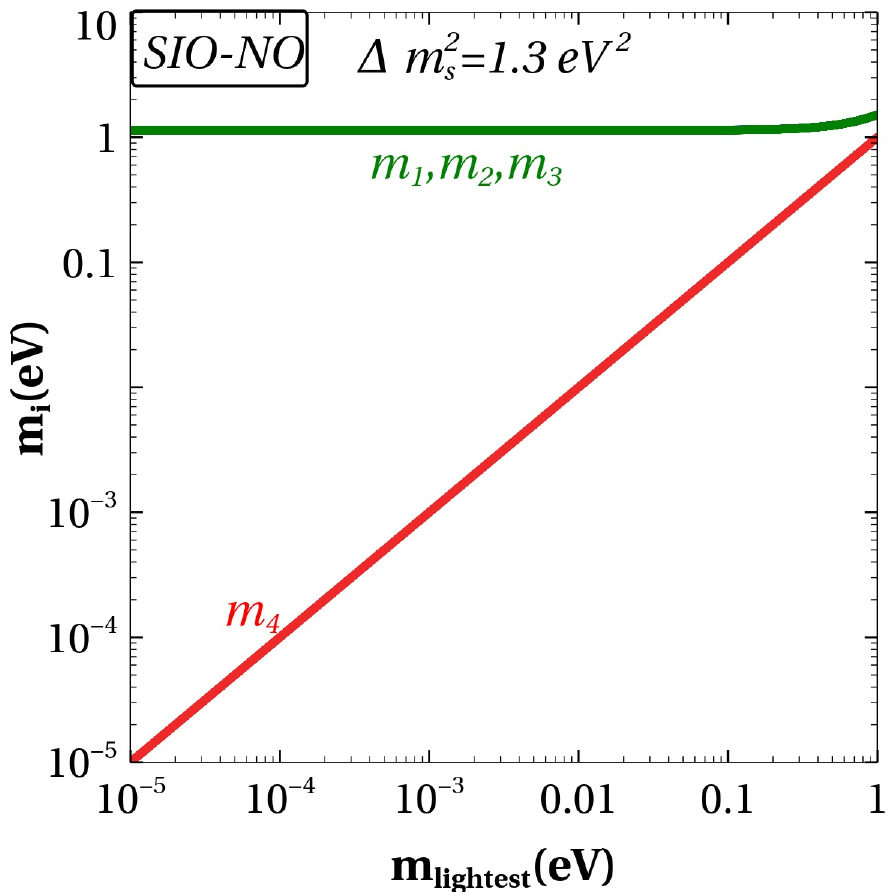}
    \caption{variation of masses with respect to the lightest neutrino mass for SIO-NO}
    \label{fig:m_SIONO}
\end{figure}

\begin{figure}[H]
    \centering
     \includegraphics[width=0.32\linewidth]{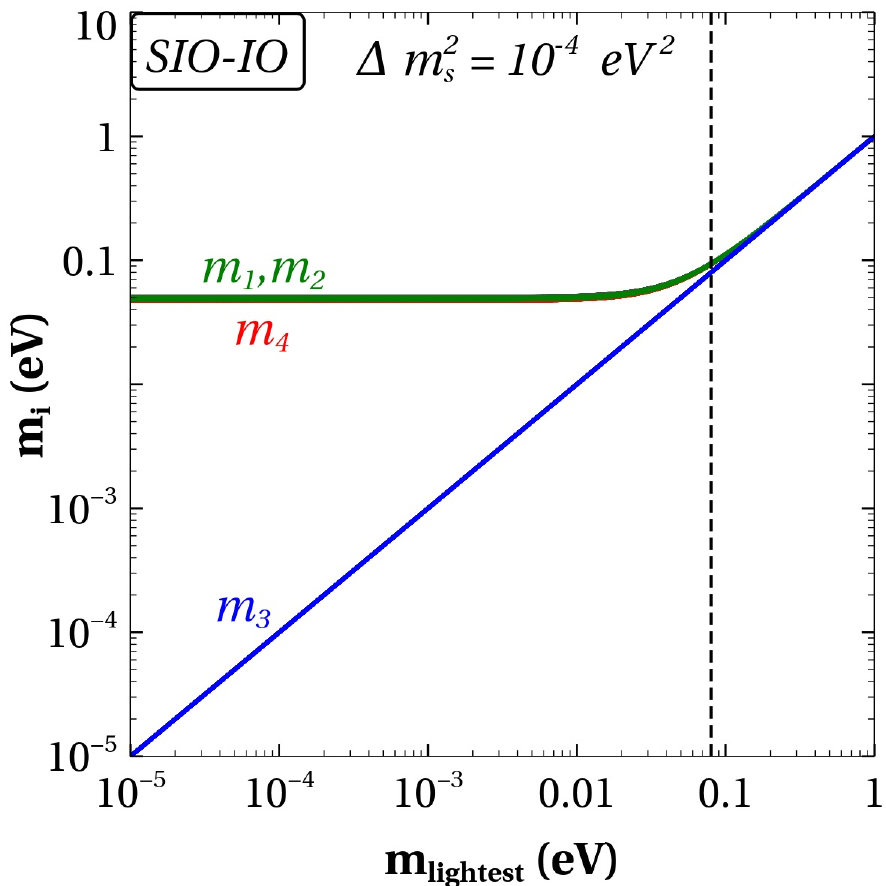}
    \includegraphics[width=0.32\linewidth]{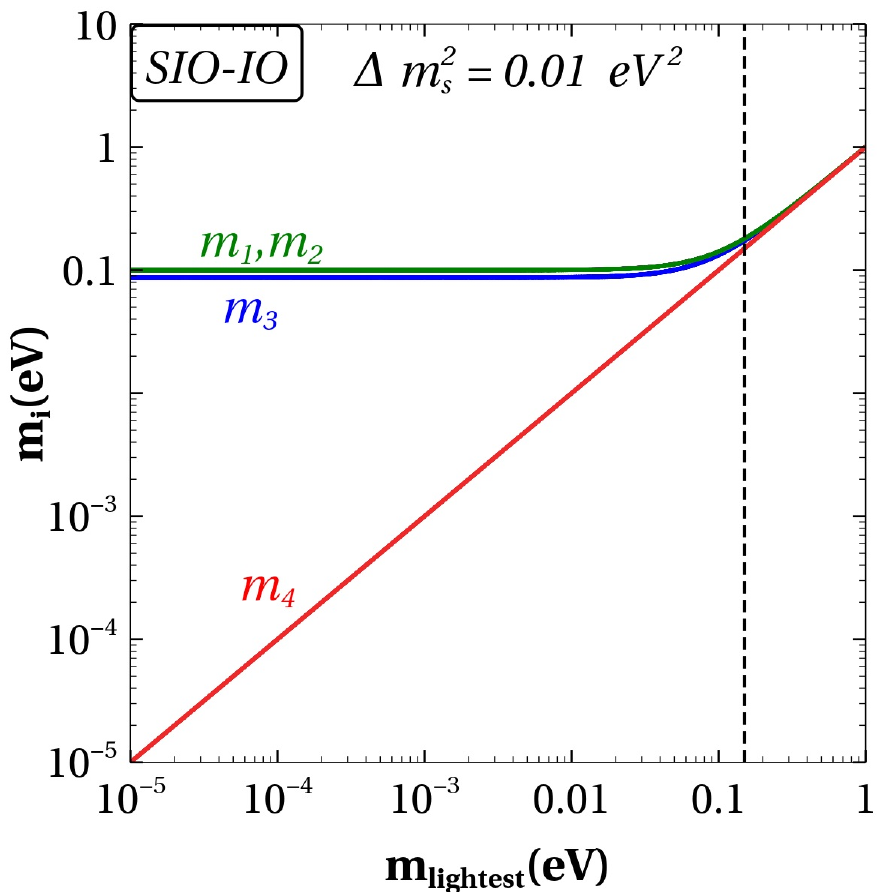}
   \includegraphics[width=0.32\linewidth]{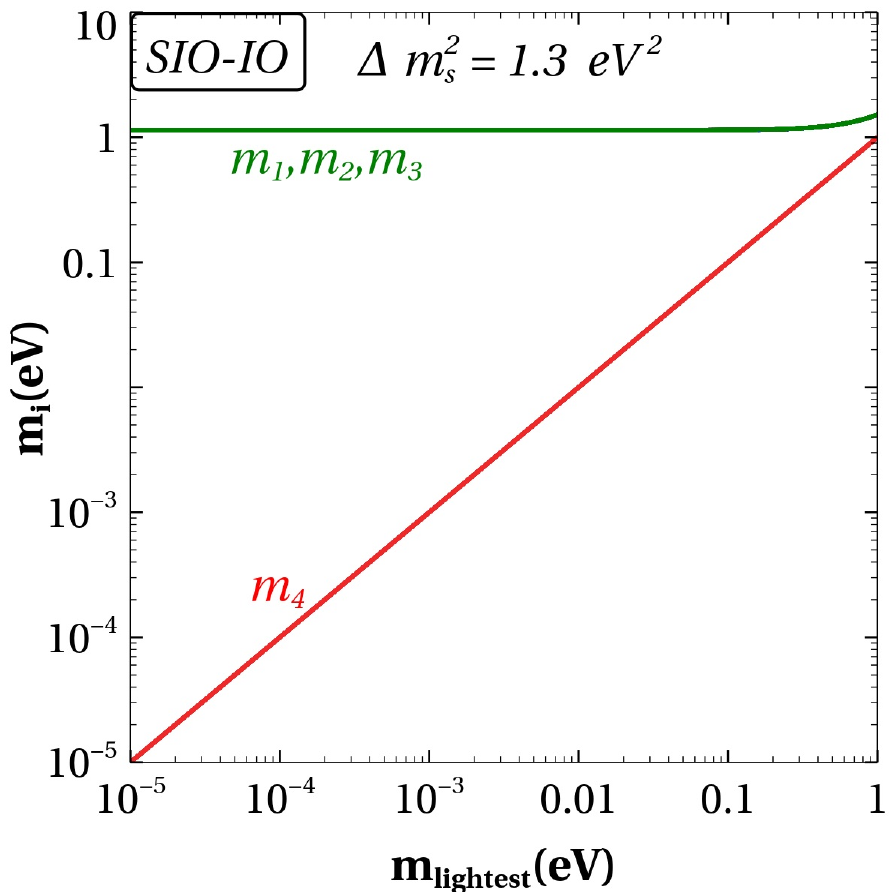}
    \caption{variation of masses with respect to the lightest neutrino mass for SIO-IO}
    \label{fig:m_SIOIO}
\end{figure}
\begin{sidewaystable}[h]
    \centering
    \begin{tabular}{|p{1.5 cm}|c|c|c|c|}
        \hline
        Mass Spectra & Region & \multicolumn{3}{c|}{Expression of $m_{\beta}^2$}\\
         \cline{3-5} 
         & &  (I) $\ms^2=10^{-4}\,\ev^2$ & (II) $\ms^2=10^{-2}\, \ev^2$ & (III) $\ms^2=1.3\, \ev^2$\\
        \hline \hline
        \multirow{4}{*}{SNO-NO} & $m_{\rm lightest} << \sqrt{\msol^2} << \sqrt{\matm^2}$ & $\msol^2 |U_{e2}|^2  + \matm^2 |U_{e3}|^2 + \ms^2 |U_{e4}|^2 $ & same as (I) & $\ms^2 |U_{e4}|^2$  \\
        \cline{2-5} & $\sqrt{\msol^2} << m_{\rm lightest} << \sqrt{\matm^2}$ & $m_{\rm lightest}^2$. & same as (I) &  $\ms^2 |U_{e4}|^2 $\\
        \hline \hline
            \multirow{3}{*}{SNO-IO} & $m_{\rm lightest} << \sqrt{\msol^2} << \sqrt{\matm^2}$ & $\matm^2 $ & $\matm^2 $ & $\matm^2 + \ms^2 \, |U_{e4}|^2 $ \\
        \cline{2-5} & $ \sqrt{\matm^2} \lesssim m_{\rm lightest} $ & $ m_{\rm lightest}^2 $ & $ m_{\rm lightest}^2 $ & $ m_{\rm lightest}^2 + \ms^2 \, |U_{e4}|^2 $ \\
        \cline{2-5} & $\sqrt{\matm^2} << m_{\rm lightest} $ & $ m_{\rm lightest}^2 $ & $ m_{\rm lightest}^2 $ & $ m_3^2 $\\
        \hline \hline
        \multirow{3}{*}{SIO-NO} & $m_{\rm lightest} << \sqrt{\msol^2} << \ms^2 $ & $\ms^2$ & $ \ms^2 $ &  $ \ms^2 $ \\
        \cline{2-5} &  $ \sqrt{\ms^2} << m_{\rm lightest} $ & $m_{\rm lightest}^2$ & $ m_{\rm lightest}^2 $ &  $ m_{\rm lightest^2} $ \\
        \hline \hline
        \multirow{2}{*}{SIO-IO} & $ m_{\rm lightest} << \sqrt{\matm^2}$ & same as SNO-IO & same as SIO-NO & same as SIO-NO\\
        \cline{2-5} &  $ m_s^2 << \sqrt{\matm^2} << m_{\rm lightest}$  & $m_{\rm lightest}^2 $ & N.A & N.A.\\
        \cline{2-5} &  $ \sqrt{\matm^2} << m_s^2 << m_{\rm lightest}$  & N.A. & $ m_{\rm lightest}$ &  $m_{\rm lightest}$\\
        \hline
    \end{tabular}
    \caption{The table depicts expressions of $m_{\beta}^{2}$ in third, fourth, and fifth columns corresponding to $\ms^2=10^{-4}, \, 10^{-2}, \, 1.3\, \ev^2$ respectively for various mass spectra given in the first column under different regions shown in column two.}
    \label{tab:mbeta2}
\end{sidewaystable}

\begin{sidewaystable}[h]
    \centering
    \begin{tabular}{|p{1.5cm}|p{2cm}|c|c|c|}
        \hline
        Mass Spectra & Region & \multicolumn{3}{c|}{Expression of $m_{\beta\beta}$}\\
         \cline{3-5} 
         & &  (I) $\ms^2=10^{-4}\,\ev^2$ & (II) $\ms^2=10^{-2}\, \ev^2$ & (III) $\ms^2=1.3\, \ev^2$\\
        \hline \hline
        \multirow{4}{*}{SNO-NO} & $m_1\approx 0$ & $\sqrt{\msol^2}U_{e2}^2 e^{i\alpha} + \sqrt{\matm^2}U_{e3}^2 e^{i\beta}+ \sqrt{\ms^2}U_{e4}^1 e^{i\gamma}$ & $\sqrt{\msol^2}U_{e2}^2 e^{i\alpha} + \sqrt{\matm^2}U_{e3}^2 e^{i\beta}$ & same as (I) \\
        \cline{2-5} & $m_1\approx m_2 \approx m_3$ & N.A. & $m_1\left[U_{e1}^2 + U_{e2}^2 e^{i\alpha} \right]$ & same as (II)\\
        \cline{2-5} & $m_1 \approx m_2 \approx m_4$ & $m_1\left[U_{e1}^2 + U_{e2}^2 e^{i\alpha} + U_{e4}^2 e^{i\gamma} \right]$ & N.A. & N.A. \\
        \cline{2-5} & $m_1\approx m_2 \approx m_3 \approx m_4$ & $m_1\left[U_{e1}^2 + U_{e2}^2 e^{i\alpha} + U_{e4}^2 e^{i\gamma} \right]$ & $m_1\left[U_{e1}^2 + U_{e2}^2 e^{i\alpha} \right]$ & N.A.\\
        \hline \hline
        \multirow{3}{*}{SNO-IO} & $m_3 \approx 0$ & $\sqrt{\matm^2} \left[ U_{e1}^2 + U_{e2}^2 e^{i\alpha} + U_{e4}^2 e^{i\gamma} \right] $ & $\sqrt{\matm^2} \left[ U_{e1}^2 + U_{e2}^2 e^{i\alpha} \right]$ & $\sqrt{\matm^2} \left[ U_{e1}^2 + U_{e2}^2 e^{i\alpha} \right] + \sqrt{\ms^2} U_{e4}^2 e^{i\gamma}$ \\
        \cline{2-5} & $m_1 \approx m_2 \approx m_3$ & $m_3 \left[  U_{e1}^2 + U_{e2}^2 e^{i\alpha} + U_{e4}^2 e^{i\gamma} \right]$ & $m_3 \left[  U_{e1}^2 + U_{e2}^2 e^{i\alpha} \right]$ & same as (II)\\
        \cline{2-5} & $m_1 \approx m_2 \approx m_3 \approx m_4$ & $m_3 \left[  U_{e1}^2 + U_{e2}^2 e^{i\alpha} + U_{e4}^2 e^{i\gamma} \right] $ & $m_3 \left[  U_{e1}^2 + U_{e2}^2 e^{i\alpha} \right]$ & N.A.\\
        \hline \hline
        \multirow{3}{*}{SIO-NO} & $m_4 \approx 0$ & $\sqrt{\ms^2} \left[  U_{e1}^2 + U_{e2}^2 e^{i\alpha} \right] + \sqrt{\matm^2} U_{e3}^2 e^{i\beta}$ & $ \sqrt{\ms^2} \left[  U_{e1}^2 + U_{e2}^2 e^{i\alpha} \right] $ & same as (II)\\
        \cline{2-5} & $m_1 \approx m_2 \approx m_4$ & $m_4 \left[ U_{e1}^2 + U_{e2}^2 e^{i\alpha} +  U_{e4}^2 e^{i\gamma} \right]$ & $m_4 \left[ U_{e1}^2 + U_{e2}^2 e^{i\alpha} \right]$ & N.A.\\
        \cline{2-5} & $m_1 \approx m_2 \approx m_3 \approx m_4$ &  $m_4 \left[ U_{e1}^2 + U_{e2}^2 e^{i\alpha} +  U_{e4}^2 e^{i\gamma} \right]$ & $m_4 \left[ U_{e1}^2 + U_{e2}^2 e^{i\alpha} \right]$ & N.A.\\
        \hline \hline
        \multirow{2}{*}{SIO-IO} & $m_{\rm lightest} \approx 0$ & $\sqrt{\matm^2} \left[ U_{e1}^2 + U_{e2}^2 e^{i\alpha} +  U_{e4}^2 e^{i\gamma} \right]$ & $ \sqrt{\ms^2} \left[ U_{e1}^2 + U_{e2}^2 e^{i\alpha} \right] $ & same as (II)\\
        \cline{2-5} & $m_1 \approx m_2 \approx m_3 \approx m_4$ & $m_3 \left[ U_{e1}^2 + U_{e2}^2 e^{i\alpha} +  U_{e4}^2 e^{i\gamma} \right]$ & $m_4 \left[ U_{e1}^2 + U_{e2}^2 e^{i\alpha} \right]$ & N.A.\\
        \hline
    \end{tabular}
    \caption{The table depicts expressions of $m_{\beta\beta}$ in third, fourth, and fifth columns corresponding to $\ms^2=10^{-4}, \, 10^{-2}, \, 1.3\, \ev^2$ respectively for various mass spectra given in the first column under different regions shown in column two.}
    \label{tab:mbetambeta}
\end{sidewaystable}
\pagebreak
\bibliographystyle{unsrt}
\bibliography{reference}

\begin{thebibliography}{10}

\bibitem{Super-Kamiokande:1998kpq}
Y.~Fukuda et~al.
\newblock {Evidence for oscillation of atmospheric neutrinos}.
\newblock {\em Phys. Rev. Lett.}, 81:1562--1567, 1998.

\bibitem{SNO:2002tuh}
Q.~R. Ahmad et~al.
\newblock {Direct evidence for neutrino flavor transformation from neutral
  current interactions in the Sudbury Neutrino Observatory}.
\newblock {\em Phys. Rev. Lett.}, 89:011301, 2002.

\bibitem{KamLAND:2002uet}
K.~Eguchi et~al.
\newblock {First results from KamLAND: Evidence for reactor anti-neutrino
  disappearance}.
\newblock {\em Phys. Rev. Lett.}, 90:021802, 2003.

\bibitem{MINOS:2006foh}
D.~G. Michael et~al.
\newblock {Observation of muon neutrino disappearance with the MINOS detectors
  and the NuMI neutrino beam}.
\newblock {\em Phys. Rev. Lett.}, 97:191801, 2006.

\bibitem{Weinberg:1979sa}
Steven Weinberg.
\newblock {Baryon and Lepton Nonconserving Processes}.
\newblock {\em Phys. Rev. Lett.}, 43:1566--1570, 1979.

\bibitem{Furry:1939qr}
W.~H. Furry.
\newblock {On transition probabilities in double beta-disintegration}.
\newblock {\em Phys. Rev.}, 56:1184--1193, 1939.

\bibitem{Shirai:2018ycl}
Junpei Shirai.
\newblock {KamLAND-Zen experiment}.
\newblock {\em PoS}, HQL2018:050, 2018.

\bibitem{GERDA}
M.~Agostini et~al.
\newblock {Final Results of GERDA on the Search for Neutrinoless Double-$\beta$
  Decay}.
\newblock {\em Phys. Rev. Lett.}, 125(25):252502, 2020.

\bibitem{KATRIN:2021uub}
M.~Aker et~al.
\newblock {Direct neutrino-mass measurement with sub-electronvolt sensitivity}.
\newblock {\em Nature Phys.}, 18(2):160--166, 2022.

\bibitem{Planck:2018vyg}
N.~Aghanim et~al.
\newblock {Planck 2018 results. VI. Cosmological parameters}.
\newblock {\em Astron. Astrophys.}, 641:A6, 2020.
\newblock [Erratum: Astron.Astrophys. 652, C4 (2021)].

\bibitem{LSND:2001aii}
A.~Aguilar et~al.
\newblock {Evidence for neutrino oscillations from the observation of
  $\bar{\nu}_e$ appearance in a $\bar{\nu}_\mu$ beam}.
\newblock {\em Phys. Rev. D}, 64:112007, 2001.

\bibitem{MiniBooNE:2020pnu}
A.~A. Aguilar-Arevalo et~al.
\newblock {Updated MiniBooNE neutrino oscillation results with increased data
  and new background studies}.
\newblock {\em Phys. Rev. D}, 103(5):052002, 2021.

\bibitem{GALLEX:1997lja}
W.~Hampel et~al.
\newblock {Final results of the Cr-51 neutrino source experiments in GALLEX}.
\newblock {\em Phys. Lett. B}, 420:114--126, 1998.

\bibitem{Abdurashitov:1996dp}
Dzh.~N. Abdurashitov et~al.
\newblock {The Russian-American gallium experiment (SAGE) Cr neutrino source
  measurement}.
\newblock {\em Phys. Rev. Lett.}, 77:4708--4711, 1996.

\bibitem{Barinov:2021asz}
V.~V. Barinov et~al.
\newblock {Results from the Baksan Experiment on Sterile Transitions (BEST)}.
\newblock {\em Phys. Rev. Lett.}, 128(23):232501, 2022.

\bibitem{deHolanda:2010am}
P.~C. de~Holanda and A.~Yu. Smirnov.
\newblock {Solar neutrino spectrum, sterile neutrinos and additional radiation
  in the Universe}.
\newblock {\em Phys. Rev. D}, 83:113011, 2011.

\bibitem{KumarAgarwalla:2019blx}
Sanjib Kumar~Agarwalla, Sabya~Sachi Chatterjee, and Antonio Palazzo.
\newblock {Physics potential of ESS$\nu$SB in the presence of a light sterile
  neutrino}.
\newblock {\em JHEP}, 12:174, 2019.

\bibitem{Agarwalla:2018nlx}
Sanjib~Kumar Agarwalla, Sabya~Sachi Chatterjee, and Antonio Palazzo.
\newblock {Signatures of a Light Sterile Neutrino in T2HK}.
\newblock {\em JHEP}, 04:091, 2018.

\bibitem{Chatterjee:2023qyr}
Animesh Chatterjee, Srubabati Goswami, and Supriya Pan.
\newblock {Probing mass orderings in presence of a very light sterile neutrino
  in a liquid argon detector}.
\newblock {\em Nucl. Phys. B}, 996:116370, 2023.

\bibitem{Chatterjee:2022pqg}
Animesh Chatterjee, Srubabati Goswami, and Supriya Pan.
\newblock {Matter effect in presence of a sterile neutrino and resolution of
  the octant degeneracy using a liquid argon detector}.
\newblock {\em Phys. Rev. D}, 108(9):095050, 2023.

\bibitem{Goswami:2005ng}
Srubabati Goswami and Werner Rodejohann.
\newblock {Constraining mass spectra with sterile neutrinos from neutrinoless
  double beta decay, tritium beta decay and cosmology}.
\newblock {\em Phys. Rev. D}, 73:113003, 2006.

\bibitem{Maki:1962mu}
Ziro Maki, Masami Nakagawa, and Shoichi Sakata.
\newblock {Remarks on the unified model of elementary particles}.
\newblock {\em Prog. Theor. Phys.}, 28:870--880, 1962.

\bibitem{Esteban:2020cvm}
Ivan Esteban, M.~C. Gonzalez-Garcia, Michele Maltoni, Thomas Schwetz, and
  Albert Zhou.
\newblock {The fate of hints: updated global analysis of three-flavor neutrino
  oscillations}.
\newblock {\em JHEP}, 09:178, 2020.

\bibitem{MINOS:2017cae}
P.~Adamson et~al.
\newblock {Search for sterile neutrinos in MINOS and MINOS+ using a
  two-detector fit}.
\newblock {\em Phys. Rev. Lett.}, 122(9):091803, 2019.

\bibitem{Acero:2022wqg}
M.~A. Acero et~al.
\newblock {White Paper on Light Sterile Neutrino Searches and Related
  Phenomenology}.
\newblock 3 2022.

\bibitem{Bennett:2020zkv}
Jack~J. Bennett, Gilles Buldgen, Pablo~F. De~Salas, Marco Drewes, Stefano
  Gariazzo, Sergio Pastor, and Yvonne Y.~Y. Wong.
\newblock {Towards a precision calculation of $N_{\rm eff}$ in the Standard
  Model II: Neutrino decoupling in the presence of flavour oscillations and
  finite-temperature QED}.
\newblock {\em JCAP}, 04:073, 2021.

\bibitem{Yaguna:2007wi}
Carlos~E. Yaguna.
\newblock {Sterile neutrino production in models with low reheating
  temperatures}.
\newblock {\em JHEP}, 06:002, 2007.

\bibitem{Abazajian:2017tcc}
Kevork~N. Abazajian.
\newblock {Sterile neutrinos in cosmology}.
\newblock {\em Phys. Rept.}, 711-712:1--28, 2017.

\bibitem{Dasgupta:2013zpn}
Basudeb Dasgupta and Joachim Kopp.
\newblock {Cosmologically Safe eV-Scale Sterile Neutrinos and Improved Dark
  Matter Structure}.
\newblock {\em Phys. Rev. Lett.}, 112(3):031803, 2014.

\bibitem{Chu:2015ipa}
Xiaoyong Chu, Basudeb Dasgupta, and Joachim Kopp.
\newblock {Sterile neutrinos with secret interactions\textemdash{}lasting
  friendship with cosmology}.
\newblock {\em JCAP}, 10:011, 2015.

\bibitem{Riess:2018uxu}
Adam~G. Riess et~al.
\newblock {New Parallaxes of Galactic Cepheids from Spatially Scanning the
  Hubble Space Telescope: Implications for the Hubble Constant}.
\newblock {\em Astrophys. J.}, 855(2):136, 2018.

\bibitem{Pan-STARRS1:2017jku}
D.~M. Scolnic et~al.
\newblock {The Complete Light-curve Sample of Spectroscopically Confirmed SNe
  Ia from Pan-STARRS1 and Cosmological Constraints from the Combined Pantheon
  Sample}.
\newblock {\em Astrophys. J.}, 859(2):101, 2018.

\bibitem{Hagstotz:2020ukm}
Steffen Hagstotz, Pablo~F. de~Salas, Stefano Gariazzo, Martina Gerbino,
  Massimiliano Lattanzi, Sunny Vagnozzi, Katherine Freese, and Sergio Pastor.
\newblock {Bounds on light sterile neutrino mass and mixing from cosmology and
  laboratory searches}.
\newblock {\em Phys. Rev. D}, 104(12):123524, 2021.

\bibitem{Project8:2022wqh}
A.~Ashtari Esfahani et~al.
\newblock {The Project 8 Neutrino Mass Experiment}.
\newblock In {\em {Snowmass 2021}}, 3 2022.

\bibitem{Cirigliano:2018hja}
Vincenzo Cirigliano, Wouter Dekens, Jordy De~Vries, Michael~L. Graesser,
  Emanuele Mereghetti, Saori Pastore, and Ubirajara Van~Kolck.
\newblock {New Leading Contribution to Neutrinoless Double-\ensuremath{\beta}
  Decay}.
\newblock {\em Phys. Rev. Lett.}, 120(20):202001, 2018.

\bibitem{Cirigliano:2019vdj}
V.~Cirigliano, W.~Dekens, J.~De~Vries, M.~L. Graesser, E.~Mereghetti,
  S.~Pastore, M.~Piarulli, U.~Van~Kolck, and R.~B. Wiringa.
\newblock {Renormalized approach to neutrinoless double- $\beta$ decay}.
\newblock {\em Phys. Rev. C}, 100(5):055504, 2019.

\bibitem{Scholer:2023bnn}
Oliver Scholer, Jordy de~Vries, and Luk\'a\v{s} Gr\'af.
\newblock {\ensuremath{\nu}DoBe \textemdash{} A Python tool for neutrinoless
  double beta decay}.
\newblock {\em JHEP}, 08:043, 2023.

\end{thebibliography}

\end{document}